**Unsteady Thin-Airfoil Theory Revisited: An Approximate Analytical Solution and the Self-Similar Wagner Effect in Viscous Flows**

Tianshu Liu[1a], Jianfeng Lin[2], Shizhao Wang[2]

1: Department of Mechanical and Aerospace Engineering

Western Michigan University, Kalamazoo, MI 49008, USA

2: LNM, Institute of Mechanics

Chinese Academy of Sciences, Beijing, China

(09/02/2026)

a: The corresponding author, tianshu.liu@wmich.edu, 269-276-3426

**Abstract**

An approximate analytical solution for the unsteady lift of a thin airfoil with a general unsteady motion is derived from a viscous-flow perspective, where the wake vortex-sheet strength is given in an explicit convolution-type expression as an approximate solution of the Wagner integral equation. For validation, this analytical solution is applied to the Wagner and Theodorsen problems, giving the explicit integral forms of the Wagner and Theodorsen functions as the reduced cases. Further, this analytical solution is applied to the starting flow with a finite timescale in the generalized Wagner problem, revealing the self-similarity of the re-normalized circulatory lift coefficient and its equivalence to the re-normalized Wagner function in a finite time domain. More importantly, the self-similar Wagner effect is found in numerical simulation of the flow over a starting flat-plate airfoil at low Reynolds numbers even when the flow is moderately separated. This self-similarity represents the Reynolds-number-invariance.

## 1. Introduction

In classical aerodynamics, the thin-airfoil theory (TAT) was developed to predict the lift of an airfoil in a steady flow, where a vortex sheet is used as an inviscid limiting model of the flow on a thin airfoil imposed with the Kutta condition at the trailing edge to model the viscous-flow effect over the airfoil (Munk 1923, 1925; Glauert 1926; Lighthill 1951). The relationship between the boundary-layer vorticity and the vortex-sheet strength was discussed by Sears (1956, 1976). The unsteady thin-airfoil theory (UTAT) is a natural extension of the TAT to unsteady flows, which was first developed by Wagner (1925), Theodorsen (1935), Küssner (1936), Garrick (1938), Jones (1938), and von Kármán and Sears (1938). The UTAT has been extended and applied to unsteady flows associated with aeroelastic, flapping and flexible wings (Leishman 1988, 2006; McCune & Tavares 1993; Pullin & Wang 2004; Peters 2008). To elucidate the physical foundation of the UTAT, Liu et al. (2015) re-formulated the UTAT from the viscous-flow lift formula with the Lamb vector term (the vortex force) and the fluid acceleration term. In unsteady aerodynamics, the Wagner problem and the Theodorsen problem are two canonical cases, which have influenced the later development of theoretical and numerical methods in the field.

Wagner (1925) studied the response of the lift of an impulsively starting flat-plate airfoil at a small angle of attack (AoA) (a step change of the effective AoA), which is referred to as the Wagner problem. Wagner gave an integral equation for the wake vortex-sheet strength and presented the Wagner function as a solution of the integral equation to describe the monotonically-increasing response of the unsteady lift known as the Wagner effect. However, the Wagner function cannot be given by an explicit analytical form (the tabularized data were originally given by Wagner). Sears (1940) gave the asymptotic solutions of the Wagner function for small and large normalized times evaluating some contour integrals in the inverse Carson transform. Küssner (1938) gave a slowly convergent series expression of the Wagner function. Peters (2008) evaluated the Wagner function using the inverse Fourier transform of the Theodorsen function that was first introduced by Garrick (1938). The Wagner function can be calculated by using the numerical inverse Laplace transform (Dawson & Brunton 2022). To simplify the unsteady lift analysis, the Wagner function was approximated by the correlations (Kármán & Sears 1938; Garrick 1938; Jones 1938). The multiple-exponential approximation originally proposed by Jones (1938) is particularly interesting, since it inspires the dynamical system modeling of the Wagner function (Dawson & Brunton 2022). Furthermore, the Wagner function was examined experimentally (Beckwith & Babinsky 2009; Babinsky et al. 2016; Stevens et al. 2017; Goyal & Nedić 2025). The Wagner problem for a range of AoAs has been studied as a testing case for evaluating numerical methods (Ford & Babinsky 2013; Xia & Mohseni 2013, 2017; Li & Wu 2015; Bai et al. 2019; DeVoria & Mohseni 2021; Pohly & Kang 2023).

Theodorsen (1935) studied a harmonically oscillating flat-plate airfoil in a constant uniform incoming flow or a stationary airfoil in an incoming flow with a harmonically oscillating vertical velocity. The circulatory lift is characterized by the Theodorsen function, a complex function of the reduced frequency expressed as a quotient of the Hankel functions. The Theodorsen function in the frequency domain is related to the Wagner function in the time domain by a pair of the Fourier transforms (Garrick 1938). In other words, the Theodorsen function is a representation of the Wagner effect in the frequency domain. In many applications, the Theodorsen lift formula is used since the unsteady lift is given by an explicit frequency-domain expression with the kinematic parameters such as the heaving amplitude, pitching angle, and their time derivatives (Feng & Wang 2024; Zhao, Li & Yang 2025).

For a general unsteady motion of a thin airfoil, the time-space formulation based on the indicial function (a multiple-exponential approximation of the Wagner function) was proposed by Garrick (1938) as a practical method for engineering applications. When the unsteady aerodynamic model is treated as a general linear system, the unsteady lift is given by the convolution of the indicial function and the time derivative of the vertical velocity in the spirit of Duhamel's principle. The indicial response theory has been extended and used in studies of aeroelasticity and wing engineering (Leishman 1988, 2006; Bisplinghoff, Ashley & Halfman 1996; Hansen et al. 2004; Bergami et al. 2013; Jones et al. 2022). Essentially, the application of Duhamel's principle provides a method linking the time-dependent AoA (or the normal velocity perturbation) as an input and the unsteady lift as an output through a convolution formulation with the Wagner function as a Green's function.

Kármán and Sears (1938) derived a thin-airfoil lift formula that has the three distinct terms: the quasi-steady lift, the added-mass lift, and the lift associated with the wake effect. The unsteady lift was directly calculated by taking the time derivative of the vortex impulse (the total momentum induced by the vortices), leading to the more general and physically elucidating result. The vortex-impulse force formulation is generally applicable to complex viscous flows, including the leading- and trailing-edge vortices (Wu et al. 2006, 2018). In this sense, the Kármán-Sears lift formula is applicable to a range of flows, including moderately separated flows (McCune & Tavares 1993; Liu et al. 2015). In contrast to the theories of Wagner and Theodorsen, the Kármán-Sears lift formula has not been widely applied to engineering problems, although it is in a concise explicit form of the three terms with the clear physical meanings. To resolve this issue, an explicit analytical expression for the wake vortex-sheet strength should be given as a solution of the Wagner integral equation. Further, a general question is whether the classical results of the UTAT such as the Wagner and Theodorsen functions are applicable to viscous flows particularly at low Reynolds numbers. These topics will be discussed in this paper.

The objectives of this work are twofold. Firstly, an approximate analytical solution of the Wagner integral equation is sought for the wake vortex-sheet strength to calculate the lift associated with the wake effect. The Kármán-Sears lift formula combined with this solution gives the unsteady lift of a thin airfoil with a general motion mode, including the quasi-steady lift, the added-mass lift, and the lift associated with the wake effect. The explicit integral forms of the Wagner and the Theodorsen functions are given as the reduced cases of the analytical solution. Secondly, this solution is applied to the starting flow with a finite rising timescale in the generalized Wagner problem, revealing the self-similarity of the re-normalized lift coefficient in a range of the rising timescales and its equivalence to the re-normalized Wagner function. Essentially, this self-similarity represents the Reynolds-number-invariance, which is examined by numerical simulation for the Wagner problem at low Reynolds numbers to elucidate the applicability of the Wagner function in viscous flows.

The paper is organized as follows. In Section 2, the thin-airfoil lift formula is given in the viscous-flow framework, and the Kármán-Sears lift formula is recovered as a reduced case where the wake effect on the lift is modeled. In Section 3, the explicit convolution-type expression of the wake vortex-sheet strength is obtained as a solution of the Wagner integral where the kernel is approximated and then the unsteady lift associated with the wake effect is calculated. In Section 4, the analytical solution is applied to the classical Wagner problem, where an explicit integral expression of the Wagner function is given. Further, this solution is applied to the starting flow with a finite rising timescale in the generalized Wagner problem, revealing the self-similarity of the re-normalized lift coefficient in a range of the rising timescale and its equivalence to the re-normalized Wagner function. In Section 5, the Reynolds-number-invariance of the re-normalized vortex lift coefficient is examined by numerical simulation of the flow over a starting flat-plate airfoil at $\alpha = 5^o$ and $\alpha = 10^o$ in a range of the Reynolds numbers ($Re = 126\text{-}1000$). The power-law relation between the rising timescale and the Reynolds number is given. The numerical results elucidate the self-similar Wagner effect in viscous flows when the flow separation is moderate at small AoAs. Finally, the conclusions are made. In Appendix A, the analytical solution is applied to the generalized Theodorsen problem where the general form of the effective AoA is expressed in a Fourier series. The explicit integral form of the Theodorsen function is given as a reduced case. Furthermore, this formal solution is expressed in a convolution-like form in the time domain that resembles the indicial response formulation. In Appendix B, the computational setup and code validation are described.

## 2. Thin-Airfoil Lift Formula

For an incompressible viscous flow, when a rectangular control surface in which an airfoil is enclosed is sufficiently large, the two-term lift formula, which was derived by Wang et al. (2013) from the Navier-Stokes equations, is expressed as

$$L = \rho \boldsymbol{k} \cdot \int_{V_f} \boldsymbol{u} \times \boldsymbol{\omega} \, dV - \rho \boldsymbol{k} \cdot \frac{d}{dt} \int_{V_f} \boldsymbol{u} \, dV \equiv L_{vor} + L_a , \tag{1}$$

where $\boldsymbol{\omega}$ is the vorticity, $-\boldsymbol{u} \times \boldsymbol{\omega}$ is the Lamb vector, $\rho$ is the fluid density, $t$ is time, $V_f$ denotes a rectangular fluid control volume, and $\boldsymbol{k}$ is the unit vector normal to the freestream velocity. In Eq. (1), the first term $L_{vor}$ is the vortex lift, and the second term $L_a$ is the lift associated with the fluid acceleration. In a limiting case where a moving body is in a completely inviscid irrotational flow, $L_a$ is interpreted as the added-mass force projected on the direction of $\boldsymbol{k}$. It has been demonstrated that the two-term lift formula is sufficiently accurate for complex unsteady viscous flows generated by flapping wings (Wang et al. 2013, 2014, 2015a, 2015b). Due to the formal simplicity and physical clarity of Eq. (2.1), it is particularly useful to evaluate the contributions of distinct vortical structures to the lift in complex unsteady flows.

For a two-dimensional (2D) incompressible viscous flow over a thin airfoil, which can be approximately decomposed into the outer potential flow and the boundary layer, the sectional-lift formula was reduced by Liu et al. (2015) from Eq. (1), i.e.,

$$L'(t) = \rho U c \int_0^1 \gamma(\bar{x},t)\, d\bar{x} + \rho c^2 \frac{d}{dt}\int_0^1 (\bar{x}_{ref} - \bar{x})\gamma(\bar{x},t)\, d\bar{x} \equiv L'_{vor} + L'_a , \tag{2}$$

where $\bar{x} = (x - x_{LE})/c$ is the normalized chordwise coordinate from the leading edge , $\bar{x}_{ref} = (x_{ref} - x_{LE})/c$ is the normalized reference location, $U$ is the incoming flow velocity (considered as a constant velocity in this work), $c$ is the chord length, and $x_{LE}$ and $x_{TE}$ are the leading-edge and trailing-edge locations, respectively. Since Eq. (2) is mathematically reduced from Eq. (1), a vortex sheet is considered as an idealized model of a boundary layer (or near-wall shear layer) for the thin airfoil.

In a 2D flow, the first term (the vortex lift $L'_{vor}$) in Eq. (2) represents the Kutta-Joukowski (K-J) theorem $L'_{vor} = \rho U \Gamma(t)$, where the circulation is given by

$$\Gamma = c\int_0^1 \gamma(\bar{x},t)\, d\bar{x} , \tag{3}$$

Here, the quantity $\gamma(\bar{x},t)$ is defined as

$$\gamma(\bar{x},t) = \int_0^{\delta^+} \omega_z^+(\bar{x},n,t)\, dn + \int_0^{\delta^-} \omega_z^-(\bar{x},n,t)\, dn , \tag{4}$$

where $\omega_z$ is the spanwise vorticity, $n$ is the normal coordinate directing outward from the airfoil surface, $\delta$ denotes the boundary-layer thickness, and the superscripts '$+$' and '$-$' denote the quantities on the upper and lower surfaces of the thin airfoil, respectively. Broadly, $\delta$ is interpreted as a characteristic thickness of a viscous region where the flow separation is moderate, and $\gamma(\bar{x},t)$ is a lumped model of the vorticity distribution on the airfoil surface. In the limiting case where the boundary layer becomes very thin as the Reynolds number is increased, $\gamma(\bar{x},t)$ is interpreted as the vortex-sheet strength in the classical UTAT.

The second term $L'_a$ in Eq. (2) is the added-mass lift expressed as the time rate of the vortex moment. In the derivation of $L'_a$ in Eq. (2), integration by parts is carried out such that the reference location $x_{ref}$ is introduced as a parameter to be determined when the mean value theorem is applied. It is interesting that $L'_a$ in Eq. (2) is reduced to the time rate of the vortex impulse (or vortex moment) in unsteady thin airfoil theory. Therefore, $x_{ref}$ physically represents a pivot-point position of the vortex-sheet moment. The more detailed derivations and interpretations can be found in the previous paper (Liu et al. 2015).

The vortex force is the major source of the lift, which is expressed as a volume integral of the Lamb vector in a viscous flow concentrated in the boundary-layer domain. Formally, the vortex lift in a steady flow is given by the K-J theorem, i.e., $L'_{vor} = \rho U \Gamma_g$, where $\Gamma_g$ is the generalized circulation defined based on the Lamb vector and $\Gamma_g = \Gamma$ is valid for an isolated vorticity region (Wang et al. 2013). The lift calculated based on the surface pressure is related to the vortex lift by $L' = \rho U \Gamma_g - \rho[\chi]_-^+$ (Liu, Wang & He 2017), where $\chi = [u\omega_z]_0^\delta$ is the advective vorticity flux integrated across the boundary layer with the thickness $\delta$, and $[\chi]_{-}^{+}$ is the jump of $\chi$ across the upper and lower boundaries in the wake. Therefore, for a viscous flow, the K-J theorem holds only when the condition $[\chi]_{-}^{+} = [u\omega_z]_{-}^{+} = 0$ is satisfied in the wake. This condition can also be expressed as the zero Lamb vector difference, i.e., $\Delta l = [u\omega_z]_{-}^{+} = 0$, where $l = u\omega_z$ is the Lamb vector component in the 2D boundary layer. In other words, the positive and negative values of the advective vorticity flux ($u\omega_z$) in the boundary layers on the upper and lower surfaces should be cancelled out in the wake. This condition was first found by Taylor (1926) and then refined by Sears (1956, 1976), which is referred to as the Taylor-Sears condition for the generation of the circulation. The Kutta condition with zero pressure difference at the trailing edge is a reduced case of the Taylor-Sears condition. It is found that the Taylor-Sears condition is valid even for moderately separated flows (Zhu et al. 2015; Liu et al. 2015, 2017; Bilbao-Ludena & Papadakis 2025; Salazar & Liu 2025). A further question is whether the Taylor-Sears condition (or the Kutta condition) holds in unsteady flows, which will be examined in the Wagner problem at low Reynolds numbers in Appendix A.

In classical aerodynamics, the formal solution of the thin-airfoil equation is sought for $\gamma(x,t)$, and then the wake-induced term is determined (Katz & Plotkin 1991). To express the wake effect explicitly, the vortex-sheet strength of the airfoil can be expressed as a decomposition $\gamma = \gamma_0 + \gamma_1$, where $\gamma_0$ is the quasi-steady part without considering the effect of the wake vortex sheet and $\gamma_1$ is the unsteady part induced by the wake. The formal solutions for $\gamma_0$ and $\gamma_1$ were given by Liu et al. (2015) solving a Cauchy integral equation of the first kind (Estrada & Kanwal 2000). The relation between $\gamma_1$ and the strength of the wake vortex sheet $\gamma_w$ is given by a convolution-like integral, where the Green's function is related to the evolution of the wake vortex sheet. In a special case where a rigid wake vortex sheet is aligned and traveled with the freestream velocity, a closed-form expression of the Green's function was given by Kármán and Sears (1938). Therefore, Eq. (2) is reduced to the Kármán-Sears lift formula, i.e.,

$$\begin{aligned} L'(t) &= \rho U \Gamma_0 + \rho c^2 \frac{d}{dt}\int_0^1 (0.5-\bar{x})\gamma_0(\bar{x},t)\,d\bar{x} \\ &+ \rho U c\int_1^\infty \gamma_w(\bar{\xi},t)\frac{0.5\,d\bar{\xi}}{\sqrt{\bar{\xi}(\bar{\xi}-1)}} \equiv L_0'(t)+L_1'(t)+L_2'(t) \end{aligned}, \tag{5}$$

where $\Gamma_0(t)$ is the quasi-steady circulation. In Eq. (5), the first term denoted by $L_0'$ is the quasi-steady K-J lift (the vortex lift), the second term denoted by $L_1'$ is the added-mass lift, and the third term denoted by $L_2'$ describes the lift associated with the wake effect, which explicitly depends on the strength distribution of the wake vortex sheet. Conventionally, $L_{cir}' = L_0' + L_2'$ is referred to as the circulatory lift directly associated with the vortex sheet and the wake (the shear layers), while the added-mass lift $L_1'$ is called the non-circulatory lift, although it is also related to $\gamma_0$. The circulatory lift $L_{cir}'$ and the added-mass lift $L_1'$ in the UTAT correspond to $L_{vor}'$ (the vortex lift) and $L_a'$ (the lift associated with the fluid acceleration) in Eq. (1), respectively.

## 3. Approximate Analytical Solution

According to the Kelvin circulation conservation theorem, when the wake travels downstream with the velocity $U$, the total circulation condition is

$$\Gamma_0(t)+\Gamma_w(t)+\Gamma_1(t)=0, \tag{6}$$

where $\Gamma_0(t)$, $\Gamma_1(t)$, and $\Gamma_w(t)$ are the quasi-steady circulation, the wake-induced circulation, and the wake circulation, respectively, which are defined as

$$\Gamma_0(t) = c\int_0^1 \gamma_0(\bar{x},t)\,d\bar{x}\ , \tag{7}$$

$$\Gamma_1(t) = c\int_0^1 \gamma_1(\bar{x},t)\,d\bar{x}\ , \tag{8}$$

$$\Gamma_w(t) = c\int_1^{1+Ut/c} \gamma_w(\bar{x},t)\,d\bar{x}\ . \tag{9}$$

The circulation conservation holds approximately even for a viscous flow over an airfoil with a narrow wake when a contour for calculating the circulation is sufficiently large (Wu et al. 2006). Eq. (6) gives the following integral equation for $\gamma_w$, i.e.,

$$\Gamma_0(t) = -c\int_1^{1+Ut/c} \gamma_w(\bar{\xi},t)\,Q(\bar{\xi},t)\,d\bar{\xi}\ , \tag{10}$$

where $Q(\bar{\xi},t)$ is the kernel depending on the evolution of the wake (Liu et al. 2015). For a specific form $Q = \sqrt{\bar{\xi}/(\bar{\xi}-1)}$ considered by Kármán and Sears (1938), Eq. (10) recovers the Wagner integral equation (Wagner 1925). In this case, by using the translation variable $s = t^* - \bar{\xi} + 1$ with the non-dimensional time $t^* = Ut/c$ (or the wake traveling distance normalized by the chord length), Eq. (10) has a convolution form

$$\Gamma_0(t^*) = -c\int_0^{t^*} \gamma_w(s)\,Q_1(t^* - s)\,ds\ , \tag{11}$$

where $Q_1(s) = \sqrt{(s+1)/s}$ . According to Eq. (11), $\Gamma_0(t^*)$ as the integral of the boundary-layer vorticity on the airfoil is proportional to the convolution integral of $\gamma_w$ as an accumulating effect of the vorticity transported in the wake. In Eq. (11), $\gamma_w$ is formally interpreted as a driving term (cause), resulting in $\Gamma_0$. Therefore, it is an inverse problem to determine $\gamma_w$ from $\Gamma_0$ by solving the Volterra integral equation of convolution type.

The application of the Laplace transform to Eq. (11) gives a formal solution, i.e.,

$$\gamma_w(s) = -c^{-1}\mathcal{L}^{-1}\left[\frac{\mathcal{L}(\Gamma_0)}{\mathcal{L}(Q_1)}\right] = -c^{-1}\Gamma_0 * \mathcal{L}^{-1}\left[\frac{1}{\mathcal{L}(Q_1)}\right], \tag{12}$$

where $*$ denotes the convolution operator, and $\mathcal{L}$ and $\mathcal{L}^{-1}$ denote the Laplace transform and inverse Laplace transform, respectively. The Laplace transform $\mathcal{L}(Q_1)$ can be expressed as the Whittaker function (Oberhettinger & Badii 1973). On the other hand, Sears (1938) gave $\mathcal{L}(Q_1)$ in terms of the modified Bessel functions, which is equivalent to the expression of the Whittaker function (Lucietti 2003). Unfortunately, the analytical inverse Laplace transform $\mathcal{L}^{-1}\left[1/\mathcal{L}(Q_1)\right]$ cannot be exactly expressed in an explicit form, although certain asymptotic expressions could be obtained. Therefore, the exact analytical inverse Laplace transform of Eq. (12) is a deadlock although the numerical inverse Laplace transform can be used (Dawson & Brunton 2022). To circumvent this issue, an approximate solution is sought.

To find an approximate explicit solution, the following approximation with a relative error of about 1% is used, i.e.,

$$Q_1(s)=\sqrt{\frac{s+1}{s}}\approx 1+s^{-1/2}\exp\left(-as^{1/2}\right), \tag{13}$$

where $a=0,85$. The corresponding Laplace transform is (Oberhettinger & Badii 1973)

$$\mathcal{L}(Q_1)\approx p^{-1}+\pi^{1/2}p^{-1/2}\exp\left(\frac{a^2}{2p}\right)\mathrm{erfc}\left(\frac{1}{2}ap^{-1/2}\right), \tag{14}$$

where $p$ is the variable in the Laplace transform. A further approximation is

$$R(p)=1/\mathcal{L}(Q_1)\approx c_1\sqrt{p}-c_2\left[\exp\left(-c_3p^{1/2}\right)-\exp\left(-c_4p^{1/2}\right)\right], \tag{15}$$

where $c_1=0.56$, $c_2=0.16$, $c_3=0.2$, and $c_4$ is a sufficiently large positive number ($c_4=10^8$ in this case). At $p=0$, the approximation gives $R(p)=0$. Figure 1 shows the approximations of $Q_1(s)$ and $R(p)=1/\mathcal{L}(Q_1)$ by Eqs. (13) and (15) in comparisons with the exact functions. Eq. (15) has a relative error of about 0.1% in a range of $p$ in the Laplace transform domain.

Substitution of Eq. (15) to Eq. (12) and application of the inverse Laplace transform yield a convolution-type expression of the wake vortex-sheet strength, i.e.,

$$\begin{aligned}\gamma_w(s) &= \frac{c_2}{2c\sqrt{\pi}}\int_0^s \Gamma_0(t^*)S_0(s-t^*)dt^* - \frac{c_1}{c\sqrt{\pi}}\int_0^s \Gamma_0'(t^*)S_1(s-t^*)dt^* \\ &= \frac{c_2}{2c\sqrt{\pi}}\int_0^s \Gamma_0(s-t')S_0(t')dt' - \frac{c_1}{c\sqrt{\pi}}\int_0^s \Gamma_0'(s-t')S_1(t')dt', \\ &\equiv \gamma_{w0}(s)+\gamma_{w1}(s)\end{aligned} \tag{16}$$

where $\Gamma_0'(t^*) = d\Gamma_0/dt^*$ is the time derivative of the quasi-steady circulation. The Green's functions $S_0(t)$ and $S_1(t)$ in Eq. (16) are defined as

$$S_0(t) = c_3G(t,c_3) - c_4G(t,c_4), \tag{17}$$

$$S_1(t) = t^{-1/2}, \tag{18}$$

where

$$G(t,h) = t^{-3/2}\exp\left(-\frac{h^2}{2t}\right). \tag{19}$$

In Eq. (16), the initial condition $\Gamma_0(0)=0$ is imposed. Eqs. (16)-(19) are the key results in this work, which the following developments are based on. Eq. (16) gives a decomposition $\gamma_w(s)=\gamma_{w0}(s)+\gamma_{w1}(s)$, where the first and second terms $\gamma_{w0}(s)$ and $\gamma_{w1}(s)$ in the right-hand-side (RHS) are the quasi-steady and unsteady terms, respectively. In $\gamma_{w0}(s)$, $S_0(0)=0$ and $S_0(t)\to 0$ as $t\to\infty$. There is the maximum of $S_0(t)$ at $t=0.0133$. For a steady circulation $\Gamma_0 = const.$, $\gamma_{w0}(s)$ as the integral of $S_0(t)$ is a monotonically increasing function, which approaches an asymptotic value as $s\to\infty$. This behavior is related to the Wagner function. Interestingly, the unsteady term $\gamma_{w1}(s)$ has the same mathematical form as the solution of the inverse heat transfer problem in hypersonic flows where the surface heat flux (corresponding to $\gamma_{w1}$ here) is determined from the given transient surface temperature (corresponding to $\Gamma_0$) on an infinite base (Liu et al. 2010). This form of $\gamma_{w1}$ is considered as a special solution of the Abel integral equation.

Further, by introducing the traveling variables $s = t^* - \bar{\xi} + 1 = t^* - \eta$ and $\eta = \bar{\xi} - 1$, the lift associated with the wake effect, the third term in the RHS of Eq. (5), can be formally written as

$$L_2'\left(t^*\right) = \rho U c \int_0^{t^*} \gamma_w\left(s\right) P\left(t^* - s\right) ds = \rho U c \int_0^{t^*} \gamma_w\left(t^* - \eta\right) P\left(\eta\right) d\eta \,, \tag{20}$$

where the Green's function is

$$P\left(\eta\right) = \frac{1}{2\sqrt{\eta\left(\eta + 1\right)}} \,. \tag{21}$$

In Eq. (16), $\gamma_w = 0$ for $\eta \leq 0$, and $\gamma_w \neq 0$ for $\eta > 0$ in the wake. The approximate analytical solution (simply called the analytical solution hereafter) for the unsteady lift of a moving thin airfoil is given by a combination of Eqs. (5) and (16)-(21). This analytical solution is a key result of this work.

Operationally, for a given velocity $w_0\left(\bar{x}, t^*\right)$ normal to the chord line of a moving airfoil, the quasi-steady vortex-sheet strength $\gamma_0\left(x, t^*\right)$ is determined by solving the thin-airfoil equation (Katz & Plotkin 1991), and the quasi-steady circulation $\Gamma_0\left(t^*\right)$ is calculated using Eq. (7). In a lumped model, the effective AoA is given by $\alpha_{eff}\left(t^*\right) = -\bar{w}_0\left(t^*\right)/U$, where $\bar{w}_0\left(t^*\right)$ is the chord-averaged normal velocity. According to the K-J theorem $L' = \rho U \Gamma_0$ and the classical result $C_l = 2\pi\alpha_{eff}$, the quasi-steady circulation can be given by $\Gamma_0\left(t^*\right) = \pi U c\, \alpha_{eff} = -c\,\pi \bar{w}_0\left(t^*\right)$. Then, $\gamma_w\left(s\right)$ is evaluated by substituting $\Gamma_0\left(t^*\right)$ to Eq. (16), and $L_2'\left(t^*\right)$ is calculated by using Eq. (20). The total unsteady lift $L'\left(t^*\right)$ is calculated by using Eq. (5). For validation, this analytical solution is applied to the classical Wagner problem in Section 4.1 and Theodorsen problem in Appendix A, which leads to the explicit integral forms of the Wagner and Theodorsen functions. More importantly, this analytical solution can be applied to a general unsteady motion of a thin airfoil, particularly the generalized Wagner problem with a finite rising timescale in Section 4.2, to elucidate the self-similarity or Reynolds-number-invariance of the re-normalized circulatory lift coefficient and evaluate the applicability of the Wagner function in viscous flows (see Section 5).

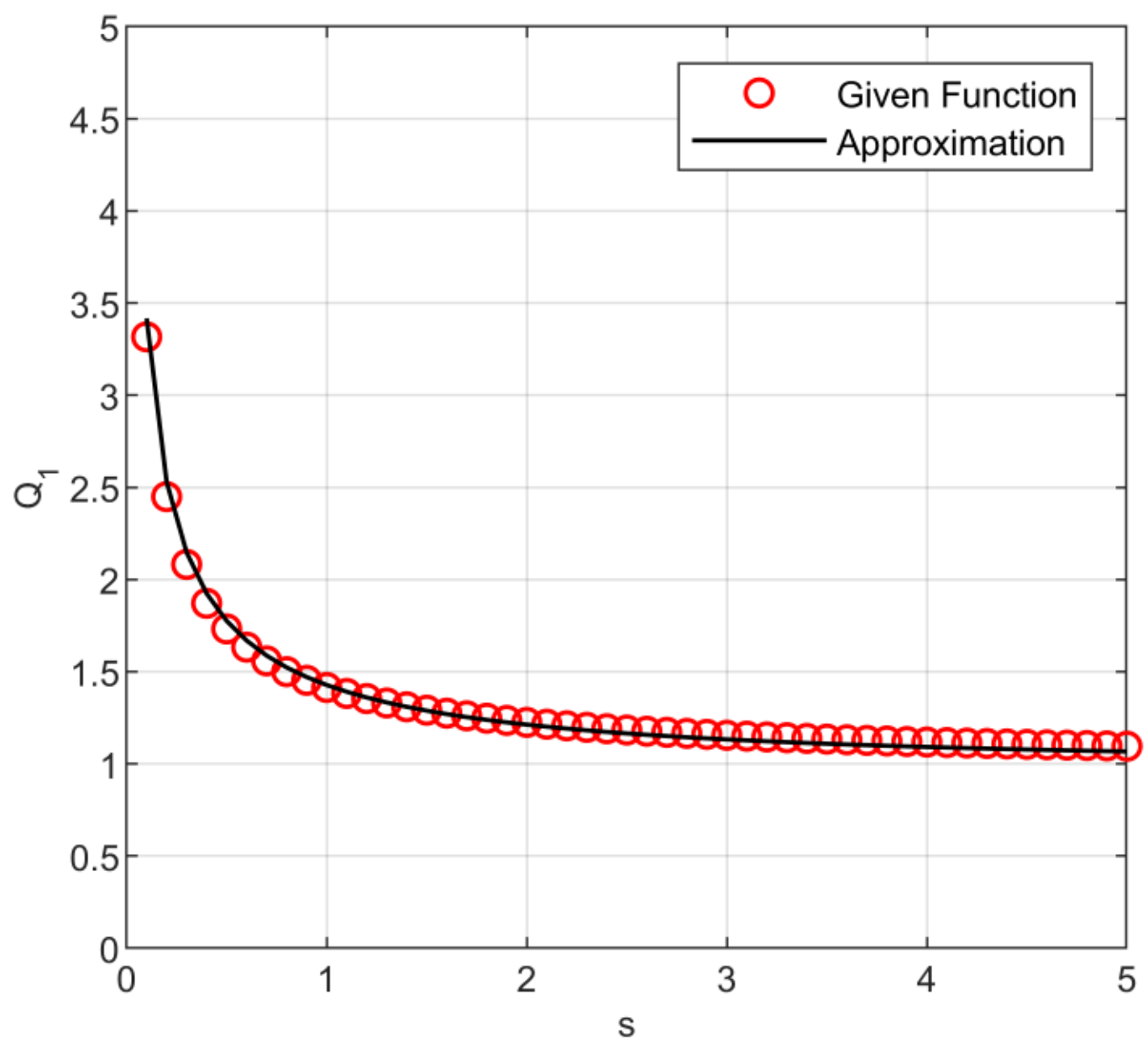

(a)

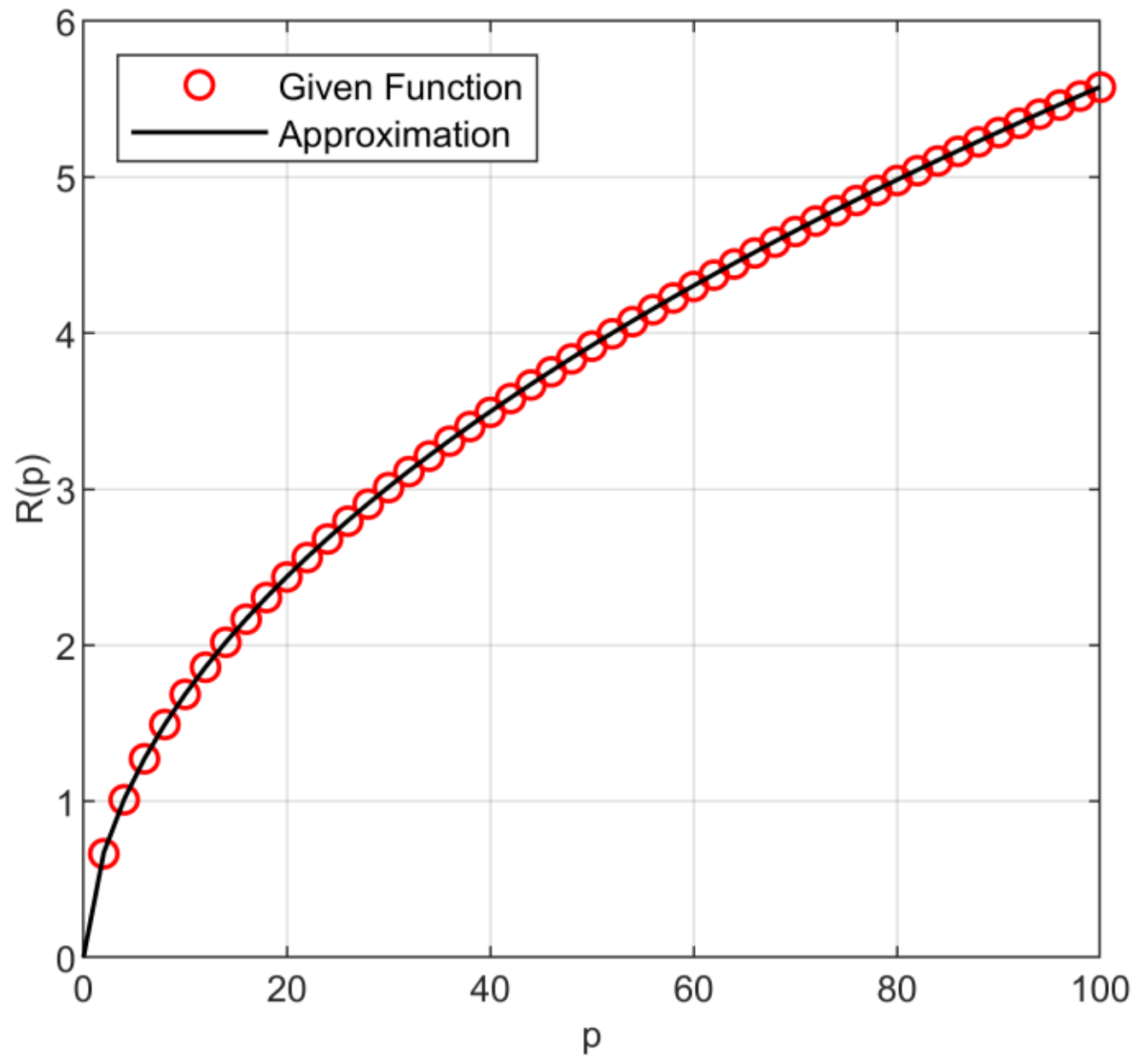

(b)

Figure 1. Approximations of (a) $Q_1(s)$ and (b) $R(p)=1/\mathcal{L}(Q_1)$, where the given functions are the exact functions to be approximated.

## 4. Impulsive Motion

### 4.1. Classical Wagner's Problem

For validation, the analytical solution is first applied to the classical Wagner problem in which a thin airfoil suddenly gains the circulation in a step function (Wagner 1925). This problem is related to the response of a thin airfoil to sudden starting, pitching, heaving or gust. From a physical viewpoint, the step change implies that the vortex sheet and circulation are established instantaneously, which is considered as a limiting case as the Reynolds number approaches to infinity (see Section 4.2). Therefore, a thin airfoil has the vortex-sheet strength given by $\gamma_0(\bar{x},t^*)=\gamma_{00}(\bar{x})H(t^*)$, where $\gamma_{00}(\bar{x})$ is the steady-state vortex-sheet strength and $H(t^*)$ is the Heaviside function defined as $H(t^*)=0$ for $t^*\leq 0$ and $H\left(t^*\right)=1$ for $t^*>0$ ] The quasi-steady circulation is $\Gamma_0\left(t^*\right)=\Gamma_{00}H\left(t^*\right)$, where the constant circulation is

$$\Gamma_{00}=c\int_0^1\gamma_{00}\left(\overline{x}\right)d\overline{x}\,. \tag{22}$$

For $\Gamma_0'\left(t^*\right)=d\Gamma_0/dt^*=\Gamma_{00}\delta\left(t^*\right)$, where $\delta\left(t^*\right)$ is the Dirac-delta function, Eq. (16) is given by an approximate form

$$\frac{\gamma_w\left(s\right)}{\Gamma_{00}}\approx-\frac{c_1}{c\sqrt{\pi}}\frac{1}{\sqrt{s}}+\frac{c_2}{2c\sqrt{\pi}}\left[c_3\int_0^s G\left(t^*,c_3\right)dt^*-c_4\int_0^s G\left(t^*,c_4\right)dt^*\right]. \tag{23}$$

The integral in Eq. (23) is expressed as

$$\int_0^s G\left(t^*,h\right)dt^*=\left(2/h^2\right)F\left(s,h\right), \tag{24}$$

where $F\left(s,h\right)$ is defined as

$$F\left(s,h\right)=s^{1/2}\,exp\left(-\frac{h^2}{2s}\right)-I\left(s,h\right), \tag{25}$$

where the integral $I\left(s,h\right)$ is defined as

$$I\left(s,h\right)=\int_0^s\frac{1}{2}\left(t^*\right)^{-1/2}exp\left(-\frac{h^2}{2t^*}\right)dt^*\,. \tag{26}$$

For a very large $c_4$, Eq. (23) becomes

$$\begin{aligned}\frac{\gamma_w\left(s\right)}{\Gamma_{00}}&\approx-\frac{c_1}{c\sqrt{\pi}}\frac{1}{\sqrt{s}}+\frac{c_2}{2c\sqrt{\pi}}\left[\left(\frac{2}{c_3}\right)F\left(s,c_3\right)-\left(\frac{2}{c_4}\right)F\left(s,c_4\right)\right]\\&\approx-\frac{c_1}{c\sqrt{\pi}}\frac{1}{\sqrt{s}}+\frac{c_2}{c\sqrt{\pi}c_3}F\left(s,c_3\right)\end{aligned}. \tag{27}$$

The approximation of $I\left(s\right)$ with the relative error of 2% is expressed by a composite power-law function, i.e.,

$$I\left(s,c_3\right)\approx 0.78s^{0.58}-\sum_{k=1}^{2}a_k\left(s^{n_k}+d_k\right)Tr\left(s,s_k,\Delta s_k\right), \tag{28}$$

where the transition function is defined as

$$Tr\left(s,s_k,\Delta s_k\right)=\frac{1}{2}+\frac{1}{2}tanh\left(\frac{s-s_k}{\Delta s_k}\right), \tag{29}$$

and the parameters are

$a_1=0.084$, $a_2=0.315$, $n_1=0.05$, $n_2=0.09$, $d_1=0.38$, $d_2=0.04$,

$$s_1=20,\ s_2=35,\ \Delta s_1=10,\ \Delta s_2=15\,. \tag{30}$$

Figure 2 shows the integral $I\left(s,c_3\right)$ and its approximation.

Substitution of Eq. (27) to Eq. (20) yields the unsteady lift associated with the wake effect, i.e.,

$$L_2'\left(t^*\right)=-\rho U\Gamma_{00}\Phi\left(t^*\right). \tag{31}$$

By using the traveling variables $s=t^*-\overline{\xi}+1=t^*-\eta$ and $\eta=\overline{\xi}-1$, t and introducing the similar variable $\overline{\eta}=\eta/t^*$, the lift deficiency function in Eq. (31) is expressed as

$$\Phi\left(t^*\right)=A_1\int_0^1\frac{d\overline{\eta}}{\overline{\eta}^{1/2}\left(1-\overline{\eta}\right)^{1/2}\left(1+t^*\overline{\eta}\right)^{1/2}}-A_2\int_0^1\sqrt{t^*}\,\frac{F\left[t^*\left(1-\overline{\eta}\right)\right]}{\overline{\eta}^{1/2}\left(1+t^*\overline{\eta}\right)^{1/2}}d\overline{\eta}\,, \tag{32}$$

where $A_1=c_1/2\sqrt{\pi}=0.158$ and $A_2=c_2/2c_3\sqrt{\pi}=0.226$. The definite integral form of $\Phi\left(t^*\right)$ is ready to evaluate using the numerical integration.

According to Eq. (5), the unsteady lift of a suddenly moving (impulsively starting) airfoil for $t^*\geq 0$ is given by

$$L'\left(t^*\right)=L_0'\left(t^*\right)+L_1'\left(t^*\right)+L_2'\left(t^*\right)=\rho U\Gamma_{00}Wa\left(t^*\right)+\rho UM\left(\gamma_{00}\right)\delta\left(t^*\right), \tag{33}$$

where the Wagner function is

$$Wa\left(t^*\right)=\frac{L'\left(t^*\right)}{\rho U\Gamma_{00}}=1-\Phi\left(t^*\right), \tag{34}$$

and the vortex moment is defined as

$$M\left(\gamma_{00}\right)=c\int_0^1\left(0.5-\overline{x}\right)\gamma_{00}\left(\overline{x}\right)d\overline{x}\,. \tag{35}$$

The first term in the RHS of Eq. (33) is the circulatory lift with the wake effect [ $L'_{cir} = \rho U \Gamma_{00} Wa\left(t^*\right)$ ], and the second term is the added-mass [ift [ $\rho UM\left(\gamma_{00}\right)\delta\left(t^*\right)$ ]. Therefore, the lift history of an impulsive airfoil demonstrates a Dirac-delta pulse response initially and then follows the Wagner function to approach to the steady-state value. In Eqs. (32) and (34), the lift deficiency function $\Phi\left(t^*\right)$ and the Wagner function $Wa\left(t^*\right)$ are directly given by the analytical solution as a special case.

The lift coefficient for $t^* \geq 0$ is

$$C_l\left(t^*\right) = \frac{L'\left(t^*\right)}{q_\infty c} = \hat{\Gamma}_{00} Wa\left(t^*\right) + \hat{M}\left(\gamma_{00}\right)\delta\left(t^*\right), \tag{36}$$

where $\hat{\Gamma}_{00} = \Gamma_{00} / Uc$ is the non-dimensional steady circulation, $\hat{M}\left(\gamma_{00}\right) = M\left(\gamma_{00}\right) / Uc$ is the non-dimensional vortex moment, and $q_\infty = \rho U^2 / 2$ is the freestream dynamics pressure. For $t^* > 0$ after the short transient stage (the Dirac-delta function), according to Eq. (36), the circulatory lift coefficient $C_{lcir} = L'_{cir} / q_\infty c$ is presented as a self-similar relation that is independent of the AoA ( $\alpha$ ), i.e.,

$$\frac{C_{lcir}\left(t^*\right)}{C_{l\infty}} = Wa\left(t^*\right), \tag{37}$$

where $C_{l\infty} = C_{lcir}\left(t^* \to \infty\right)$ is the steady-flow lift coefficient. This self-similarity can be extended to the generalized Wagner problem, as indicated in Section 4.2.

Figure 3 shows the integral form of the Wagner function $Wa\left(t^*\right)$, where $t^* = Ut / c$ is the non-dimensional time (or the wake traveling distance normalized by the chord length). For comparison, Figure 3 also includes the numerical data from Wagner (1925) and Sears (1940), the asymptotic solution given by Sears (1940) for large $t^*$, and the correlations given by Kármán and Sears (1938), Garrick (1938) and Jones (1938). The exact solution of the Wagner function in $t^* = 0\text{-}10$ is also shown in Fig. 3 for comparison, which was calculated by Dawson & Brunton (2022) using the numerical inverse Laplace transform. The asymptotic behavior of $Wa\left(t^*\right)$ for large $t^*$ depends sensitively on the higher-order terms of $I\left(s\right)$ for large $s$ in Eq. (28). It is noted that in the classical literature of the unsteady aerodynamics, the lift deficiency function is often expressed as a function of the time variable $\sigma = 2Ut / c = 2t^*$ where the half-chord length issued for normalization since it is more convenient in the use of the conformal mapping (e.g., the Joukowski transformation).

The bound circulation of the airfoil is

$$\Gamma_b\left(t^*\right) = \Gamma_0\left(t^*\right) + \Gamma_1\left(t^*\right) = -\Gamma_w\left(t^*\right) = -c\int_0^{t^*} \gamma_w\left(s\right) ds \,. \tag{38}$$

Substitution of Eq. (27) to Eq. (38) yields

$$\frac{\Gamma_b\left(t^*\right)}{\Gamma_{00}} = \frac{2c_1}{\sqrt{\pi}}\sqrt{t^*} - \frac{c_2}{\sqrt{\pi}\, c_3}\int_0^{t^*} F\left(s\right) ds \,. \tag{39}$$

Figure 4 shows the normalized bound circulation $\Gamma_b\left(t^*\right) / \Gamma_{00}$ in comparisons with the correlations given by Ford & Babinsky (2013) and Li & Wu (2015) and the experimental data by Ford & Babinsky (2014). These correlations were extracted the numerical data for an impulsive airfoil with the leading- and trailing-edge vortices. The circulation of an accelerating flat-plate airfoil was estimated based on particle image velocimetry (PIV) measurements in an early development of vortices ($t^* = 1\text{-}2$) (Beckwith & Babinsky 2009; Ford & Babinsky 2014; Babinsky et al. 2016; Stevens et al. 2017; Goyal & Nedić 2025). The measured circulation given by Ford & Babinsky (2014) agrees with the result given by the analytical solution in the early stage in Fig. 4. Overall, the Wagner theory describes the lift history even when the more complex leading-edge vortex (LEV) and trailing-edge vortex (TEV) are generated (Xia & Mohseni 2013, 2017).

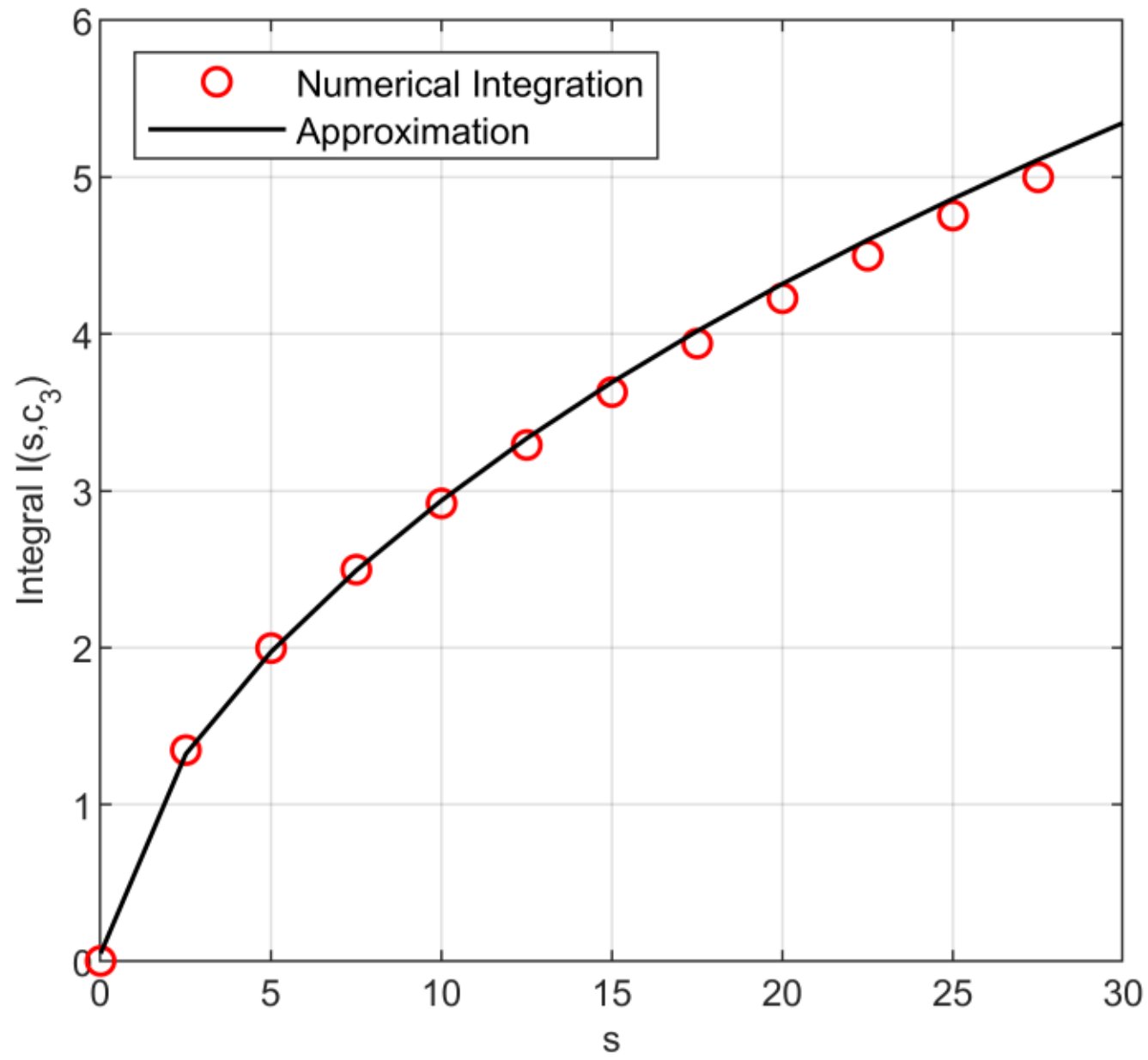

Figure 2. The integral $I(s,c_3)$ and its approximation.

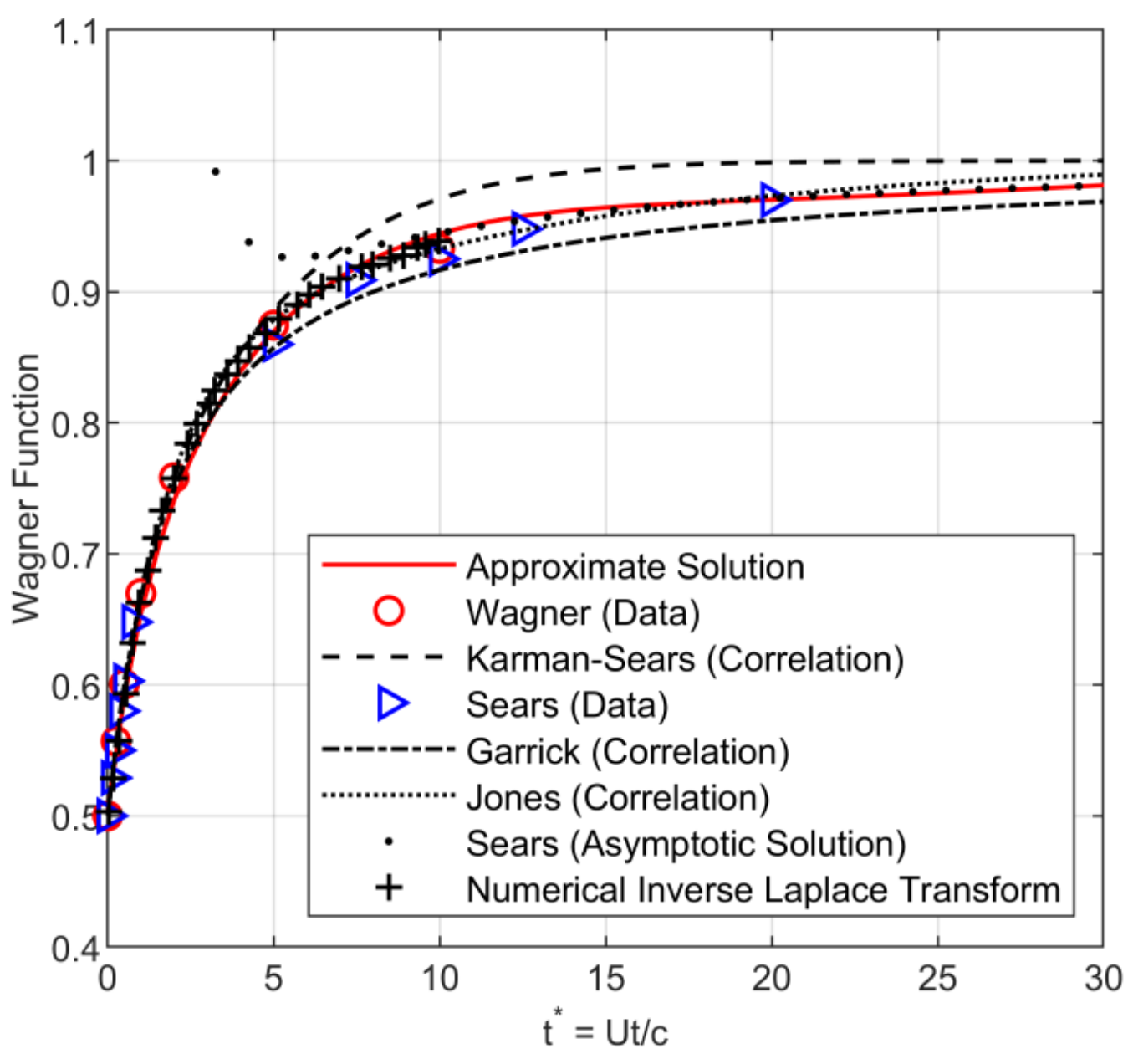

Figure 3. The Wagner function $W(t^*)$ given by the approximate analytical solution in comparison with the existing data, correlations and numerical inverse Laplace transform solution.

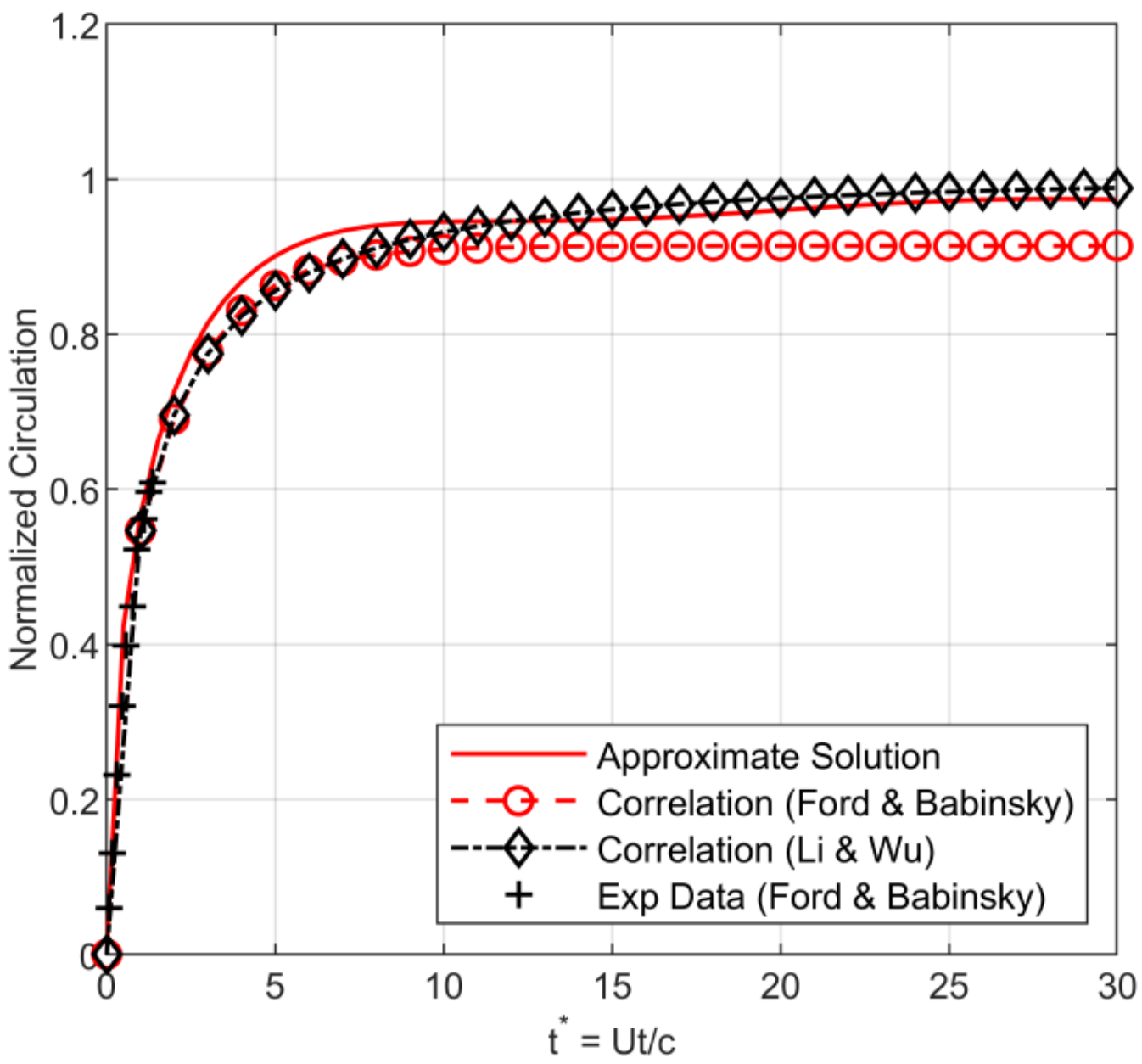

Figure 4. The normalized bound circulation $\Gamma_b\left(t^*\right)/\Gamma_{00}$ with the correlations given by Ford & Babinsky (2013) and Li & Wu (2015) and the experimental data given by Ford & Babinsky (2014).

### 4.2. Wagner's Problem with Finite Rising Timescale

In the classical Wagner problem, a step function of $\Gamma_0\left(t^*\right)$ is considered as an idealized case, and it implies that $\Gamma_0\left(t^*\right)$ is established instantaneously. However, in the real starting flow, the development of $\Gamma_0\left(t^*\right)$ has a finite rising timescale since the circulation results from the formation of the boundary layer as a viscous vorticity diffusion process driven by the boundary vorticity flux (Zhu et al. 2015; Salazar & Liu 2025). The rising timescale is related to the viscous diffusion timescale of the vorticity from a solid surface. In a simplified analysis, for the starting flow of a flat-plate airfoil at a small AoA, the exact solution of the Navier-Stokes equations in the Stokes' first problem can be used as a reference (Schlichting 1979). The Stokes' solution indicates that the timescale of the developing boundary layer on the accelerating flat plate is $\tau_{rise} \sim \delta^2$, where $\delta$ is the mean boundary-layer thickness. Further, $\delta$ is related to the Reynolds number (denoted by $Re$ as a generic Reynolds number) by a power-law relation $\delta \sim Re^{-m}$, where $m$ is a positive exponent that depends on the properties of the boundary layer such as the pressure gradient, separation and turbulence. Therefore, an estimate is $\tau_{rise} \sim \delta^2 \sim Re^{-2m}$, which is also the rising timescale of the circulation that is the integral of the vorticity in the boundary-layer domain according to Eqs. (3) and (4). For example, for a flat-plate laminar boundary layer with the zero pressure gradient, the similarity solution of the boundary-layer equation gives $m = 1/2$ such that $\tau_{rise} \sim Re^{-1}$. For a viscous layer with moderate separation, $m$ should be determined in experiments and computations. In this sense, the effect of the rising timescale on the unsteady lift globally simulates the effects of the Reynolds number and other relevant viscous-flow properties. To build a connection between the classical inviscid Wagner theory and the starting viscous flow over a thin airfoil, we consider the generalized Wagner problem with a finite rising timescale based on the application of the analytical solution.

To simulate the formation of the circulation, the transient process of $\Gamma_0(t^*)$ is modeled by a piece-wise function defined as

$$\frac{\Gamma_0(t^*)}{\Gamma_{00}} = \begin{cases} \sin^2\left(\dfrac{\pi t^*}{2\tau_{rise}}\right), & t^*/\tau_{rise} \le 1 \\ 1, & t^*/\tau_{rise} > 1 \end{cases}, \tag{40}$$

where $\Gamma_{00}$ is the steady-state circulation, and $\tau_{rise}$ is a non-dimensional rising timescale. Eq. (40) [or Eq. (47)] is a suitable model describing the starting flow in aircraft takeoff and wind tunnel startup (Liu 2021). The lift associated with the wake effect, $L_2'(t^*)$, is calculated by substituting Eq. (40) into Eqs. (16)-(20). Figure 5 shows the circulatory lift coefficient $C_{lcir}\left(t^*,\tau_{rise}\right)=L'_{cir}/q_\infty c$ as a function of the normalized time $t^*$ for different values of $\tau_{rise}$. In general, immediately after a spike of $C_{lcir}$ near $t^*=0$, there is a local minimum $C_{lcir\,min}=min\left(C_{lcir}\right)$ at a moment $t^*_{min}$. For $t^*>t^*_{min}$, $C_{l,cir}$ increases monotonically with $t^*$ from $C_{lcir\,min}$, as predicted by the Wagner function, which is known as the Wagner effect (or the wake effect on the lift). As indicated in Fig. 6, $t^*_{min}$ follows a power-law relation (a proportional relation) $t^*_{min}=\tau_{rise}$. Figure 7 shows $C_{lcir\,min}\left(\tau_{rise}\right)$ and $C_{l,cir\,max}\left(\tau_{rise}\right)$ as a function of $t^*_{min}$, where $C_{lcir\,max}\left(\tau_{rise}\right)$ is the maximum value of $C_{l,cir}$ at a moment $t^*_{max}$ ($t^*_{max}=10$ in this case) in the time domain. As shown in Fig. 7, as $t^*_{min}$ increases, $C_{lcir\,min}$ increases, while $C_{l,max}$ decreases slightly. The effect of $t^*_{min}$ on $C_{lcir}\left(t^*,\tau_{rise}\right)$ is mainly characterized by $C_{lcir\,min}$ and $C_{lcir\,max}$. The overall behavior of $C_{lcir}\left(t^*,\tau_{rise}\right)$ is generic for a starting thin-airfoil in viscous flows (see Section 5). In particular, the monotonic increase of $C_{lcir}\left(t^*,\tau_{rise}\right)$ after $t^*_{min}$ is referred to as the Wagner effect that is also observed in viscous flows.

To absorb the effect of $\tau_{rise}$ on the circulatory lift coefficient $C_{l,cir}\left(t^*,\tau_{rise}\right)$ in a finite time domain $\left[t^*_{min},t^*_{max}\right]$, we introduce the re-normalized circulatory lift coefficient

$$\bar{C}_{lcir}\left(\bar{t}^*\right)=\frac{C_{lcir}\left(\bar{t}^*,\tau_{rise}\right)-C_{lcir\,min}\left(\tau_{rise}\right)}{C_{lcir\,max}\left(\tau_{rise}\right)-C_{lcir\,min}\left(\tau_{rise}\right)}, \tag{41}$$

where the re-normalized time is

$$\bar{t}^*=\frac{t^*-t^*_{min}\left(\tau_{rise}\right)}{t^*_{max}\left(\tau_{rise}\right)-t^*_{min}\left(\tau_{rise}\right)}. \tag{42}$$

Figure 8 shows $\bar{C}_{lcir}\left(\bar{t}^*\right)$ as a function of $\bar{t}^*$, indicating that the data for different values of $\tau_{rise}$ are largely collapsed to a self-similar form. Here, we consider the scaling of $C_{lcir}\left(t^*,\tau_{rise}\right)$ in a finite time domain $\left[t^*_{min},t^*_{max}\right]$ because the Wagner effect in a viscous unsteady flow over an impulsive airfoil with a relatively high AoA is observed for $t^* < t^*_{max}$ before the flow separation becomes dominant for $t^* > t^*_{max}$. For a given airfoil, $t^*_{max}$ depends on the AoA and the Reynolds number. For an attached flow at high Reynolds numbers, when $C_{lcir\,max}(\tau_{rise})$ approaches monotonically to a limiting value (the steady-state value) as $t^*_{max} \to \infty$, Eqs. (41) and (42) still hold as a limiting case.

Similar to the classical Wagner problem with a step change of $\Gamma_0(t^*)$, the Wagner function in this case of a finite rising time is defined as

$$Wa(t^*, \tau_{rise}) = \frac{L'_{cir}(t^*, \tau_{rise})}{\rho U \Gamma_{00}} = 1 - \Phi(t^*, \tau_{rise}), \tag{43}$$

where the lift deficiency function is $\Phi(t^*, \tau_{rise}) = -L'_2(t^*, \tau_{rise}) / \rho U \Gamma_{00}$. We know

$$Wa(t^*, \tau_{rise}) = \frac{1}{2}\left(\frac{Uc}{\Gamma_{00}}\right) C_{lcir}(t^*, \tau_{rise}). \tag{44}$$

Similarly, the re-normalized Wagner function in a finite time domain $[t^*_{min}, t^*_{max}]$ is defined as

$$\overline{Wa}(\bar{t}^*) = \frac{Wa(\bar{t}^*, \tau_{rise}) - min(Wa)}{max(Wa) - min(Wa)}. \tag{45}$$

Therefore, an equivalence relation is

$$\bar{C}_{lcir}(\bar{t}^*) = \overline{Wa}(\bar{t}^*). \tag{46}$$

Eqs. (41)-(46) have two consequences. First, $\bar{C}_{l,cir}(\bar{t}^*)$ is independent of $\tau_{rise}$, and $\bar{C}_{lcir}(\bar{t}^*)$ is self-similar in this sense. Secondly, $\bar{C}_{lcir}(\bar{t}^*)$ is equivalent to $\overline{Wa}(\bar{t}^*)$. As shown in Fig. 8, the data of $\bar{C}_{lcir}(\bar{t}^*)$ computed using the analytical solution at different values of $\tau_{rise}$ are collapsed toward $\overline{Wa}(\bar{t}^*)$, exhibiting the self-similarity that is independent of $\tau_{rise} \sim Re^{-2m}$ in a finite time domain. As indicated in Section 5 in the low-Reynolds-number flows over a starting flat-plate airfoil, the power-law relation $\tau_{rise} \sim Re^{-2m}$ indeed exists. For example, $\tau_{rise} \sim Re^{-1/5}$ for $\alpha = 5^o$ and $\alpha = 10^o$ at which the flow is separated on the upper surface of the airfoil. Therefore, this self-similarity is generally referred to as the Reynolds-number-invariance or the $Re$ -invariance, indicating the applicability of the re-normalized Wagner function in viscous flows. In other words, in the sense of re-normalization, the Wagner effect in viscous flows is self-similar since the $Re$ -effect does not appear explicitly in the scaling. This theoretical finding will be examined by numerical simulation in Section 5. In the limiting case as $Re \to \infty$, $\tau_{rise} \sim Re^{-2m} \to 0$, $min(Wa) \to 0.5$ and $max(Wa) \to 1$. Thus, the Wagner function is recovered as an inviscid limit, i.e., $\bar{Wa}(\bar{t}^*) \to 2Wa(t^*) - 1$. The self-similarity of the circulatory lift coefficient in the Wagner problem at low Reynolds numbers will be examined by numerical simulation in Section 5.

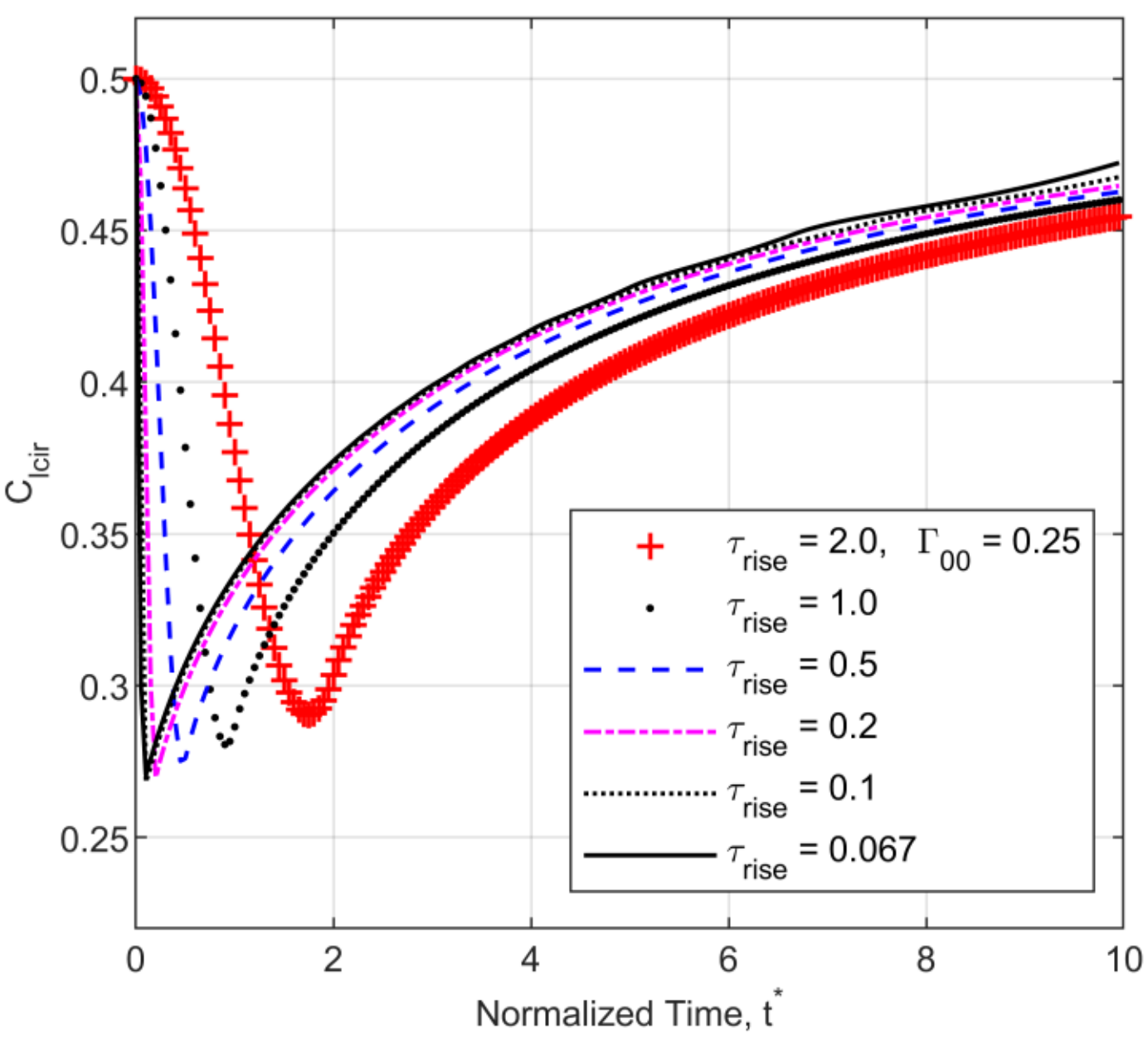

Figure 5. The circulatory lift coefficient $C_{lcir}\left(t^*, \tau_{rise}\right)$ as a function of the non-dimensional time $t^*$ in the starting flow for different values of $\tau_{rise}$.

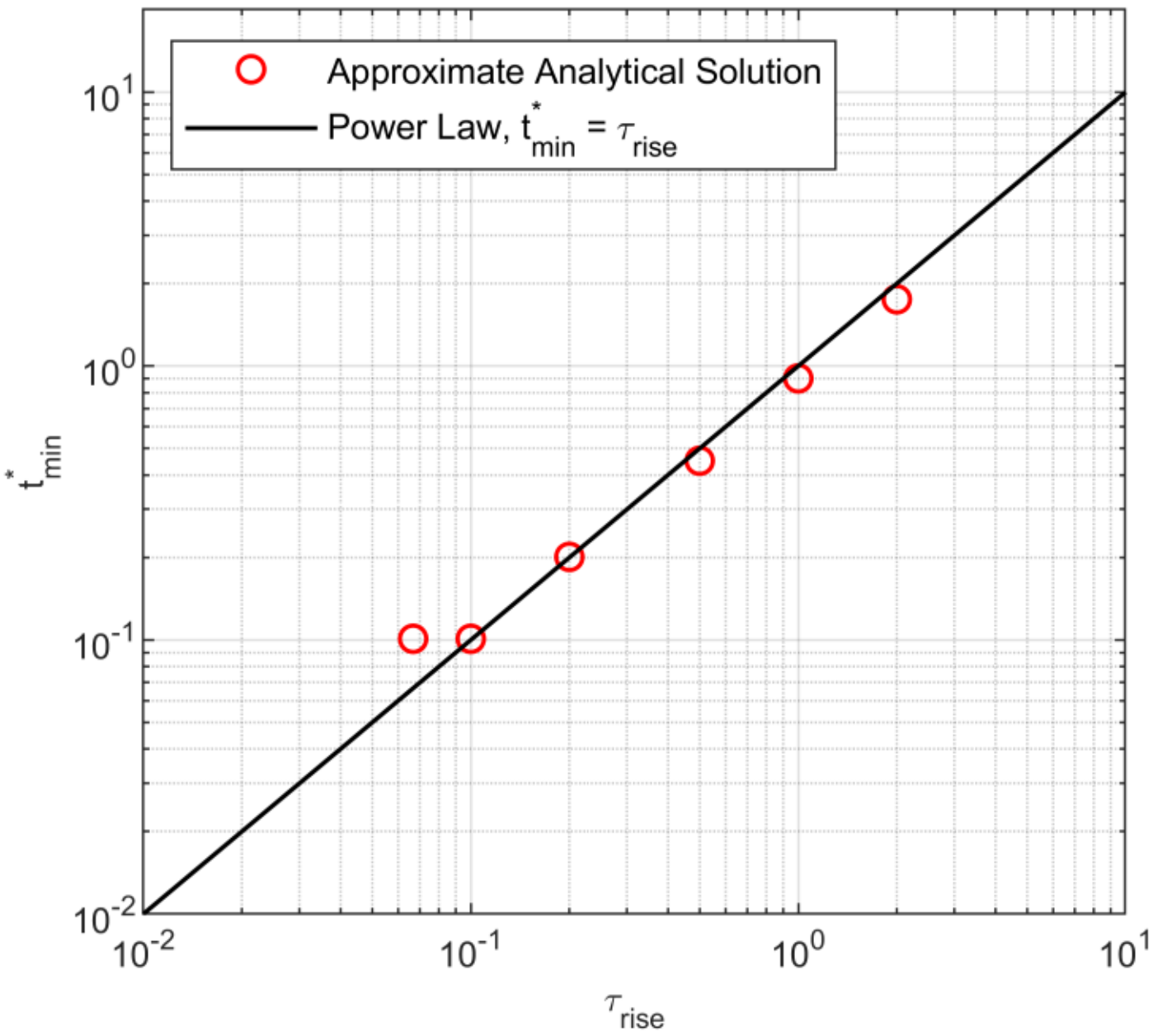

Figure 6. The non-dimensional time $t^*_{min}$ at which $C_{l,cir\,min} = min\left(C_{l,cir}\right)$ is achieved as a function of the timescale $\tau_{rise}$, where the power-law relation $t^*_{min} = \tau_{rise}$ is plotted as a reference.

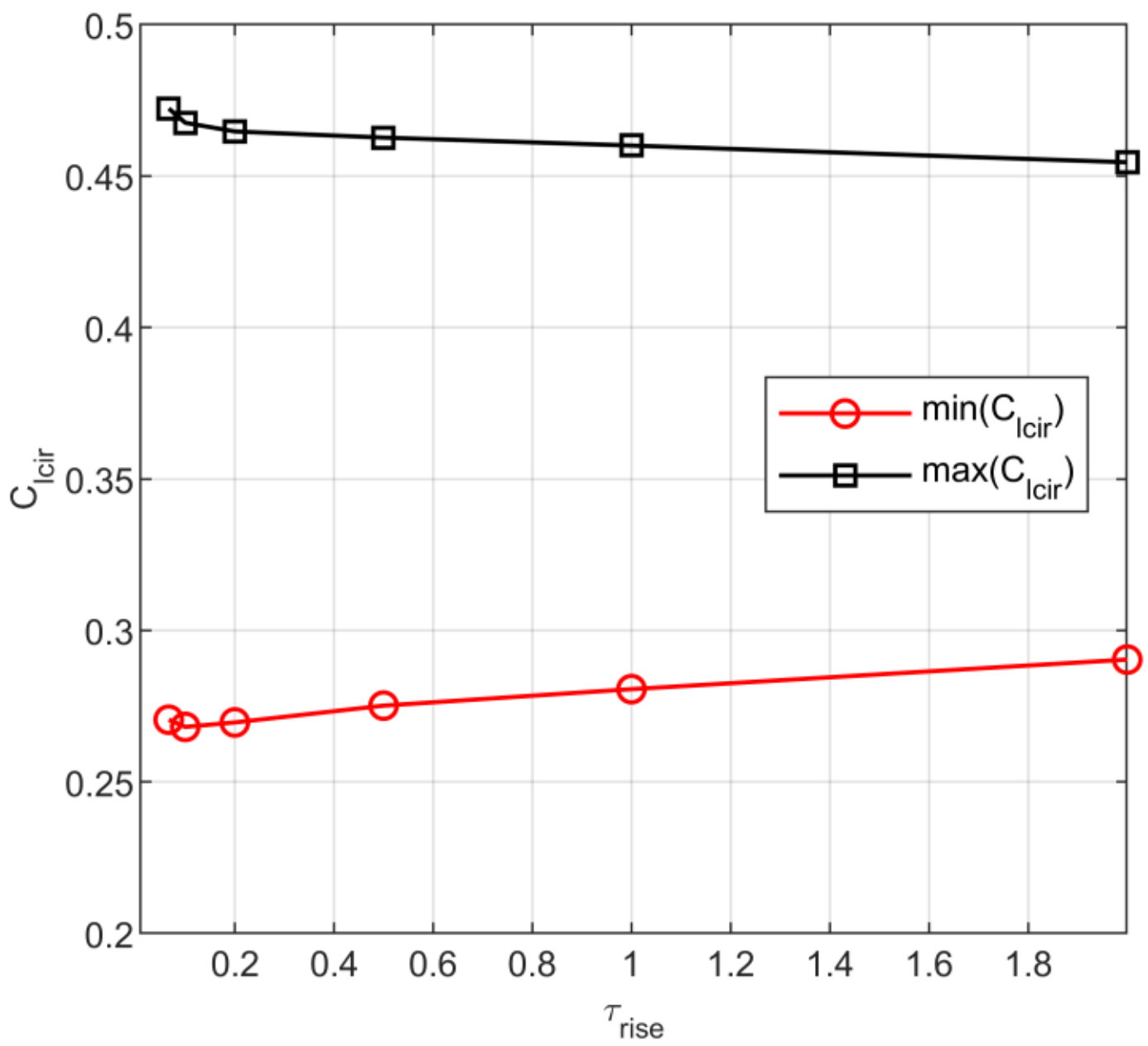

Figure 7. $C_{lcir,min} = min\left(C_{lcir}\right)$ and $C_{lcir\,max} = max\left(C_{lcir}\right)$ (the value of $C_{lcir}$ at $t^* = 10$) as a function of $\tau_{rise}$.

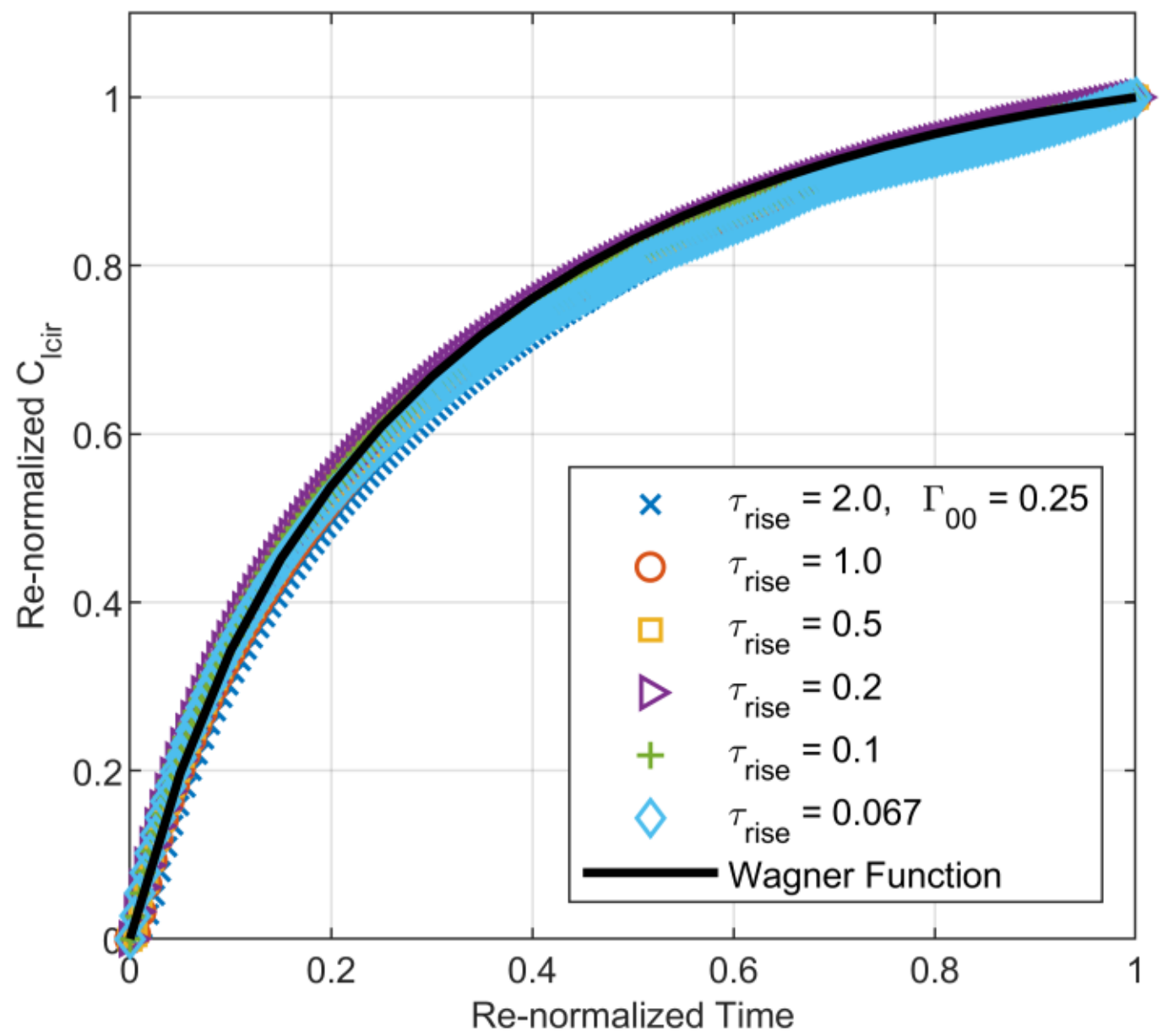

Figure 8. The re-normalized circulatory lift coefficient $\bar{C}_{lcir}\left(\bar{t}^*\right)$ as a function of the re-normalized time $\bar{t}^*$ for different values of $\tau_{rise}$, where the re-normalized Wagner function $\bar{W}a\left(\bar{t}^*\right)$ is plotted as a reference.

## 5. Wagner's Problem at Low Reynolds Numbers

### 5.1. Flow Characteristics

The theoretical analysis in Section 4.2 indicates the self-similarity (the $Re$-invariance) of the re-normalized circulatory lift. coefficient $\bar{C}_{lcir}\left(\bar{t}^*\right)$ and its equivalence to $\bar{W}a\left(\bar{t}^*\right)$. To examine this finding, numerical simulation was conducted by solving the Navier-Stokes equations to calculate the lift and the flow fields of a starting flat-plate airfoil in a range of the AoAs ($\alpha$) at low Reynolds numbers ($Re = 126\text{-}1000$), where the Reynolds number $Re = Uc/\nu$ is based on the chord length $c$. The Wagner problem at low Reynolds numbers is particularly relevant to the studies of biological flight (Liu, Wang & Liu 2023; Liu et al. 2023). Since the effect of $Re$ on the unsteady lift is significant in this problem, it is a suitable case to examine the $Re$-invariance of $\bar{C}_{lcir}\left(\bar{t}^*\right)$ and its equivalence to $\bar{W}a\left(\bar{t}^*\right)$. In a viscous flow, the vortex lift coefficient $C_{lvor}$ is considered, which corresponds to $C_{l,cir}$ that is reduced from $C_{lvor}$ in the UTAT (see Sections 3 and 4).

Numerical simulation of the flow over a starting flat-plate airfoil at low Reynolds numbers was performed using the open-source finite-volume code OpenFOAM (Weller et al. 1998). The incompressible Navier-Stokes equations were solved using the PIMPLE algorithm, a combined PISO-SIMPLE approach suitable for transient flows with moving meshes. The flat-plate airfoil was set into translational motion along the streamwise direction. The velocity of the airfoil had a smoothed piecewise function

$$\frac{U\left(t^*\right)}{U_s} = \begin{cases} sin^2\left(\dfrac{\pi t^*}{2\tau_{rise}}\right), & t^* / \tau_{rise} \leq 1 \\ 1, & t^* / \tau_{rise} > 1 \end{cases}, \tag{47}$$

where $\tau_{rise} = 0.1$ and $U_s$ is a steady velocity. The computational setup and code validation are described in Appendix B.

The typical cases of $\alpha = 5^o$ and $\alpha = 10^o$ are considered. Figure 9 shows the normalized vorticity contour maps ($\omega_z c / U_s$) around the flat-plate airfoil at $t^*$ = 0.5, 2, 4 and 6 for $\alpha = 5^o$ and $\alpha = 10^o$ at $Re = 1000$. Figure 10 shows the corresponding Lamb vector ($l = u\omega_z$) contour maps around the flat-plate airfoil. At $t^* = 0.5$, the boundary layers are formed on the upper and lower surfaces of the airfoil, and a starting vortex or a trailing-edge vortex (TEV or TV) is generated and shed downstream for both $\alpha = 5^o$ and $\alpha = 10^o$. The development of the starting vortex and wake is consistent with the previous numerical simulation and experiment measurements (Zhu et al. 2015; Salazar & Liu 2025).

In $t^* = 2\text{-}6$, the leading-edge vortex (LEV) is observed, which grows on the upper surface. For $\alpha = 5^o$, the LEV remains stabilized on the upper surface in $t^* = 2\text{-}6$. In contrast, for $\alpha = 10^o$, the LEV grows until it cannot be stabilized on the upper surface, and then it is detached from the airfoil at $t^* = 6$ and the flow is separated openly.

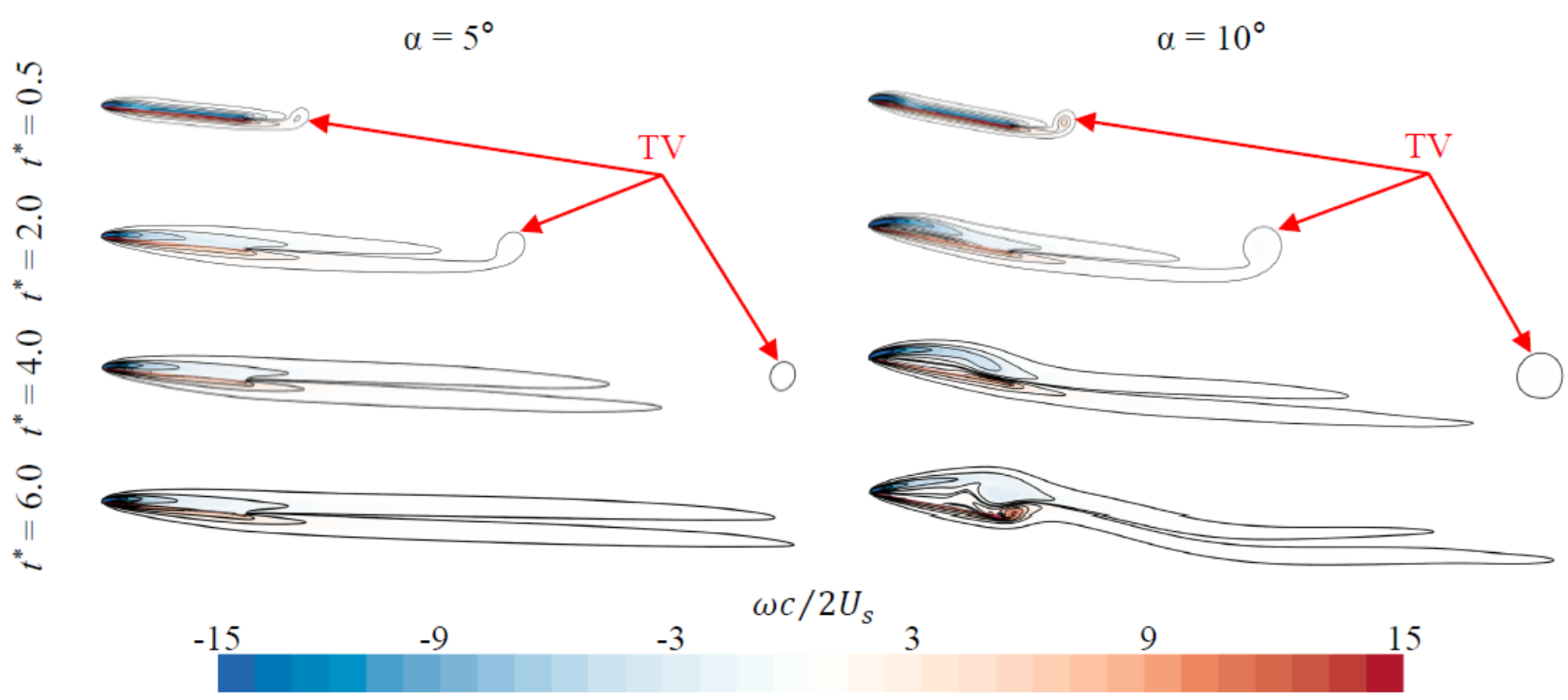

Figure 9. The normalized vorticity contour maps around the flat-plate airfoil at $t^*$ = 0.5, 2, 4 and 6 for $\alpha = 5^o$ and $\alpha = 10^o$ at $Re = 1000$.

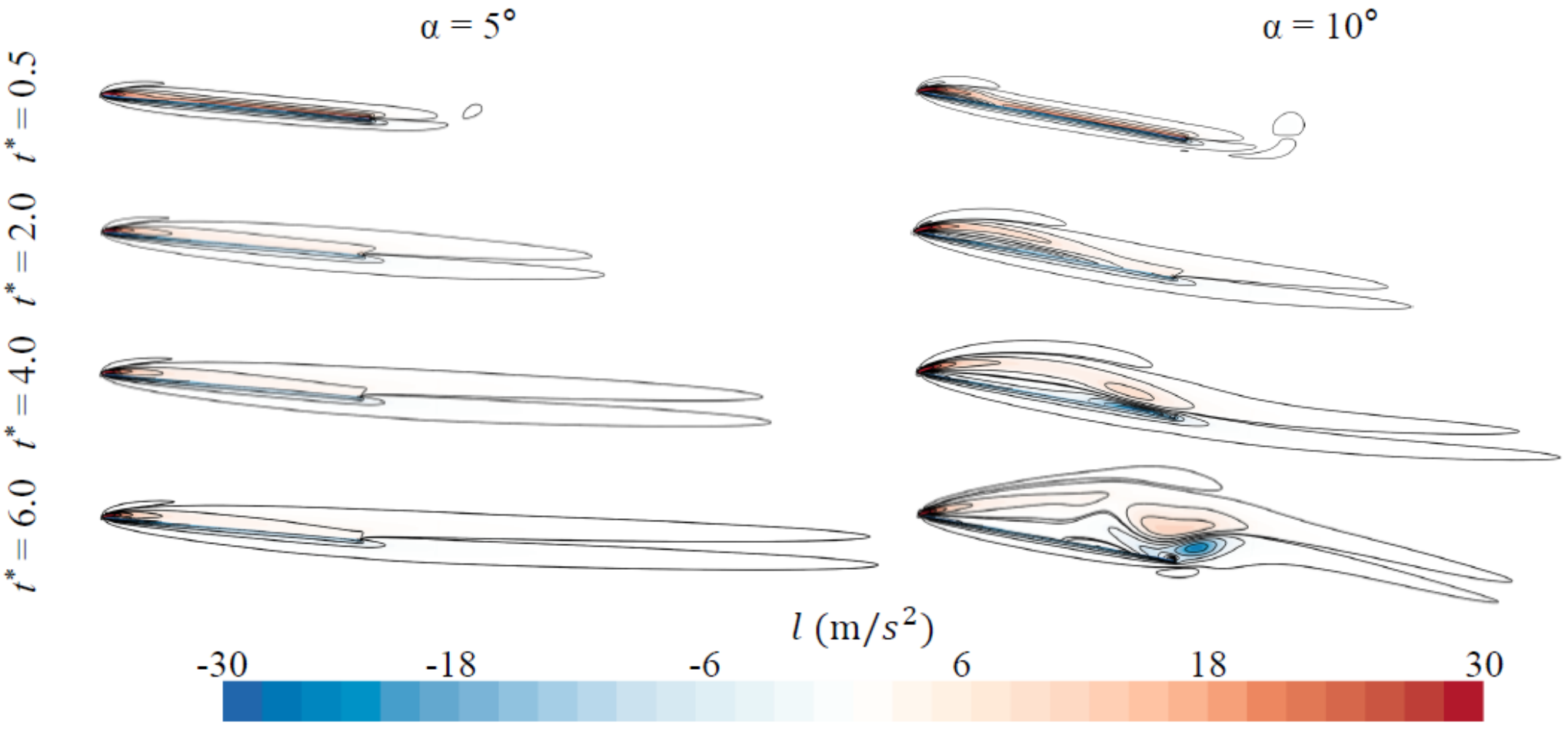

Figure 10. The Lamb vector ($l = u\omega_z$) contour maps around the flat-plate airfoil at $t^*$ = 0.5, 2, 4 and 6 for $\alpha = 5^o$ and $\alpha = 10^o$ at $Re = 1000$.

The vortex-sheet strength $\gamma(\bar{x},t)$ is calculated by using Eq. (4) from the spanwise vorticity fields on the upper and lower surfaces. Figure 11 shows the distributions of $\gamma(\bar{x},t^*)$ at $t^*$ = 0.25 and 2 for different Reynolds numbers for $\alpha = 5^o$ and $\alpha = 10^o$, where the analytical distribution $\gamma(\bar{x})$ for a steady viscous flow is also plotted for comparison. The analytical solution for $\gamma(\bar{x})$ in the viscous flow derived by Liu et al. (2017) is expressed as

$$\gamma(\bar{x}) = (1/2)U_\infty R_q \left[ \bar{x}^{2m} - \left( 1 - \Delta C_{p,TE} R_q^{-1} \right) \bar{x}^{-2m} \right], \tag{48}$$

where $\bar{x} = x/c$ is the normalized chordwise coordinate, $R_q = \left( U_{ref} / U_\infty \right)^2 Re^{2m}$ is a parameter related to the Reynolds number, $U_\infty$ is the freestream velocity, $U_{ref} = a_1 \left( \nu / U_\infty \right)^m$ is a reference velocity with a shear parameter $a_1$, $Re = U_\infty c / \nu$ is the Reynolds number, $\Delta C_{p,TE}$ is the value of the pressure coefficient difference at the trailing edge, $m = -\alpha / (\pi - \alpha)$ is a power-law exponent, and $\alpha$ denotes the AoA. For $|m| \approx \alpha / \pi << 1$ at small $\alpha$, $\gamma(x)$ is weakly dependent of $Re_c$. At the leading edge, as $\bar{x} \to 0$, there is a singularity of $\gamma(\bar{x}) \to \bar{x}^{2m}$ and the singularity is weakened as $\alpha \to 0$, which seems physically reasonable in a viscous flow. According to Eq. (48), the Kutta condition based on pressure $\Delta C_{p,TE} = 0$ leads to the Kutta condition $\gamma(\bar{x} = 1) = 0$ for the vortex sheet in the UTAT. As indicated in Fig. 11, $\gamma(\bar{x},t^*)$ at $t^*$ = 0.25 and 2 for $Re = 126\text{-}1000$ is generally smaller than the theoretical distribution for the steady viscous flow, but it tends to asymptotically approach the theoretical one when $t^*$ and $Re$ increase. Interestingly, as indicated in Fig. 11, $\gamma(\bar{x} = 1, t^*)$ at the trailing edge is not zero, deviating from the Kutta condition $\gamma(\bar{x} = 1, t^*) = 0$ particularly for $\alpha = 10^o$ at small $t^*$ (e.g., $t^* = 0.25$). At $t^* = 1000$, $\gamma(\bar{x} = 1, t^*)$ is closer to zero.

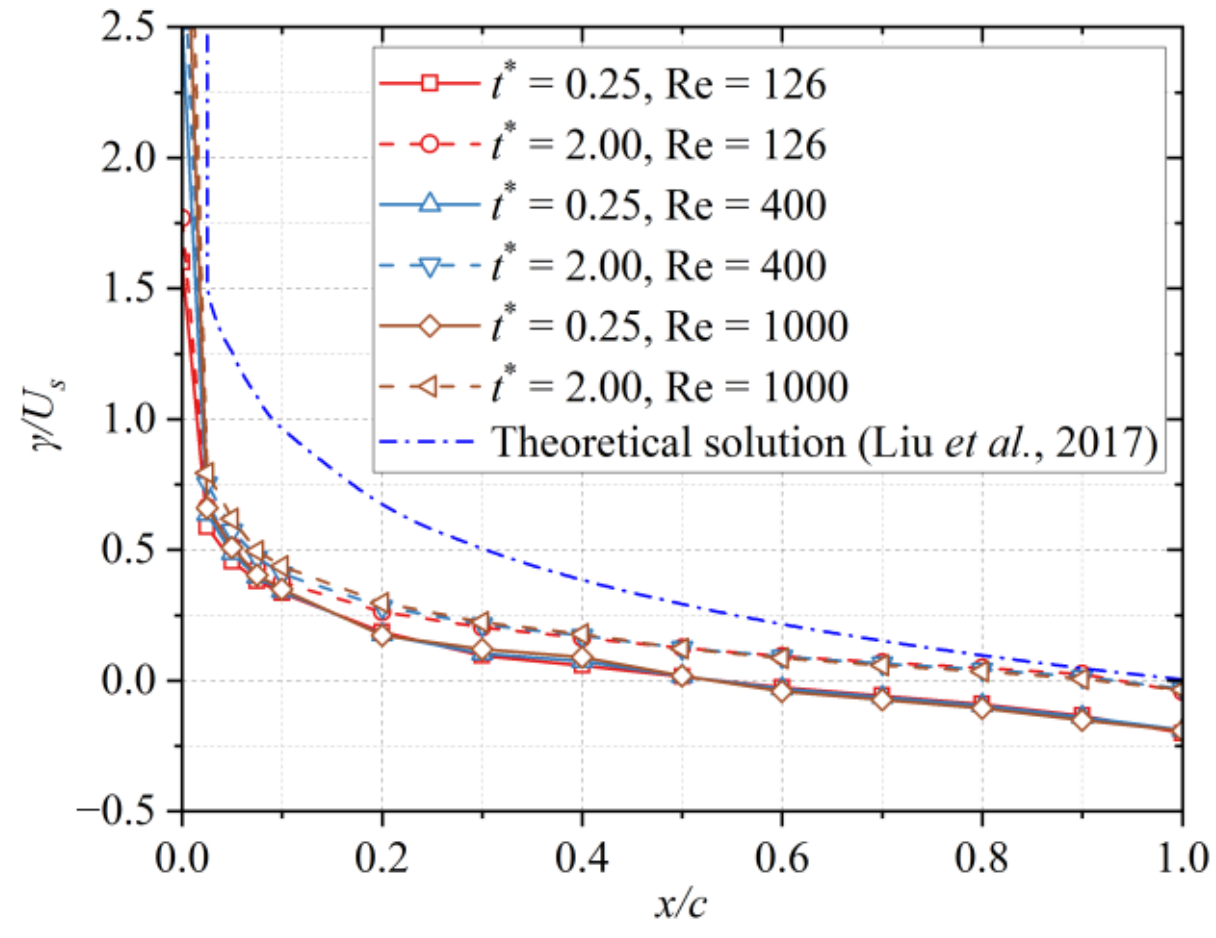

(a)

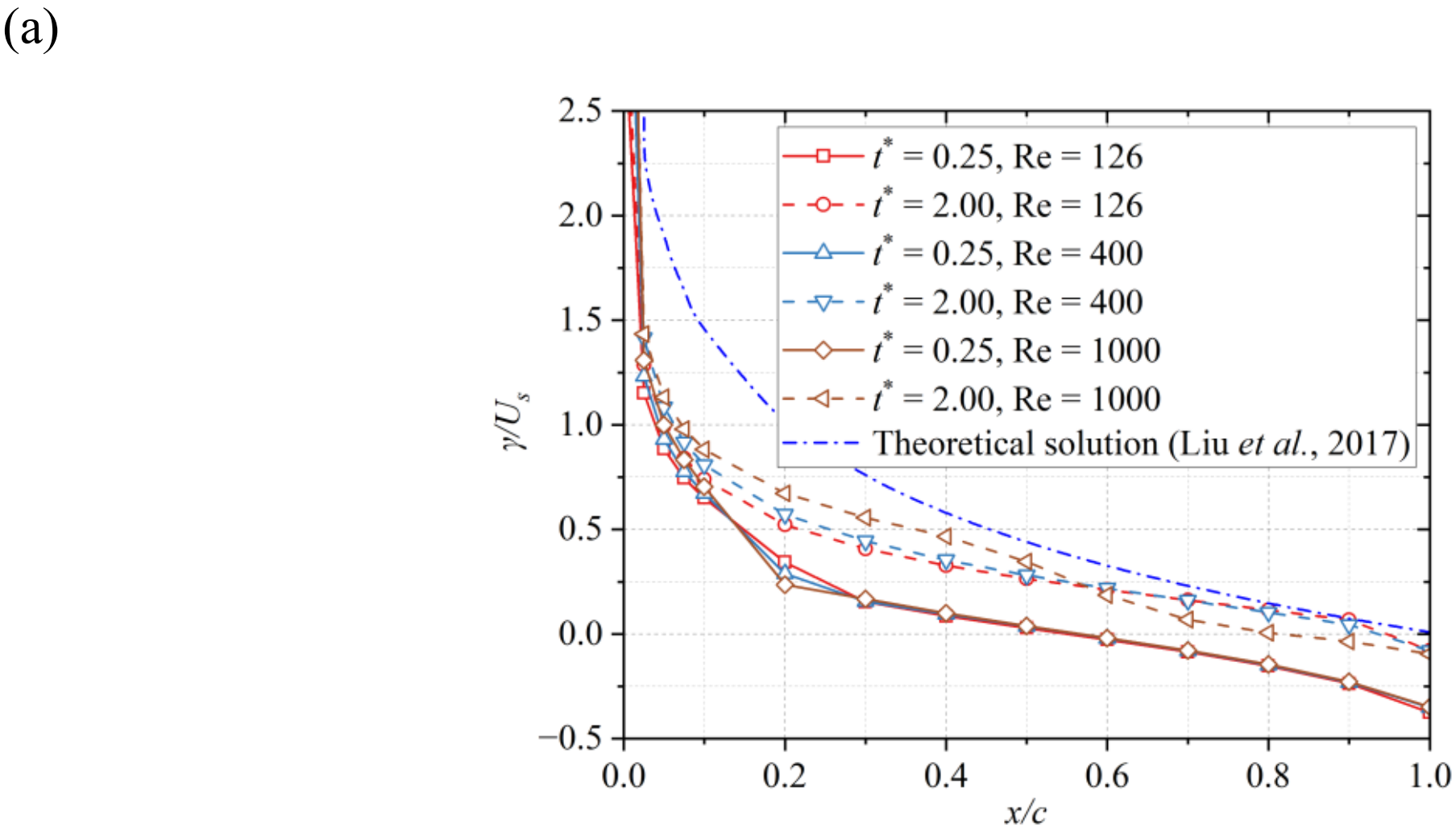

(b)

Figure 11. The vortex-sheet strength distributions of $\gamma\left(\overline{x},t^*\right)$ at $t^*$ = 0.25 and 2 for different Reynolds numbers: (a) $\alpha = 5^o$, and (b) $\alpha = 10^o$. The analytical distribution $\gamma\left(\overline{x}\right)$ for a steady viscous flow is also plotted for comparison.

For the applicability of the K-J theorem in a 2D viscous flow, Taylor (1926) provided the condition in the wake plane, which is written as

$$\int_W u\omega_z dz = 0, \tag{49}$$

where $u$ is the velocity in a boundary layer (or a shear layer), $\omega_z$ is the spanwise vorticity, and $W$ denotes the wake plane. According to Eq. (49), the positive and negative advective vorticity fluxes ($u\omega_z$) from the boundary layers on the upper and lower surfaces are cancelled out in the wake. Sears (1956, 1976) further proved that Eq. (A3) would be equivalent to the Kutta condition $\Delta C_{p,TE} = 0$. Eq. (49) is referred to as the Taylor-Sears condition providing a viscous-flow-theoretical foundation for the empirical Kutta condition. The Taylor-Sears condition is weaker than the Kutta condition. In a 2D viscous flow, the Lamb vector has one component $l = u\omega_z$. The difference between the Lamb vector integrals across the boundary layers on the upper and lower surfaces is defined as

$$\Delta l\left(\bar{x},t^*\right) = \left[u\omega_z\right]_-^+ = \int_0^{\delta^+} l^+\left(\bar{x},n,t^*\right)dn + \int_0^{\delta^-} l^-\left(\bar{x},n,t^*\right)dn, \tag{50}$$

where $n$ is the normal coordinate, $\delta$ denotes the boundary-layer thickness, and the superscripts '$+$' and '$-$' denote the quantities on the upper and lower surfaces, respectively. For a steady viscous flow, the Taylor-Sears condition is $\Delta l\left(\bar{x} = 1\right) = \left[u\omega_z\right]_-^+ = 0$. It is reported that the Taylor-Sears condition holds even in the moderately separated flow (Liu et al. 2015, 2017; Bilbao-Ludena & Papadakis 2025). Figure 12 shows the distributions of $\Delta l\left(\bar{x},t^*\right)$ at $t^*$ = 0.25 and 2 for different Reynolds numbers for $\alpha = 5^o$ and $\alpha = 10^o$. The Lamb vector difference $\Delta l\left(\bar{x},t^*\right)$ rapidly decreases to a small value around zero after a short distance from the leading edge ($x/c > 0.1$), indicating that the Taylor-Sears condition is satisfied at the TE. The Taylor-Sears condition is satisfied for $\alpha = 5^o - 10^o$ after $t^* > 0.25$, but $\Delta l\left(\bar{x} = 1,t^*\right)$ slightly deviates from zero. However, the Kutta condition is not satisfied for $\alpha = 10^o$.

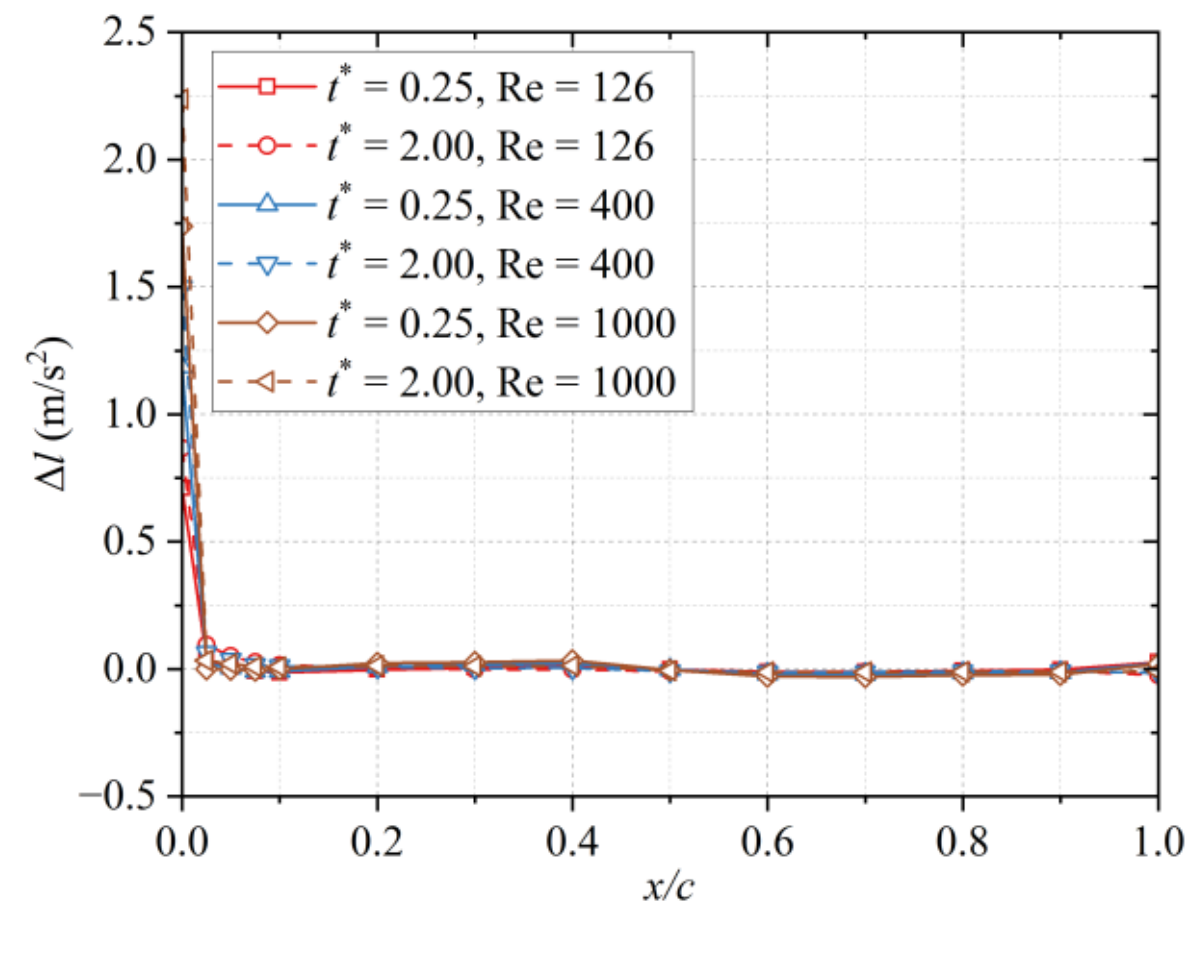

(a)

(b)

Figure 12. The distributions of the Lamb vector difference $\Delta l\left(\bar{x},t^*\right)$ at $t^*$ = 0.25 and 2 for different Reynolds numbers: (a) $\alpha = 5^o$ and (b) $\alpha = 10^o$ .

### 5.2. Lift Characteristics

Figures 13(a) and 13(b) show the total lift coefficient $C_l\left(t^*, Re\right)$ in $t^* = 0 - 2.5$ for the starting flat-plate airfoil at $\alpha = 5^o$ and $\alpha = 10^o$ for $Re = 126 - 1000$, respectively. The behavior of $C_l\left(t^*, Re\right)$ h is characterized by $C_{l\,min}\left(Re\right)$ that is achieved at a moment $t^*_{min}\left(Re\right)$ and $C_{l\,max}\left(Re\right)$ at $t^*_{max}\left(Re\right) = 2.5$. The monotonic increase of $C_l\left(t^*, Re\right)$ as $t^*$ increases is found after $t^*_{min}\left(Re\right)$ except for $Re = 126$ at $\alpha = 10^o$. Figure 14 shows the non-dimensional time $t^*_{min}\left(Re\right)$ which follows the power-law relations $t^*_{min} = 0.4\, Re^{-1/5}$ for $\alpha = 5^o$ and $\alpha = 10^o$. Based on the result in Section 4.2, we infer $t^*_{min} = \tau_{rise} \sim Re^{-1/5}$, indicating that $\tau_{rise}$ indeed represents the effect of $Re$. Figure 15 shows $C_{l\,min}\left(Re\right) = min\left(C_l\right)$ and $C_{l\,max}\left(Re\right) = max\left(C_l\right)$ for $\alpha = 5^o$ and $\alpha = 10^o$. It is noted that the total lift coefficient $C_l\left(t^*, Re\right)$ itself does not have the self-similarity discussed in Section 4.2 for the circulatory lift due to the strong influence of the lift associated with the fluid acceleration at low Reynolds numbers. Therefore, the vortex lift should be extracted from the total lift for the suitable scaling.

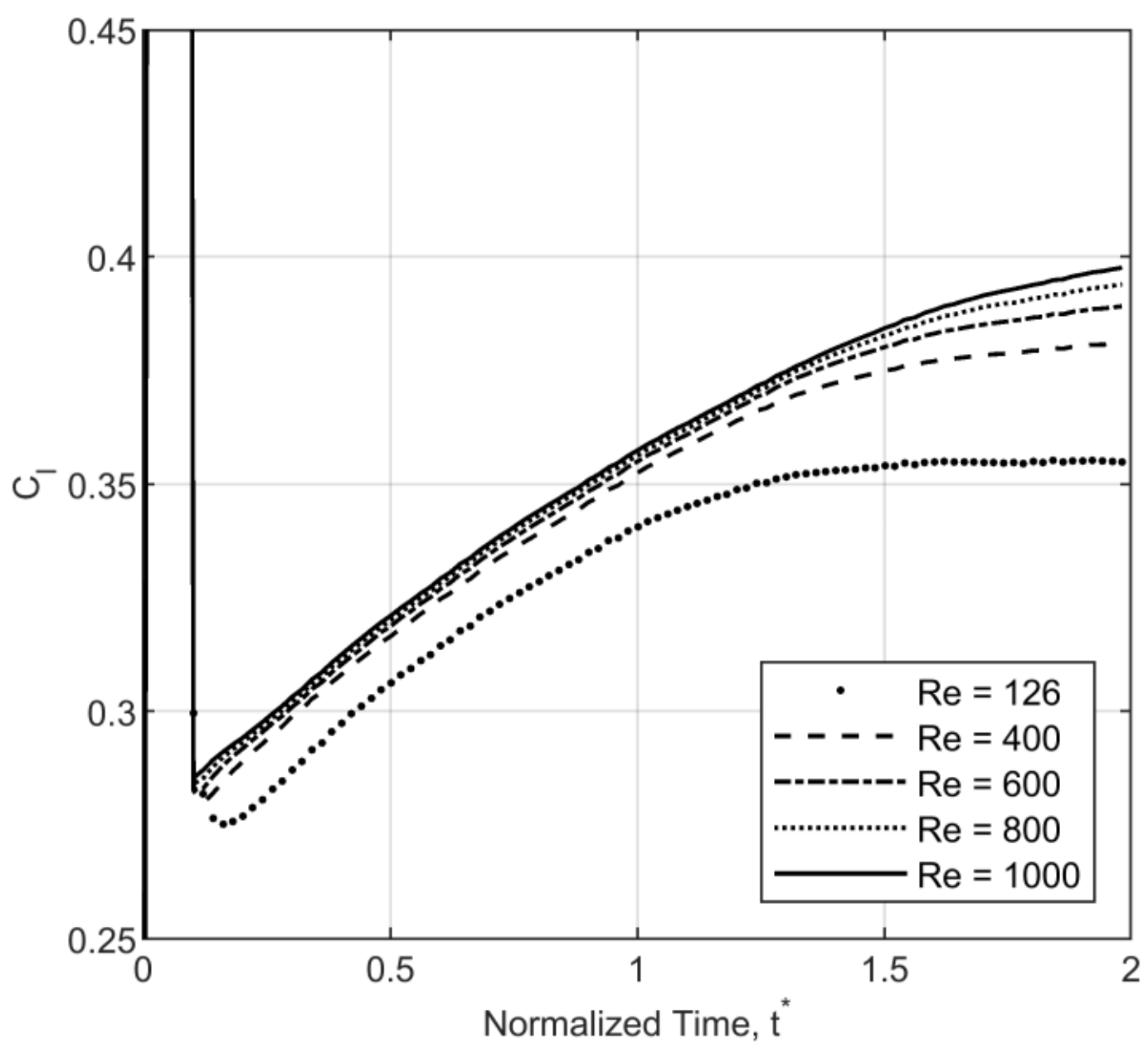

(a)

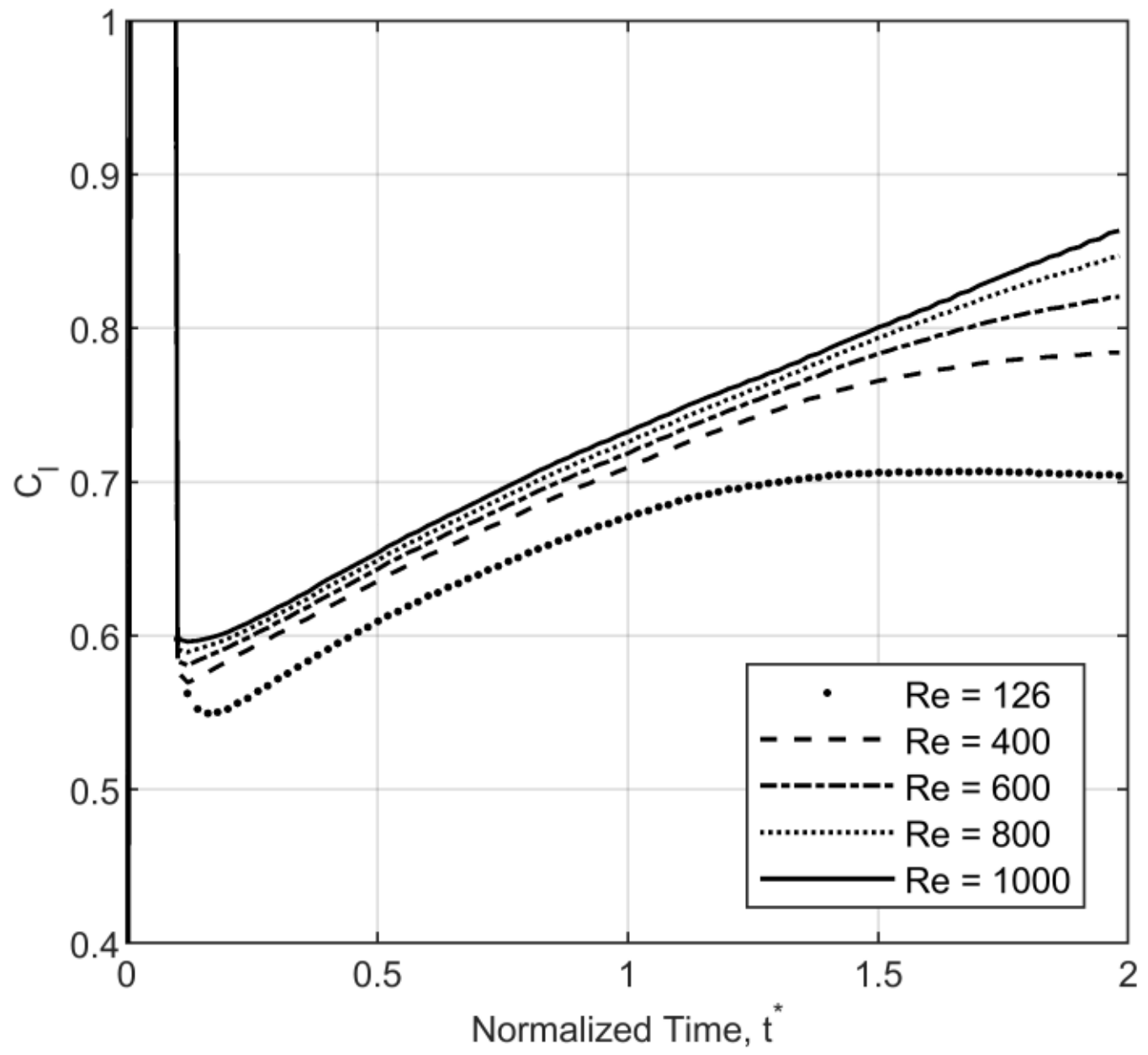

(b)

Figure 13. The lift coefficient $C_l\left(t^*, Re\right)$ of the flat-plate airfoil as a function of the non-dimensional time $t^*$ in the starting flow at different Reynolds numbers: (a) $\alpha = 5^o$, and (b) $\alpha = 10^o$.

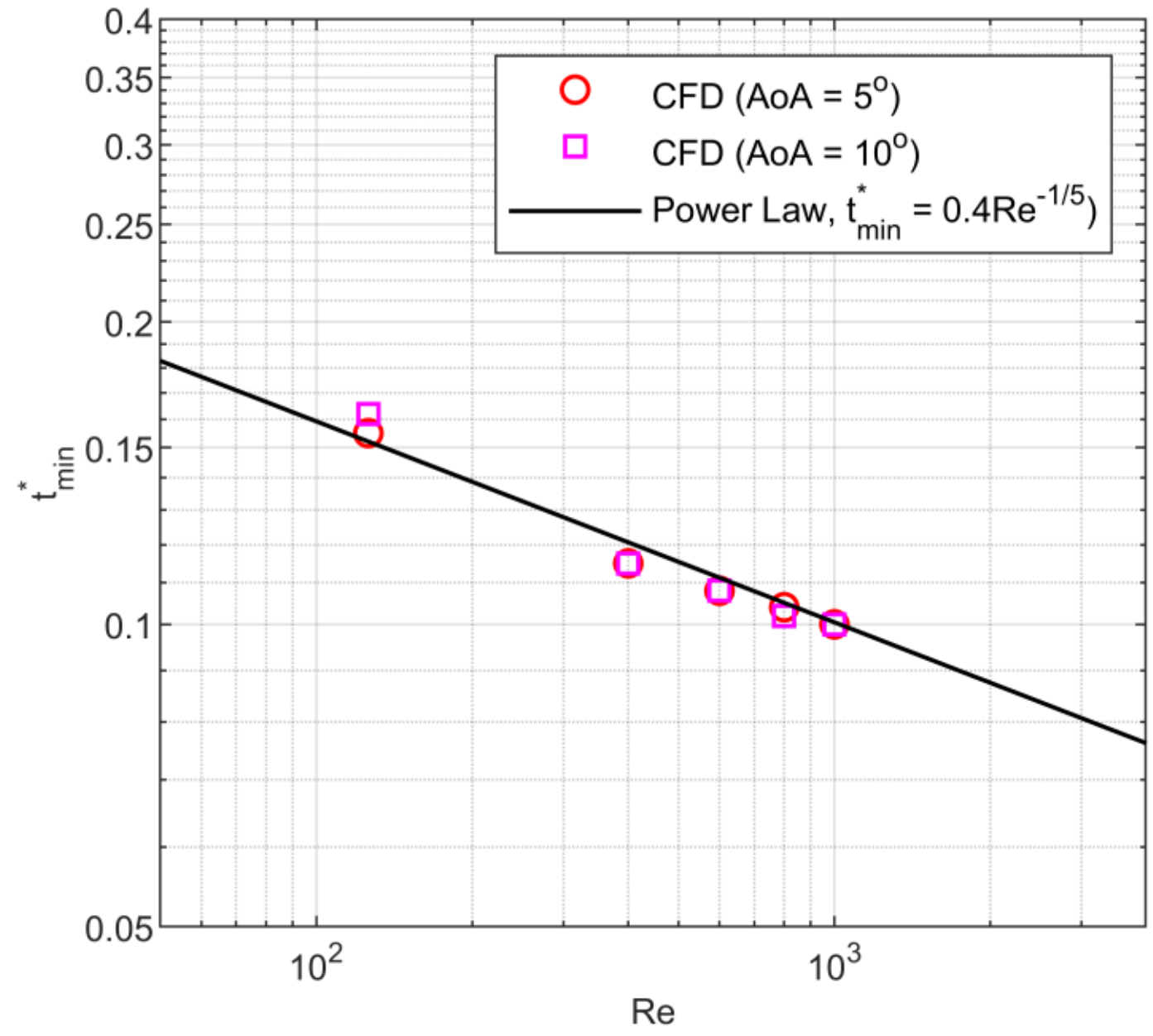

Figure 14. The non-dimensional time $t^*_{min}(Re)$ at which $C_{l,min} = min(C_l)$ is achieved in the flow over the starting flat-plate airfoil with $\alpha = 5^o$, and $\alpha = 10^o$. The power laws are plotted as references.

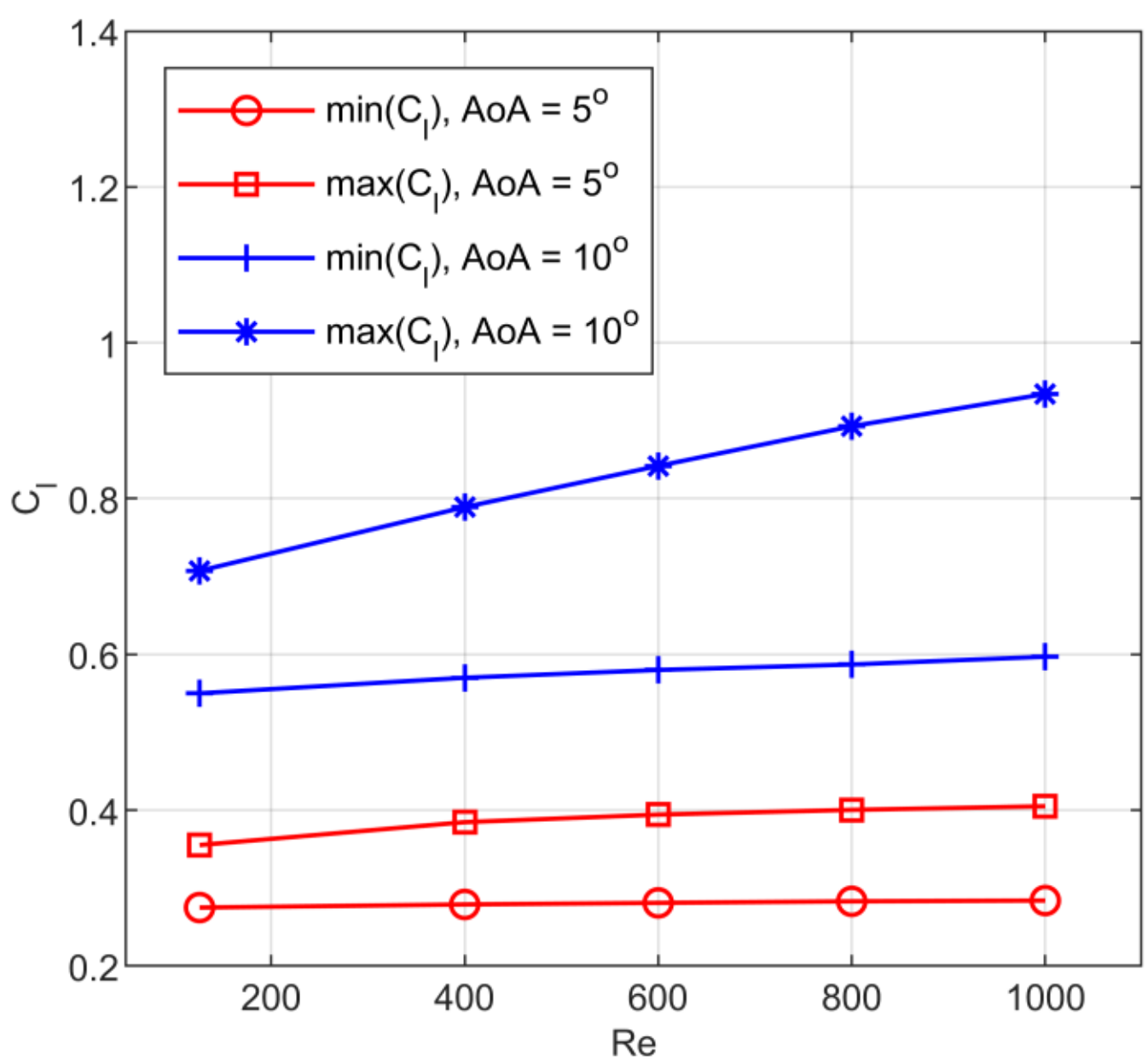

Figure 15. $C_{l\,min}(Re) = min(C_l)$ and $C_{l\,max}(Re) = max(C_l)$ in the starting flow over the flat-plate airfoil with $\alpha = 5^o$ and $\alpha = 10^o$.

According to Eq. (1), the lift coefficient can be decomposed into the two terms, i.e.,

$$C_l = C_{lvor} + C_{la}, \tag{51}$$

where $C_{l\,vor} = L'_{vor} / q_\infty c$ and $C_{l\,a} = L'_a / q_\infty c$ are the coefficients of the vortex lift and the lift associated with the fluid acceleration, respectively. The sectional lifts $L'_{vor}$ and $L'_a$ are calculated by the integrals in Eq. (1) in the rectangular flow domain. Figures 16(a) and 16(b) show $C_{l\,vor}(t^*)$ for $\alpha = 5^o$ and $\alpha = 10^o$, respectively. $C_{l\,vor}(t^*)$ monotonically increases to the maximum value $C_{l\,vor\,max} = max(C_{l\,vor})$ at $t^*_{max}$ in a finite time span, which is considered as the Wagner effect although the flow is viscous at low Reynolds numbers. Interestingly, the vortex lift coefficient $C_{l\,vor}(t^*)$ is almost independent of $Re$ for $t^* < 2$.

Since $C_{lvor}$ corresponds to $C_{lcir}$ in Section 4.2, to absorb the effect of $Re$ on $C_{lvor}\left(t^*, Re\right)$, we introduce the re-normalized vortex lift coefficient defined as

$$\bar{C}_{lvor}\left(\bar{t}^*\right) = \frac{C_{lvor}\left(\bar{t}^*, Re\right) - C_{lvor\,min}\left(Re\right)}{C_{lvor\,max}\left(Re\right) - C_{lvor\,min}\left(Re\right)}, \tag{52}$$

where the re-normalized time is

$$\bar{t}^* = \frac{t^* - t^*_{min}\left(Re\right)}{t^*_{max}\left(Re\right) - t^*_{min}\left(Re\right)}. \tag{53}$$

In this case, the re-normalized Wagner function is defined as

$$\bar{Wa}\left(\bar{t}^*\right) = \frac{Wa\left(\bar{t}^*, Re\right) - min\left(Wa\right)}{max\left(Wa\right) - min\left(Wa\right)}. \tag{54}$$

For $C_{l\,vor\,min} = 0$ and $t^*_{min} = 0$, the re-normalized vortex lift coefficient is $\bar{C}_{l\,vor}\left(\bar{t}^*\right) = C_{l\,vor}\left(\bar{t}^*, Re\right) / C_{l\,vor\,max}$, where the re-normalized time is $\bar{t}^* = t^* / t^*_{max}$.

Figures 17(a) and 17(b) show $\bar{C}_{lvor}\left(\bar{t}^*\right)$ and $\bar{Wa}\left(\bar{t}^*\right)$ for $\alpha = 5^o$ and $\alpha = 10^o$, respectively, where $C_{l\,vor\,max}$ is the value at $t^*_{max} = 2$. It is indicated that the data of $\bar{C}_{lvor}\left(\bar{t}^*\right)$ at different Reynolds numbers are collapsed onto $\bar{Wa}\left(\bar{t}^*\right)$. Even when the flow over the plate is separated in these cases, the self-similarity (the $Re$-invariance) of the re-normalized vortex lift coefficient exists, which can be described by the re-normalized Wagner function. The Wagner effect is explicitly shown in the vortex lift history, whereas it is often embedded or hidden in the total lift history due to the flow separation at large AoAs. The numerical result confirms the analysis in Section 4.2.

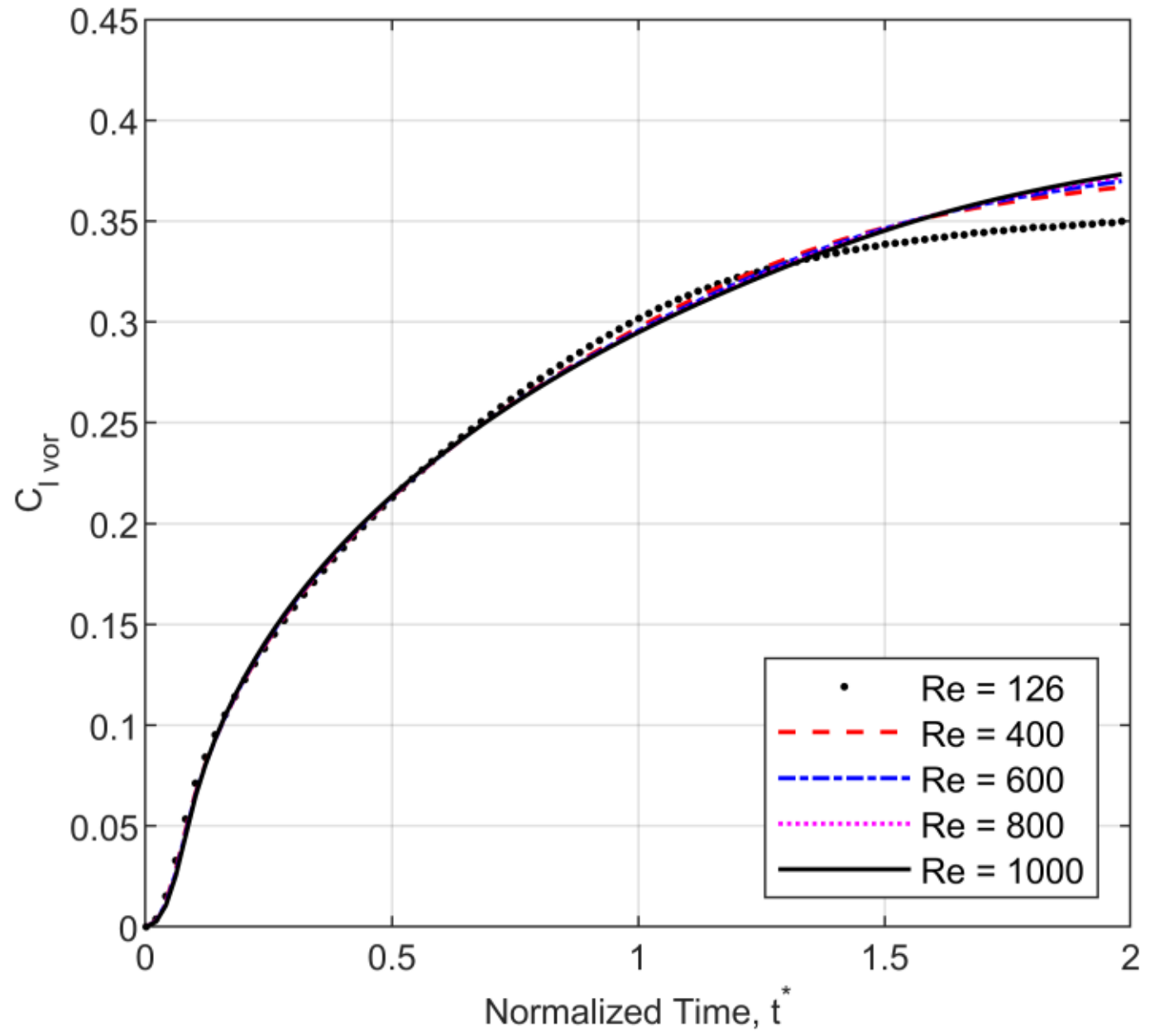

(a)

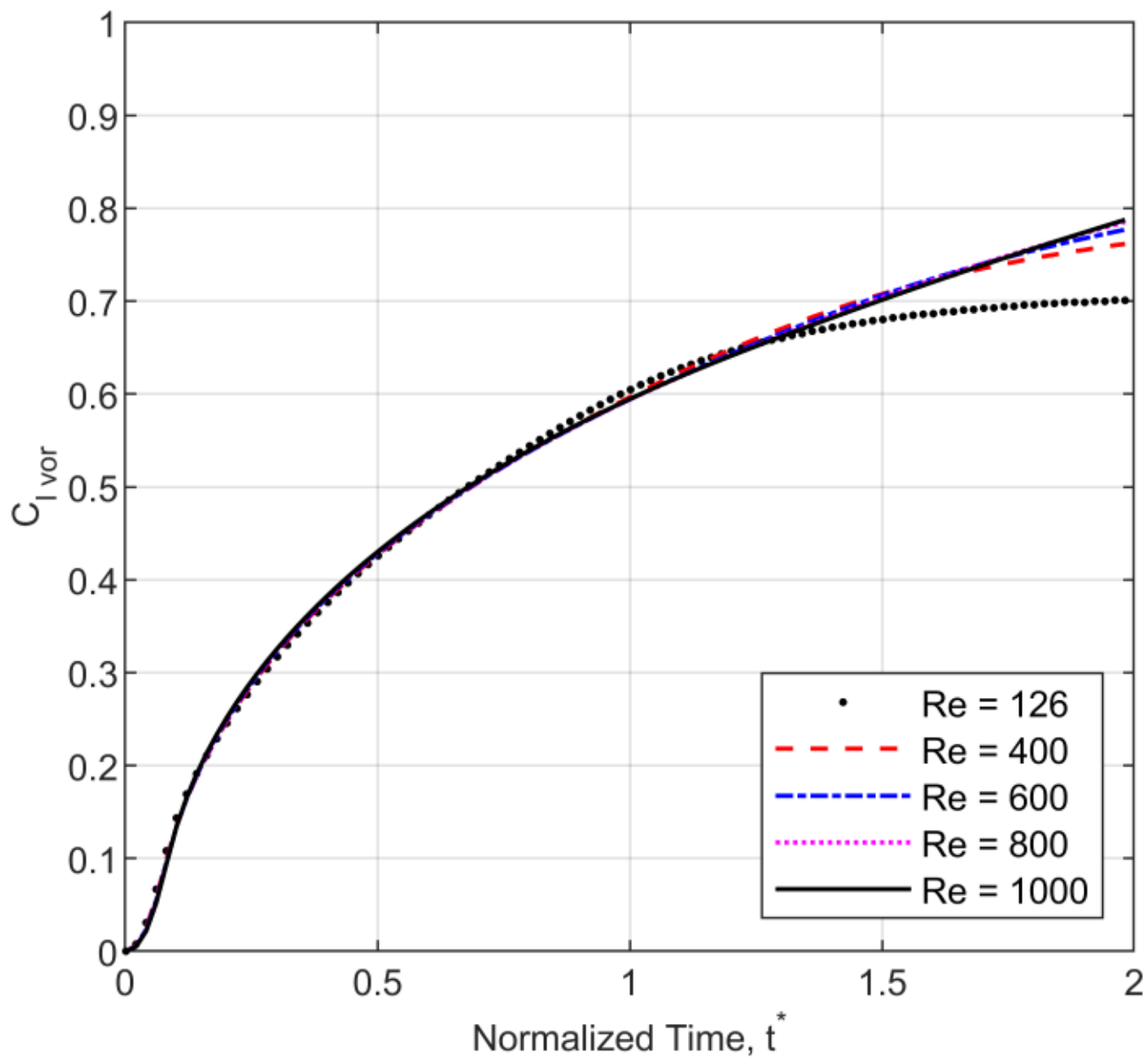

(b)

Figure 16. The vortex lift coefficient $C_{l\,vor}\left(t^{*}, Re\right)$ of the flat-plate airfoil as a function of the non-dimensional time $t^{*}$ in the starting flow at different Reynolds numbers: (a) $\alpha = 5^{o}$ and (b) $\alpha = 10^{o}$.

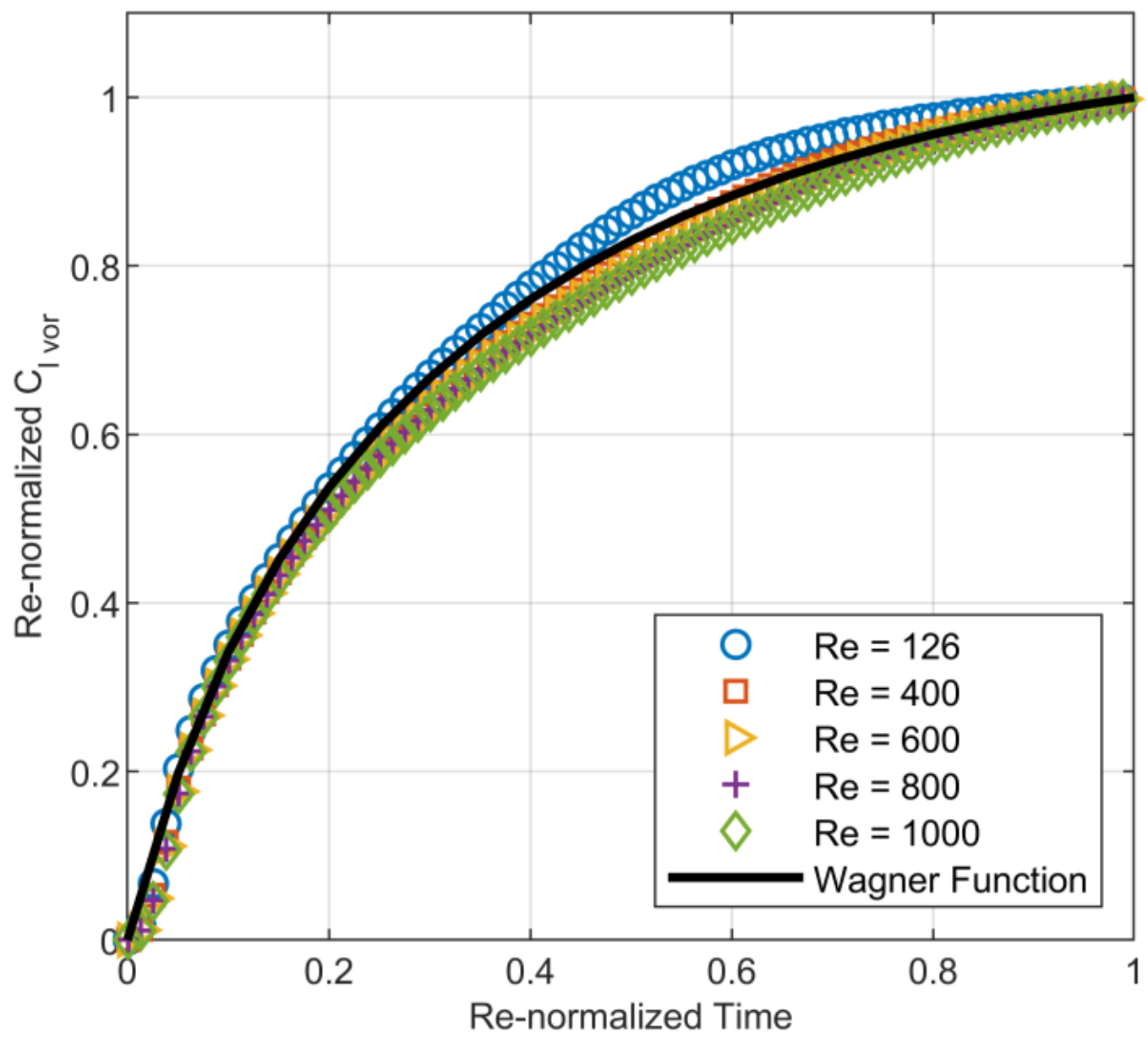

(a)

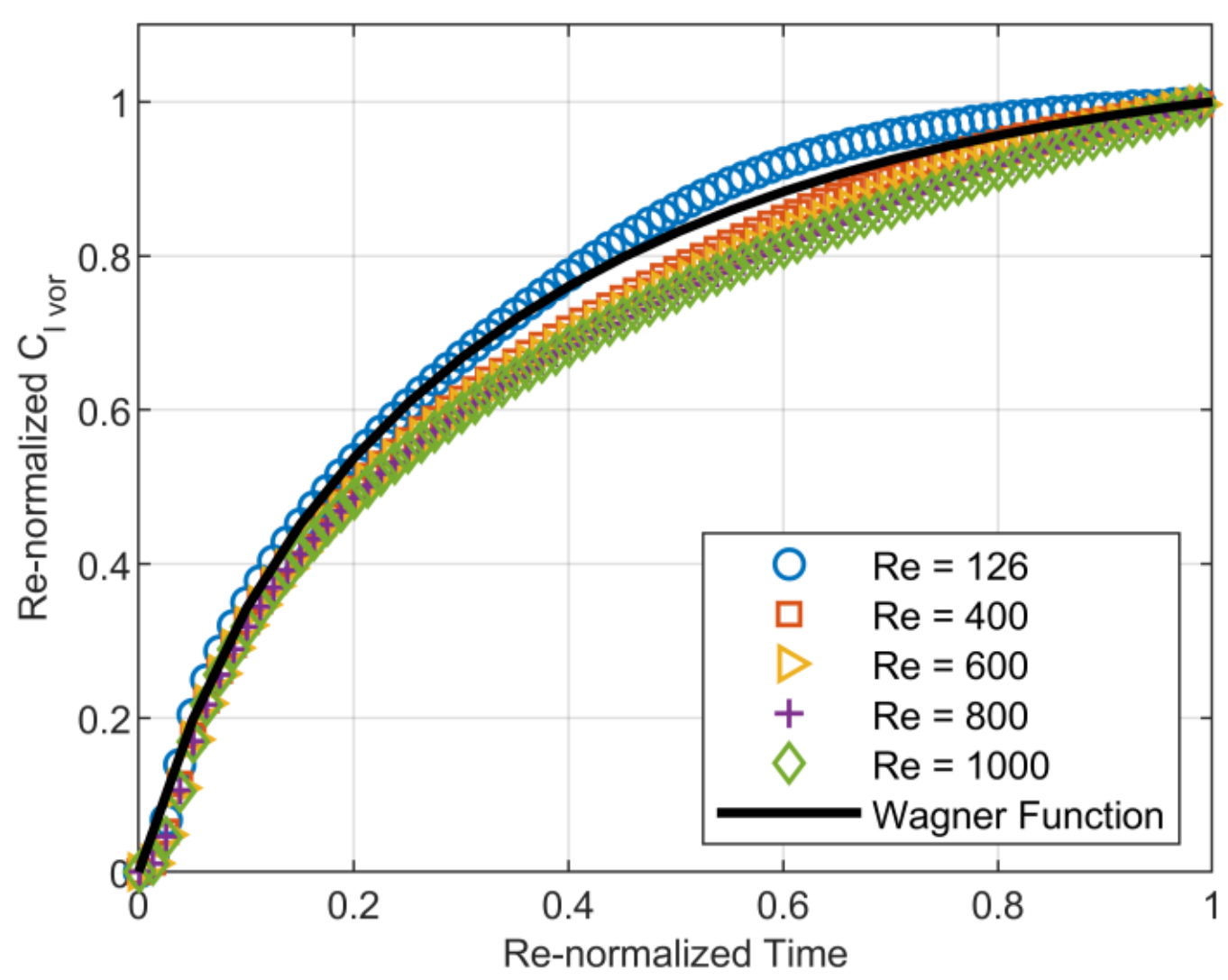

(b)

Figure 17. The re-normalized vortex lift coefficient $\bar{C}_{l\,vor}\left(\bar{t}^*\right)$ of the flat-plate airfoil as a function of the re-normalized time $\bar{t}^*$ in the starting flow at different values of $Re$: (a) $\alpha = 5^o$ and (b) $\alpha = 10^o$. The re-normalized Wagner function $\bar{W}a\left(\bar{t}^*\right)$ is plotted as a reference.

Figures 18(a) and 18(b) show $C_{l\,a}\left(t^*, Re\right)$ as a function of $t^*$ for $\alpha = 5^o$ and $\alpha = 10^o$, respectively, mainly contributing to the impulsive spike of $C_l\left(t^*, Re\right)$ near $t^* = 0$ and then decaying as $t^*$ increases. Compared to the added-mass lift with the Dirac-delta function in Eq. (36) as an inviscid-flow limit in the UTAT, $C_{l\,a}\left(t^*, Re\right)$ decays in a time domain of $t^* = 0 - 2$ and the effect of $Re$ on $C_{l\,a}\left(t^*, Re\right)$ is appreciable for $Re = 126\text{-}1000$ because the unsteady viscous flow is separated. As $Re$ increases, $C_{l\,a}\left(t^*, Re\right)$ elevates overall. $C_{l\,a}\left(t^*\right)$ monotonically decreases from the maximum value $C_{l\,a\,max} = max\left(C_{l\,a}\right)$ at $t^*_{max}$ to $C_{l\,a\,min} = min\left(C_{l\,a}\right)$ at $t^*_{min}$ in a finite time span. Figure 19 shows $C_{la,min}\left(Re\right)$ and $C_{la,max}\left(Re\right)$ in the starting flow over the flat-plate airfoil for $\alpha = 5^o$ and $\alpha = 10^o$. To absorb the effect of $Re$ on $C_{l\,a}\left(t^*, \mathrm{Re}\right)$, we introduce the re-normalized lift coefficient associated with the fluid acceleration, which is defined as

$$\bar{C}_{la}\left(\bar{t}^*\right) = \frac{C_{la}\left(\bar{t}^*, Re\right) - C_{la\,min}\left(Re\right)}{C_{la\,max}\left(Re\right) - C_{la\,min}\left(Re\right)}. \tag{55}$$

Figure 20 shows the re-normalized lift coefficient $\bar{C}_{l\,a}\left(\bar{t}^*\right)$ associated with the fluid acceleration of the flat-plate airfoil as a function of the re-normalized time $\bar{t}^*$ in the starting flow at different values of $Re$ for $\alpha = 5^o$ and $\alpha = 10^o$. Therefore, since the data of $C_{la}$ for different Reynolds numbers are collapsed, $\bar{C}_{l\,a}\left(\bar{t}^*\right)$ is self-similar. Although $C_{lvor}$ and $C_{la}$ are self-similar individually, $C_l = C_{lvor} + C_{la}$ does not enjoy the self-similarity since the characteristic scales of $C_{lvor}$ are different from those of $C_{la}$.

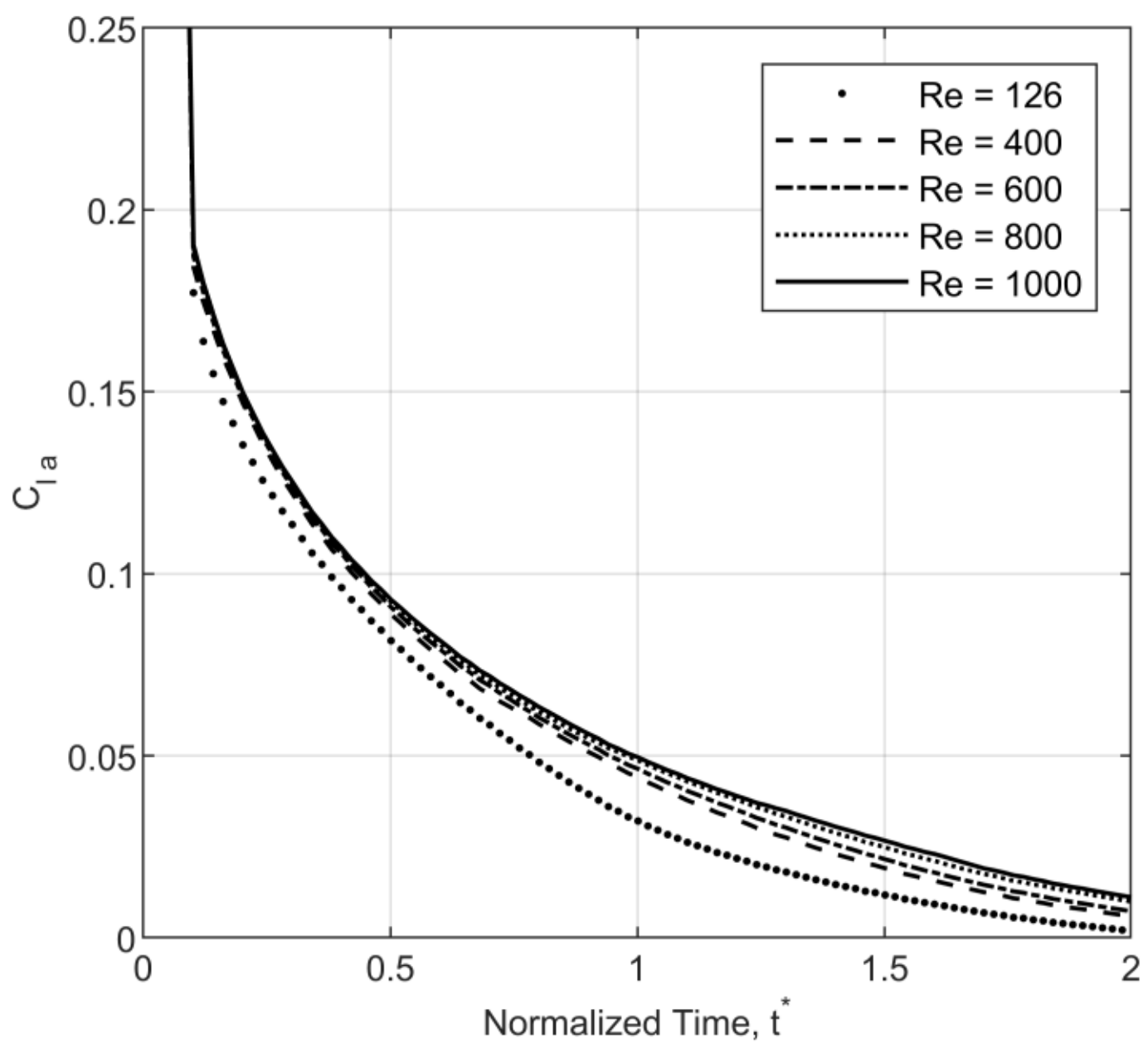

(a)

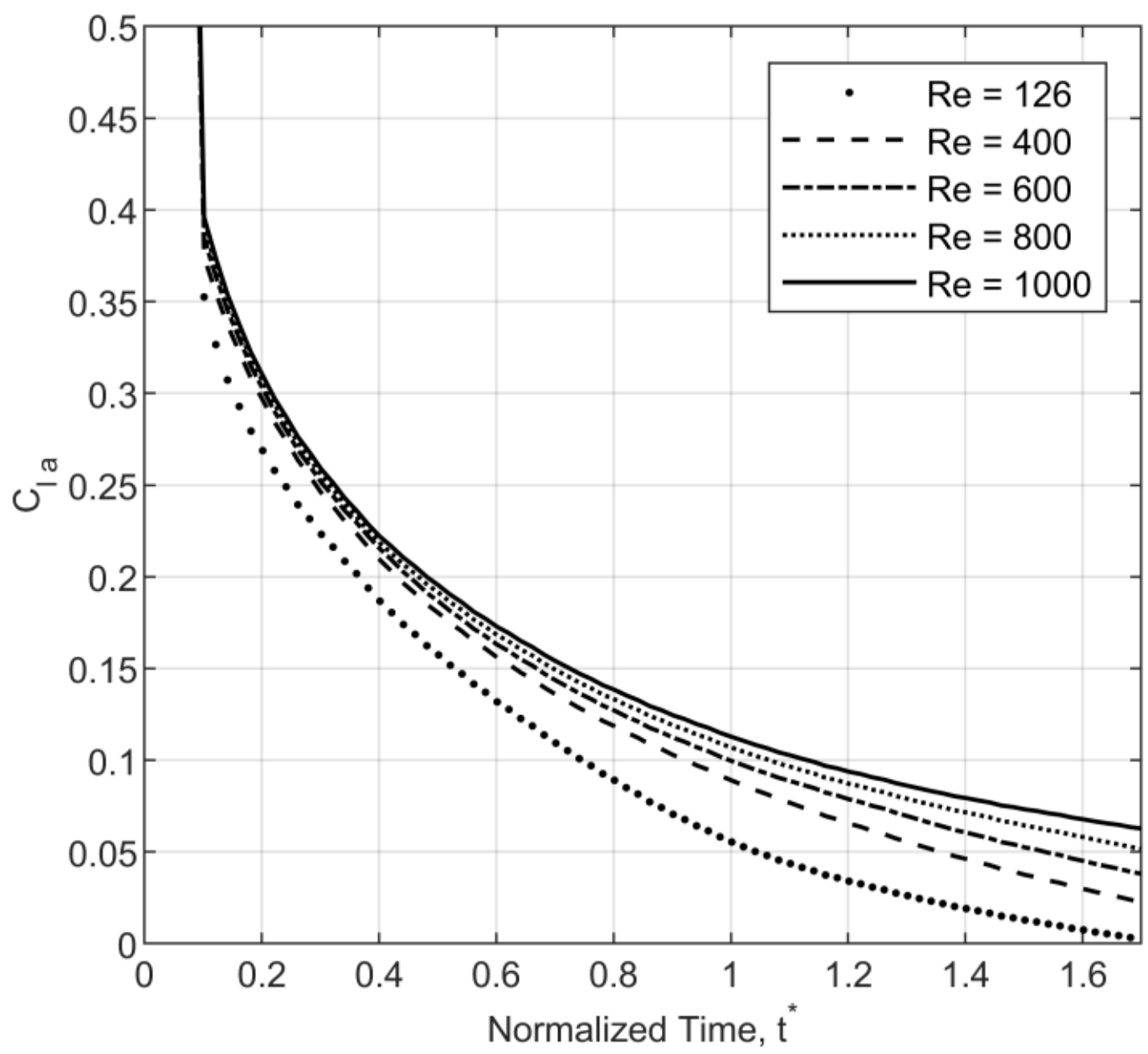

(b)

Figure 18. The lift coefficient $C_{l\,a}\left(t^{*}, Re\right)$ associated with the fluid acceleration of the flat-plate airfoil as a function of the non-dimensional time $t^{*}$ in the starting flow at different Reynolds numbers: (a) $\alpha = 5^{o}$, and (b) $\alpha = 10^{o}$.

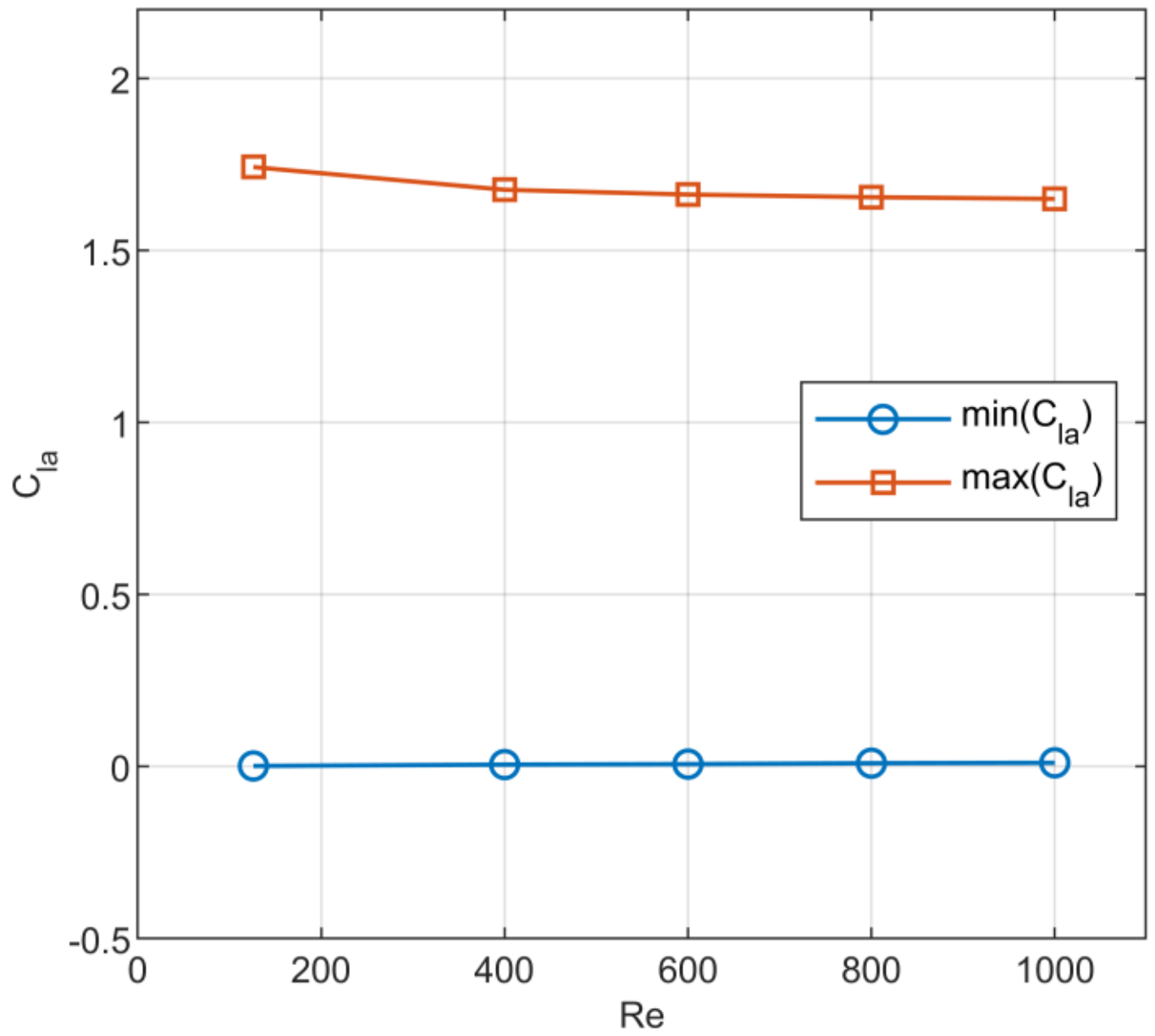

(a)

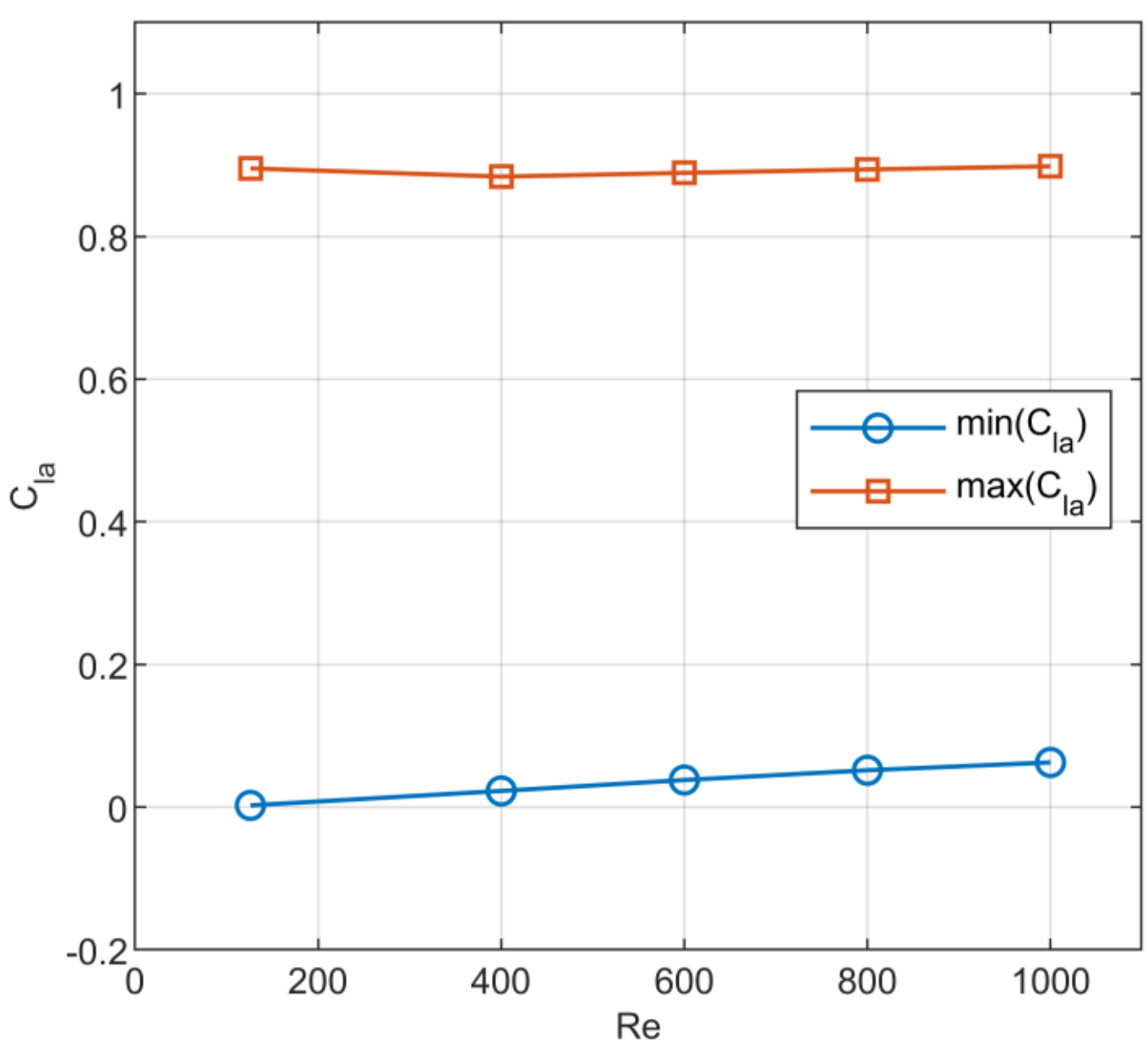

(b)

Figure 19. $C_{la,min}\left(Re\right)=min\left(C_{la}\right)$ and $C_{la,max}\left(Re\right)=max\left(C_{la}\right)$ in the starting flow over the flat-plate airfoil: (a) $\alpha=5^{o}$ and (b) $\alpha=10^{o}$.

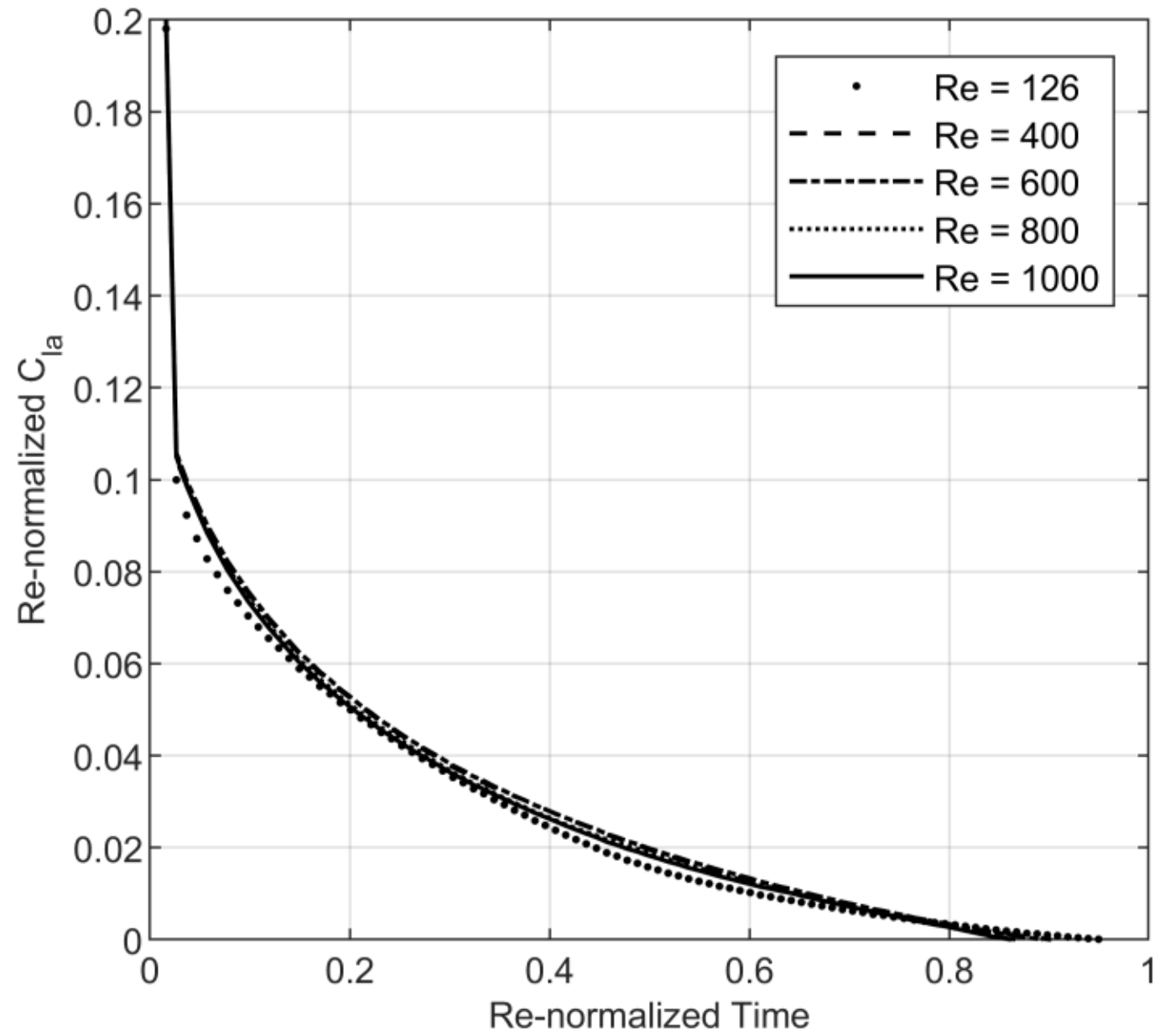

(a)

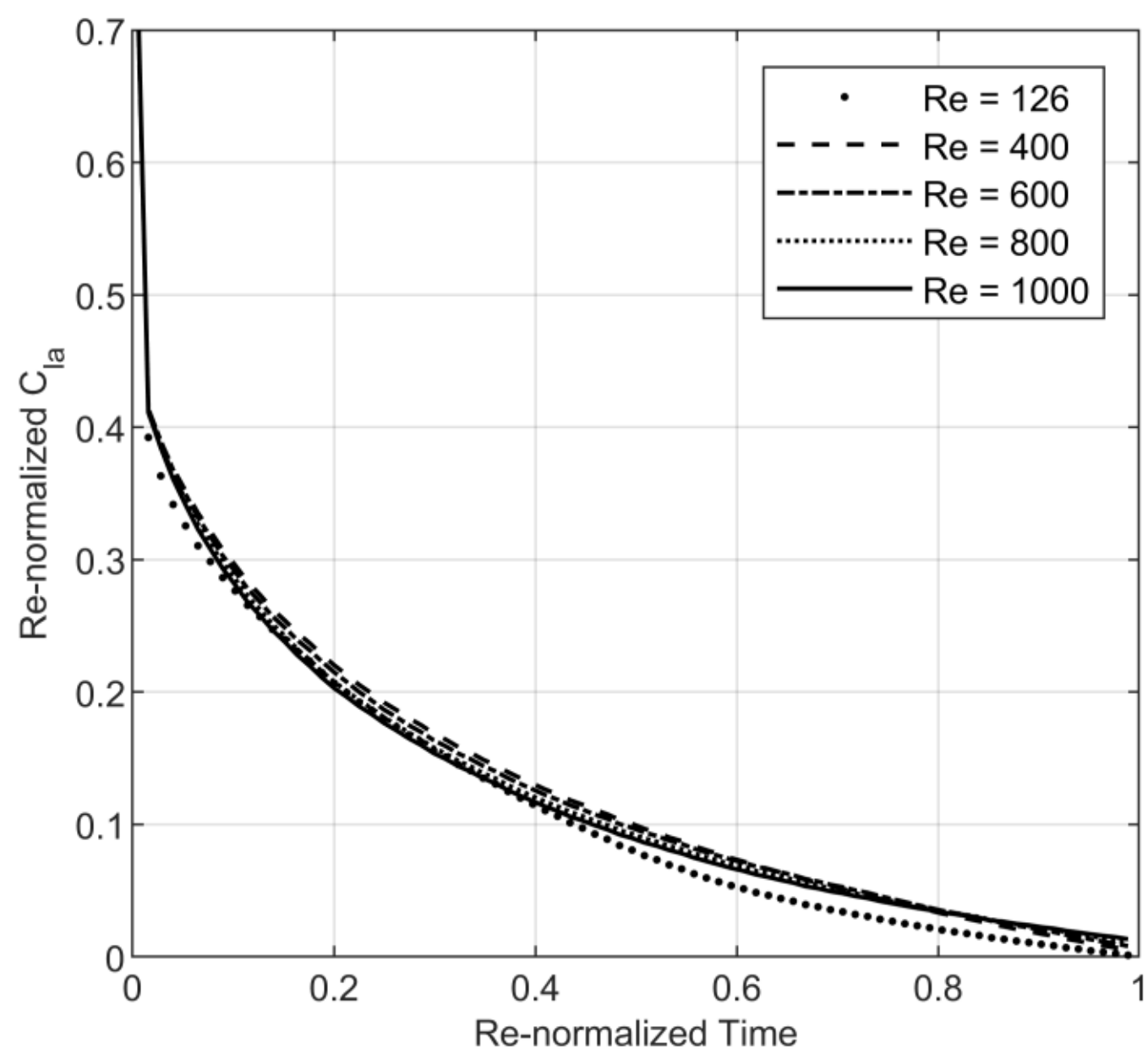

(b)

Figure 20. The re-normalized lift coefficient $\bar{C}_{l\,a}\left(\bar{t}^*\right)$ associated with the fluid acceleration of the flat-plate airfoil as a function of the re-normalized time $\bar{t}^*$ in the starting flow at different values of $Re$: (a) $\alpha = 5^o$ and (b) $\alpha = 10^o$.

## 6. Conclusions

The unsteady thin-airfoil theory (UTAT) is re-formulated in a viscous-flow framework. The explicit convolution-type expression for the wake vortex-sheet strength is obtained by solving the Wagner integral equation where the kernel is approximated by a suitable function that allows the analytical inverse Laplace transform. Therefore, the unsteady lift associated with the wake effect is expressed in an explicit analytical form, and then the total unsteady lift is calculated by using the Kármán-Sears lift formula. This analytical solution applied to the classical Wagner problem leads to the explicit integral form of the Wagner function for easy numerical integration. Further, the analytical solution is applied to the starting flow with a finite rising timescale in the generalized Wagner problem, revealing the self-similarity of the re-normalized circulatory lift coefficient $\bar{C}_{lcir}\left(\bar{t}^*\right)$ in a range of the rising timescales and its equivalence to the re-normalized Wagner function $\bar{Wa}\left(\bar{t}^*\right)$ in a finite time domain. Since the rising timescale is related to the Reynolds number ($Re$), this self-similarity represents the Reynolds-number-invariance (the $Re$-invariance), indicating the applicability of the Wagner function in viscous flows.

To examine the $Re$-invariance of the re-normalized vortex lift coefficient $\bar{C}_{lvor}\left(\bar{t}^*\right)$ [corresponds to $\bar{C}_{lcir}\left(\bar{t}^*\right)$] and its equivalence to $\bar{Wa}\left(\bar{t}^*\right)$, numerical simulation of the flow over a starting flat-plate airfoil is conducted for $\alpha = 5^o$ and $\alpha = 10^o$ at low Reynolds numbers ($Re = 126\text{-}1000$). The total lift coefficient $C_l\left(t^*, \mathrm{Re}\right)$ is decomposed into the vortex lift coefficient $C_{l\,vor}\left(t^*, \mathrm{Re}\right)$ and the lift coefficient $C_{l\,a}\left(t^*, \mathrm{Re}\right)$ associated with the fluid acceleration. In the cases of $\alpha = 5^o$ and $\alpha = 10^o$ with moderate flow separation on the upper surface of the airfoil, the data of $\bar{C}_{l\,vor}\left(t^*\right)$ at different Reynolds numbers are collapsed approximately onto $\bar{W}a\left(\bar{t}^*\right)$. The numerical results confirm the self-similarity (the $Re$-invariance) of $\bar{C}_{l\,vor}\left(t^*\right)$ in a finite time domain and the applicability of the Wagner function in viscous flows.

Furthermore, the analytical solution is applied to the generalized Theodorsen problem in which the general unsteady motion of a thin airfoil is described by a Fourier series. The formal solution of the unsteady lift is expressed in the form of the discrete inverse Fourier transform, which is considered as the generalized Theodorsen function. The convolution-like form of this formal solution resembles the indicial response formulation, elucidating the connection between the Wagner function and the Theodorsen function in a unified framework. The explicit integral form of the Theodorsen function is given as a reduced case of the formal solution.

## Appendix A. Oscillating Motion

### A1. Generalized Theodorsen's Problem

The analytical solution is applied to the general oscillation of a thin airfoil with the effective AoA expressed as a Fourier series, which represents the oscillating heaving, pitching and gust modes (Leishman 2006). This is considered as a generalized case of the Theodorsen problem (Theodorsen 1935). The formal analysis in this section presents the generalized Theodorsen theory which not only gives an explicit integral form of the Theodorsen function for a single-frequency steady oscillation as a reduced case, but also elucidates its relation to the Wagner function. In addition, it provides a connection between the generalized Theodorsen theory and the indicial response theory.

A general form of the effective AoA is expressed as a Fourier series, i.e.,

$$\alpha_{eff}\left(t^*\right)=\sum_{n=-\infty}^{\infty} d_n \, exp\left(i\, n\, \omega^* t^*\right), \tag{A1}$$

where $\omega^* = \omega c / U$ is the reduced frequency based on the chord length $c$ and $i=\sqrt{-1}$ is the imaginary number. In Eq. (A1), the Fourier coefficients are

$$d_n = \frac{1}{T}\int_0^T \alpha_{eff}\left(t'\right) exp\left(i\, n\omega^* t'_n\right) dt', \tag{A2}$$

where $T = 2\pi c / U\omega^*$ is a period, and $t'_n = t' I\left(n\right)$ is the indexed time. Here, the index function is defined as $I\left(n\right)=-1$ for $n>0$, $I\left(n\right)=1$ for $n<0$ and $I\left(n\right)=0$ for $n=0$. The quasi-steady circulation and its time derivative are, respectively,

$$\Gamma_0\left(t^*\right) = c U \pi\, \alpha_{eff}\left(t^*\right), \tag{A3}$$

$$\Gamma'_0\left(t^*\right) = c U \pi \frac{d\alpha_{eff}\left(t^*\right)}{dt^*}. \tag{A4}$$

Substitution of Eqs. (A3) and (A4) to Eq. (16) gives the wake vortex-sheet strength, i.e.,

$$\gamma_w\left(s\right) = -U\sum_{n=-\infty}^{\infty} Y\left(s, n\omega^*\right) d_n \, exp\left(i\, n\, \omega^* s\right). \tag{A5}$$

The amplitude $Y\left(s, n\omega^*\right)$ is defined as

$$Y\left(s, n\omega^*\right) = B_1\left(in\omega^*\right) J_1\left(s, n\omega^*\right) - B_0 J_0\left(s, n\omega^*\right), \tag{A6}$$

where the coefficients are $B_0 = c_2\sqrt{\pi}/2$ and $B_1 = c_1\sqrt{\pi}$, and the functions $J_0$ and $J_1$ are defined as

$$J_0\left(s, n\omega^*\right) = \int_0^s exp\left(-i\, n\omega^* t'\right) S_0\left(t'\right) dt', \tag{A7}$$

$$J_1\left(s, n\omega^*\right) = \int_0^s exp\left(-i\, n\omega^* t'\right) S_1\left(t'\right) dt'. \tag{A8}$$

Eq. (A5) indicates that $\gamma_w\left(s\right)$ is linearly contributed by all the Fourier modes of $\alpha_{eff}\left(t^*\right)$ through $Y\left(s, n\omega^*\right)$ as a transfer function. Substitution of Eq. (A5) to Eq. (20) leads to the lift associated with the wake effect, i.e.,

$$L_2'\left(t^*\right) = -\pi\rho U^2 c \sum_{n=-\infty}^{\infty} Z\left(t^*, n\omega^*\right) d_n \exp\left(i\, n\, \omega^* t^*\right), \tag{A9}$$

where the lift deficiency function for the *n*th harmonic component (the *n*th Fourier mode) is defined as

$$Z\left(t^*, n\omega^*\right) = \pi^{-1} \int_0^{t^*} \exp\left(-i\, n\omega^* \eta\right) Y\left(t^* - \eta, n\omega^*\right) P(\eta)\, d\eta\,, \tag{A10}$$

and $P(\eta) = 0.5\eta^{-1/2}(\eta+1)^{-1/2}$. Eq. (A9) indicates that $L_2'\left(t^*\right)$ is linearly contributed by all the Fourier modes of $\alpha_{eff}\left(t^*\right)$ through a transfer function $Z\left(t^*, n\omega^*\right)$ that is essentially the generalized Theodorsen deficiency function for the *n*th Fourier mode.

Therefore, the circulatory lift is given by

$$L_{cir}'\left(t^*\right) = L_0'\left(t^*\right) + L_2'\left(t^*\right) = \pi\rho U^2 c\, \alpha_{eff}\left(t^*\right)\left[1 - \hat{Z}\left(t^*, \omega^*\right)\right], \tag{A11}$$

where the lift deficiency function averaged over all the harmonic components is defined as

$$\hat{Z}\left(t^*, \omega^*\right) = \frac{\sum_{n=-\infty}^{\infty} Z\left(t^*, n\omega^*\right) d_n \exp\left(i\, n\, \omega^* t^*\right)}{\sum_{n=-\infty}^{\infty} d_n \exp\left(i\, n\, \omega^* t^*\right)}\,. \tag{A12}$$

The complex function $Z\left(t^*, n\omega^*\right)$ is the generalized Theodorsen deficiency function for the *n*th Fourier mode. For the first Fourier mode ($n = 1$) of steady oscillation as $t^* \to \infty$, $Z\left(\infty, \omega^*\right)$ is the Theodorsen lift deficiency function and $1 - Z\left(\infty, \omega^*\right)$ is the Theodorsen function. By including the added-mass lift

$$L_1'\left(t^*\right) = \frac{\pi}{4} \rho U^2 c \frac{d\alpha_{eff}\left(t^*\right)}{dt^*}, \tag{A13}$$

the lift coefficient is written as

$$C_l = \frac{L'\left(t^*\right)}{\rho U^2 c / 2} = 2\pi\alpha_{eff}\left(t^*\right)\left[1 - \hat{Z}\left(t^*, \omega^*\right)\right] + \frac{\pi}{2} \frac{d\alpha_{eff}}{dt^*}\,. \tag{A14}$$

According tp Eq. (A12), $\hat{Z}\left(t^*, \omega^*\right)$ is interpreted as the weight-averaged Theodorsen deficiency function over all the Fourier modes.

Interestingly, Eqs. (A5) and (A9) have the same linear superposition structure in the frequency domain, where $\gamma_w(s)$ and $L_2'(t^*)$ are independently contributed by all the Fourier modes of $\alpha_{eff}(t^*)$ through $Y(s,n\omega^*)$ and $Z(t^*,n\omega^*)$ as the transfer functions, respectively. Equivalently, $\gamma_w$ in Eq. (A5) and $L_2'$ in Eq. (A9) can be expressed in an integral form in the time domain. Substitution of Eq. (A2) to Eq. (A5) yields

$$\gamma_w(s) = -\frac{U}{T}\int_0^T E_\gamma(s,t')\alpha_{eff}(t')dt', \tag{A15}$$

where the Green's function for the wake vortex-sheet strength is

$$E_\gamma(s,t') = \sum_{n=-\infty}^{\infty} Y(s,n\omega^*) exp\left[i\,n\,\omega^*(s+t_n')\right]. \tag{A16}$$

$E_\gamma(s,t')$ represents the total transfer function of all the Fourier modes for $\gamma_w(s)$. Similarly, substitution of Eq. (A2) to Eq. (A9) gives the wake-related lift, i.e.,

$$L_2'(t^*) = -\frac{\pi}{T}\rho U^2 c\int_0^T E_l(t^*,t')\alpha_{eff}(t')dt', \tag{A17}$$

where the Green's function for the lift is

$$E_l(t^*,t') = \sum_{n=-\infty}^{\infty} Z(t^*,n\omega^*) exp\left[i\,n\,\omega^*(t^*+t_n')\right], \tag{A18}$$

which represents the total transfer function of all the Fourier modes for $L_2'(t^*)$ through the generalized Theodorsen deficiency function $Z(t^*,n\omega^*)$. $E_l(t^*,t')$ can be also interpreted as the interaction function of all the Fourier modes between the times $t^*$ and $t'$. Both Eqs. (A16) and (A18) are expressed in the form of the discrete inverse Fourier transform (DIFT). The circulatory lift is given by

$$\begin{aligned} L_{cir}'(t^*) &= L_0'(t^*) + L_2'(t^*) \\ &= \pi\rho U^2 c\left[\alpha_{eff}(t^*) - \frac{1}{T}\int_0^T E_l(t^*,t')\alpha_{eff}(t')dt'\right]. \end{aligned} \tag{A19}$$

By applying integration by parts, Eq. (A19) is written as

$$L'_{cir}\left(t^*\right)=\pi\rho U^2 c\left[\alpha_{eff}\left(T\right)\phi\left(t^*\right)+\int_0^T F_l\left(t^*,t'\right)\frac{d\alpha_{eff}}{dt'}dt'\right]. \quad \text{(A20)}$$

where the relevant functions are defined as (A21)

$$F_l\left(t^*,t'\right)=\frac{1}{T}\int_0^{t'}E_l\left(t^*,t''\right)dt'', \quad \text{(A22)}$$

$$\phi\left(t^*\right)=\frac{\alpha_{eff}\left(t^*\right)}{\alpha_{eff}\left(T\right)}-F_l\left(T,t^*\right). \quad \text{(A23)}$$

In Eq. (A20), the Green's function $F_l\left(t^*,t'\right)$ represents the accumulated effect of the interaction function $E_l\left(t^*,t''\right)$ that is linearly contributed by all the Fourier modes of $\alpha_{eff}\left(t^*\right)$ through $Z\left(t^*,n\omega^*\right)$. The mathematical form of Eq. (A20) resembles the indicial response formulation (Leishman 1988, 2006). In the indicial response theory, it is further assumed that $F_l\left(t^*,t'\right)$ is symmetric, i.e., $F_l\left(t^*,t'\right)=F_l\left(\left|t^*-t'\right|\right)$, and the upper limit $T$ is replaced by $t^*$. Eq. (A20) provides an alternative theoretical perspective on the indicial response formulation widely used in engineering problems (Leishman 1988, 2006; Jones et al. 2022).

For the Wagner problem, the step change $\alpha_{eff}\left(t^*\right)=\alpha_{00}H\left(t^*\right)$ is considered, where $\alpha_{00}$ is a constant magnitude and $H\left(t^*\right)$ is the Heaviside function. Eq. (A20) is reduced to

$$L'_{cir}\left(t^*\right)=\pi\rho U^2 c\alpha_{00}\left[1-F_l\left(T,t^*\right)\right]. \quad \text{(A24)}$$

Therefore, $1-F_l\left(T,t^*\right)$ is an alternative form of the Wagner function. The Wagner lift deficiency function $F_l\left(T,t^*\right)$ is related to the time-averaged the discrete inverse Fourier transform (DIFT) of the Theodorsen deficiency function $Z\left(t^*,\omega^*\right)$. This result mirrors the analysis given by Garrick (1938), Sears (1940) and Peters (2008) on the relationship between the Wagner function and the Theodorsen function. Interestingly, Sears (1940) related the step function response to the steady oscillation response by replacing the Laplace transform variable $p$ with the complex frequency $i\omega$. Thus, the Wagner function in the Laplace transform domain is directly transformed to the Theodorsen function in the frequency domain. In summary, the formal solution of the unsteady lift provides a unified framework where the Wagner function and the Theodorsen function are reduced as the special cases and they are related. In addition, the convolution-like form of the formal solution lays the theoretical foundation for the indicial response formulation.

For a steady oscillation, as $t^* \to \infty$, the explicit form of the real and imaginary parts of the complex function $Z\left(\infty, n\omega^*\right)$ is given as

$$Z\left(\infty, n\omega^*\right) = Re\left(Z\right) + i\, Im\left(Z\right), \tag{A25}$$

where

$$Re\left(Z\right) = \pi^{-1}\left[R_1\left(n\omega^*\right)D_1\left(n\omega^*\right) + R_2\left(n\omega^*\right)D_2\left(n\omega^*\right)\right], \tag{A26}$$

$$Im\left(Z\right) = \pi^{-1}\left[R_2\left(n\omega^*\right)D_1\left(n\omega^*\right) - R_1\left(n\omega^*\right)D_2\left(n\omega^*\right)\right], \tag{A27}$$

$$R_1\left(n\omega^*\right) = B_1 n\omega^* \int_0^\infty sin\left(n\omega^* t'\right) S_1\left(t'\right) dt' - B_0 \int_0^\infty cos\left(n\omega^* t'\right) S_0\left(t'\right) dt', \tag{A28}$$

$$R_2\left(n\omega^*\right) = B_1 n\omega^* \int_0^\infty cos\left(n\omega^* t'\right) S_1\left(t'\right) dt' + B_0 \int_0^\infty sin\left(n\omega^* t'\right) S_0\left(t'\right) dt', \tag{A29}$$

$$D_1\left(n\omega^*\right) = \int_0^\infty cos\left(n\omega^* \eta\right) P\left(\eta\right) d\eta, \tag{A30}$$

$$D_2\left(n\omega^*\right) = \int_0^\infty sin\left(n\omega^* \eta\right) P\left(\eta\right) d\eta. \tag{A31}$$

## A2. Classical Theodorsen's Problem

Theodorsen (1935) studied a single harmonic motion mode of a thin airfoil in a uniform flow, which is a special case of the result given in Section A1 for the first Fourier mode ($n = 1$). For a harmonically heaving motion of an thin airfoil, the position of the airfoil at the aerodynamic center (denoted by ac) and the chord-averaged velocity $\overline{w}_0\left(t^*\right)$ normal to the chord line at the ac are given by, respectively,

$$z_{ac}\left(t^*\right)=A_z\,exp\left(i\omega^* t^*\right), \tag{A32}$$

$$\overline{w}_0\left(t^*\right)=\frac{dz_{ac}}{dt}=U\left(\frac{A_z}{c}\right)\left(i\omega^*\right)exp\left(i\omega^* t^*\right), \tag{A33}$$

where $\omega^*=\omega c/U$ is the reduced frequency, and $A_z$ is the heaving amplitude. It is noted that $z_{ac}\left(t^*\right)$ and $\overline{w}_0\left(t^*\right)$ are positive when they are in the *z*-direction in the airfoil coordinate system. Using the quasi-steady vortex-sheet strength

$$\gamma_0\left(\overline{x},t^*\right)=-2\overline{w}_0\left(t^*\right)\sqrt{\frac{1-\overline{x}}{\overline{x}}}\,, \tag{A34}$$

we have the quasi-steady circulation is

$$\Gamma_0\left(t^*\right)=c\int_0^1\gamma_0\left(\overline{x},t^*\right)d\overline{x}=-c\,\pi\overline{w}_0\left(t^*\right). \tag{A35}$$

Therefore, the quasi-steady lift is

$$L_0'\left(t^*\right)=\rho Uc\int_0^1\gamma_0\left(\overline{x},t\right)d\overline{x}=\pi\rho U^2 c\left(-\frac{\overline{w}_0\left(t^*\right)}{U}\right). \tag{A36}$$

The added-mass lift is

$$L_1'\left(t^*\right)=\rho c^2\frac{d}{dt}\int_0^1\left(\frac{1}{2}-\overline{x}\right)\gamma_0\left(\overline{x},t\right)d\overline{x}=\frac{\pi}{4}\rho U^2 c\frac{d}{dt^*}\left(-\frac{\overline{w}_0}{U}\right). \tag{A37}$$

Substitution of Eq. (A35) to Eq. (16 leads to

$$\gamma_w\left(s\right)=exp\left(i\omega^* s\right)\left(\frac{A_z}{c}\right)U\left(i\omega^*\right)Y\left(s,\omega^*\right), \tag{A38}$$

where $Y\left(s,\omega^*\right)$ is given by Eq. (A6) for $n=1$. Substitution of Eq. (A38) to Eq. (20) leads to the lift associated with the wake effect, i.e.,

$$L_2'\left(t^*\right)=-\pi\rho U^2 c\left[-\frac{\overline{w}_0\left(t^*\right)}{U}\right]Z\left(\infty,\omega^*\right), \tag{A39}$$

where the lift deficiency function $Z\left(\infty,\omega^*\right)$ is given by Eqs. (A25)-(A31) for $n=1$. Therefore, the circulatory lift is given by

$$L'_{cir}\left(t^*\right)=L'_0\left(t^*\right)+L'_2\left(t^*\right)=\pi\rho U^2 c\left[-\frac{\overline{w}_0\left(t^*\right)}{U}\right]\left[1-Z\left(\infty,\omega^*\right)\right]. \tag{A40}$$

where $1-Z\left(\infty,\omega^*\right)$ is known as the Theodorsen function.

The Theodorsen function is given by

$$Th\left(\omega^*\right)=1-Z\left(\infty,\omega^*\right)=\left[1-Re\left(Z\right)\right]-i\,Im\left(Z\right), \tag{A41}$$

where $Re\left(Z\right)$ and $Im\left(Z\right)$ are given by Eqs. (A26)-(A31) for $n=1$. Figures A1(a) and A1(b) show the global and zoomed-in views of the Theodorsen function given by Eq. (A41), respectively, in comparisons with Theodorsen's exact solution and tabularized data (Theodorsen 1935; Garrick 1938). Here, the reduced frequency $\omega^*=\omega c/U$ based on the chord length is used, while the reduced frequency $k=\omega c/2U=\omega^*/2$ based on the half-chord length is used in the Theodorsen's original paper. The exact closed-form expression of the Theodorsen function is $Th\left(k\right)=H_1^{(2)}\left(k\right)\left[H_1^{(2)}\left(k\right)+iH_0^{(2)}\left(k\right)\right]^{-1}$, where $H_\gamma^{(n)}\left(k\right)$ is the Hankel function (the modified Bessel function of the third kind). In the numerical calculations of the integrals in Eqs. (A26)-(A31), the upper limit is set at $t^*=800$. The trapezoidal rule is used by partitioning the integration domain into $10^6$ position points and sufficiently small intervals are required to evaluate the singular integrals. The approximate solution in Fig. A1(b) shows some small fluctuations that are related to sine and cosine functions in Eqs. (A26)-(A31). In general, as the upper limit $t^*$ of the integrals in Eqs. (25)-(31) increases, the result will be improved. Figure A2 shows the effect of the upper limit $t^*$ of the integrals on the calculation of the Theodorsen function.

The lift of the airfoil is expressed as

$$L'(t^*) = L'_0(t^*) + L'_1(t^*) + L'_2(t^*)$$
$$= \frac{1}{2}\rho U^2 c\left[2\pi\,\alpha_{eff}(t^*)Th(\omega^*) + \frac{\pi}{2}\frac{d\alpha_{eff}(t^*)}{dt^*}\right], \tag{A42}$$

where $\alpha_{eff}(t^*) = -\bar{w}_0 / U$ is generally interpreted as the effective AoA. The non-dimensional circulation is $\Gamma_0(t^*)/cU = \pi\,\alpha_{eff}(t^*)$. The first and second terms in the RHS of Eq. (A42) are the circulatory lift and the added-mass lift, respectively. The lift coefficient is

$$C_l = \frac{L'(t^*)}{\rho U^2 c/2} = 2\pi\,\alpha_{eff}(t^*)Th(\omega^*) + \frac{\pi}{2}\frac{d\alpha_{eff}(t^*)}{dt^*}. \tag{A43}$$

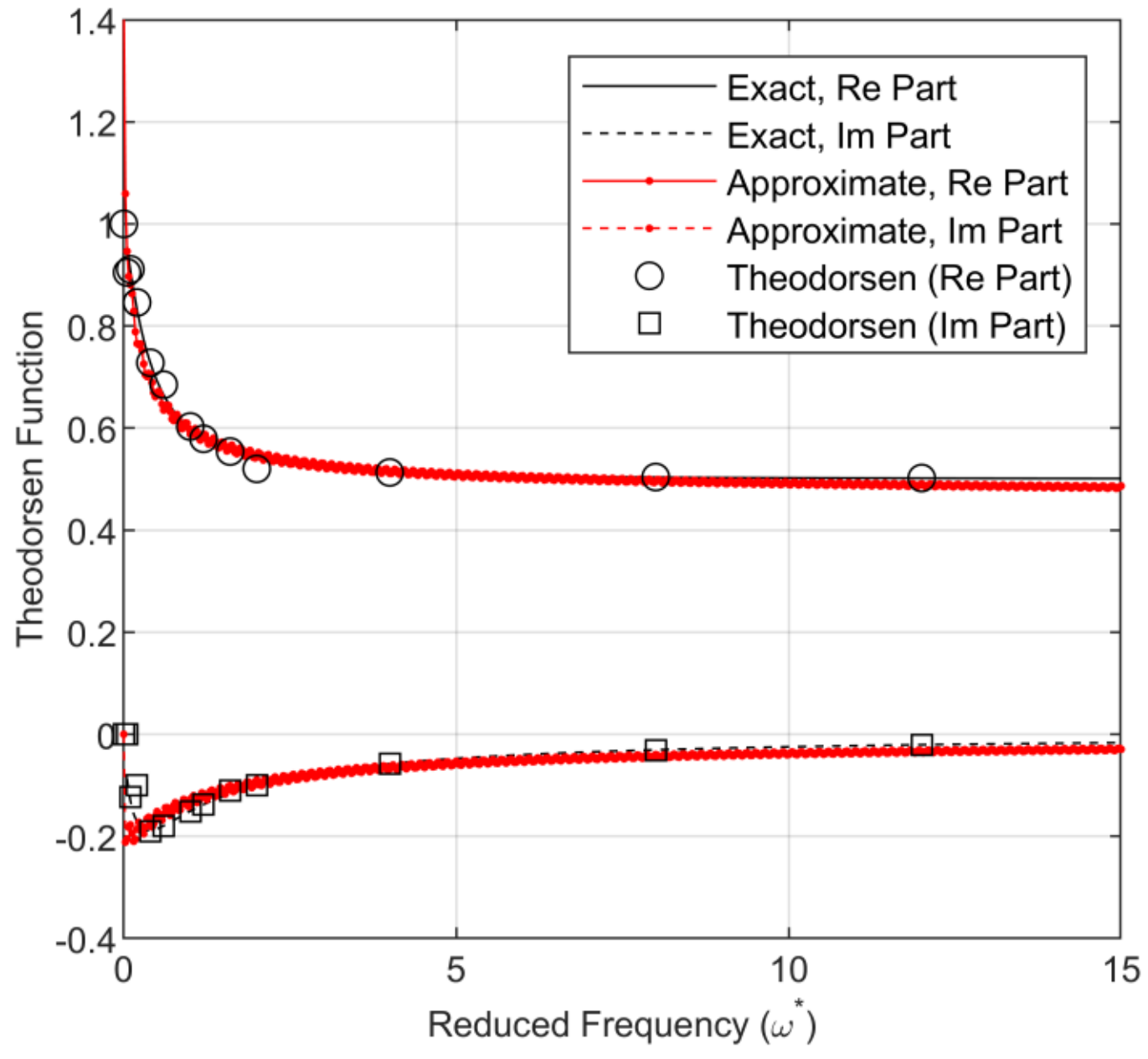

(a)

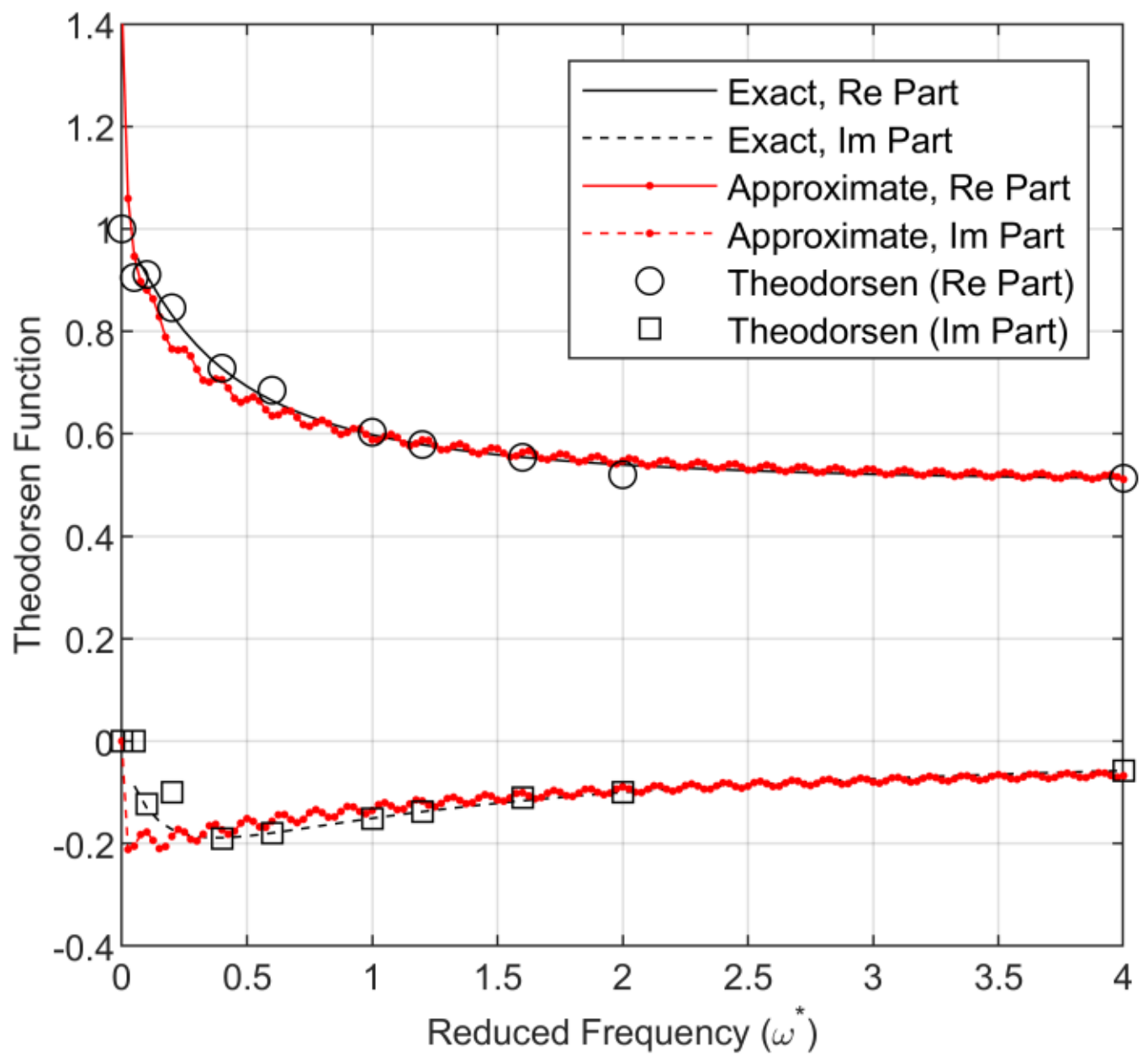

(b)

Figure A1. The Theodorsen function given by the approximate analytical solution in comparisons with Theodorsen's exact closed-form solution and tabularized data. (a) Global view in $\omega^* = 0\text{-}15$, and (b) zoomed-in view in $\omega^* = 0\text{-}4$

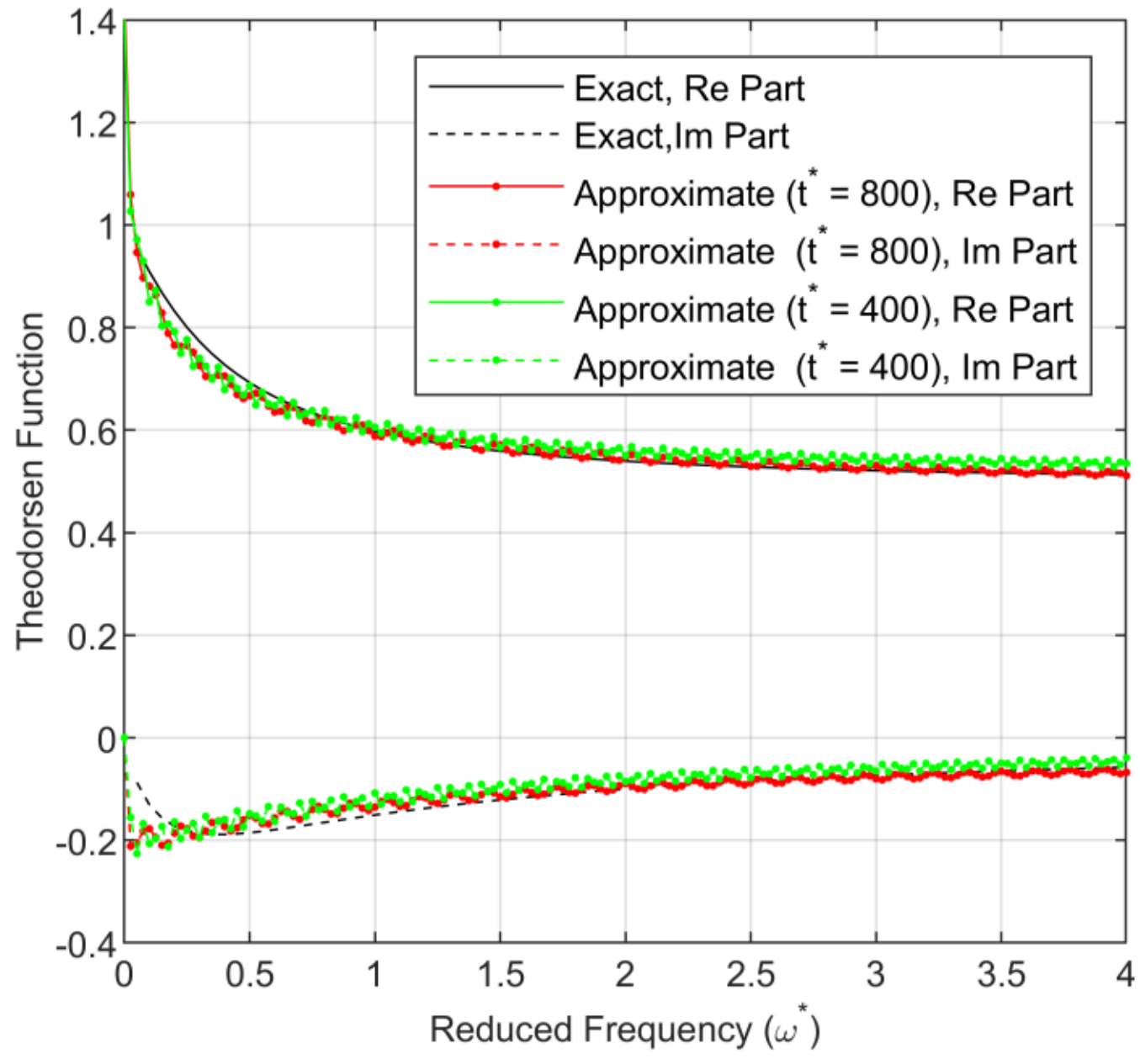

Figure A2. Effect of the upper limit $t^*$ on calculation of the Theodorsen function given by the approximate analytical solution.

## Appendix B. Computational Setup and Code Validation

In the present study, numerical simulations were performed using the open-source finite-volume code OpenFOAM (Weller et al. 1998). The incompressible Navier-Stokes equations were solved by using the PIMPLE algorithm, a combined PISO-SIMPLE approach suitable for transient flows with moving meshes. Spatial discretization utilized a second-order upwind scheme for the convective terms and a second-order central difference scheme for the diffusive terms. Pressure interpolation was performed using a linear scheme, and gradients were reconstructed using a cell-limited Gauss linear scheme to ensure boundedness. The 2D computational domain consisted of two overlapping regions: a background Cartesian mesh and a body-fitted region around the flat plate. The mesh contained approximately 13.4 million cells, with local refinement near the plate surface. A boundary-layer mesh comprising 20 layers was applied along the plate, with a first-layer thickness of the non-dimensional wall-normal coordinate $y^+ < 1$ and a growth ratio of 1.1. The overset mesh interface was handled using the built-in overset capability in OpenFOAM, with linear interpolation between donor and acceptor cells.

The flat plate was set into translational motion along the streamwise direction. The flat-plate velocity followed a smoothed piecewise function given by Eq. (47). No-slip conditions were enforced on the flat-plate surface, while symmetry conditions were applied at the inlet and outlet boundaries. The top and bottom boundaries were treated as free-slip walls. A time step of $\Delta t = 0.005$ as selected based on the CFL condition below 0.5, ensuring temporal resolution sufficient to capture the initial vortex development and subsequent shedding dynamics. Simulations were advanced for a total physical time of $t^* = 8$ to capture the initial transient and establish periodic shedding where present. Grid and time-step independence were verified through systematic refinement studies.

To establish the accuracy of the present solver and setup, two classic benchmarks of an impulsively started flat plate at the AoA ($\alpha$) of $90°$ were simulated. Figure B1 shows the normalized vorticity contour maps of an impulsive flat plate with $\alpha = 90°$ at $t^* = 1$ and $t^* = 8$ for $Re = 40$ and $Re = 126$. The drag coefficient $C_d$ and the normalized wake vortex length $s/c$ for $Re = 40$ and $Re = 126$ were computed and compared with the numerical results of Koumoutsakos & Shiels (1996). As shown in Fig. B2, the present results show good agreement with the reference data. The root-mean-squared (RMS) errors of $C_d$ for $Re = 40$ and $Re = 126$ are 0.32 and 0.176, respectively, which correspond to the relative RMS errors of about 16% and 10% based on their asymptotic steady-state values, respectively. The RMS errors of the normalized wake vortex length $s/c$ for $Re = 40$ and $Re = 126$ are 0.04 and 0.026, respectively, which correspond to the relative RMS errors of about 2% and 1% based on their median values, respectively. This close comparison confirms the capability of the present numerical framework to accurately capture the viscous, unsteady flow phenomena central to this investigation.

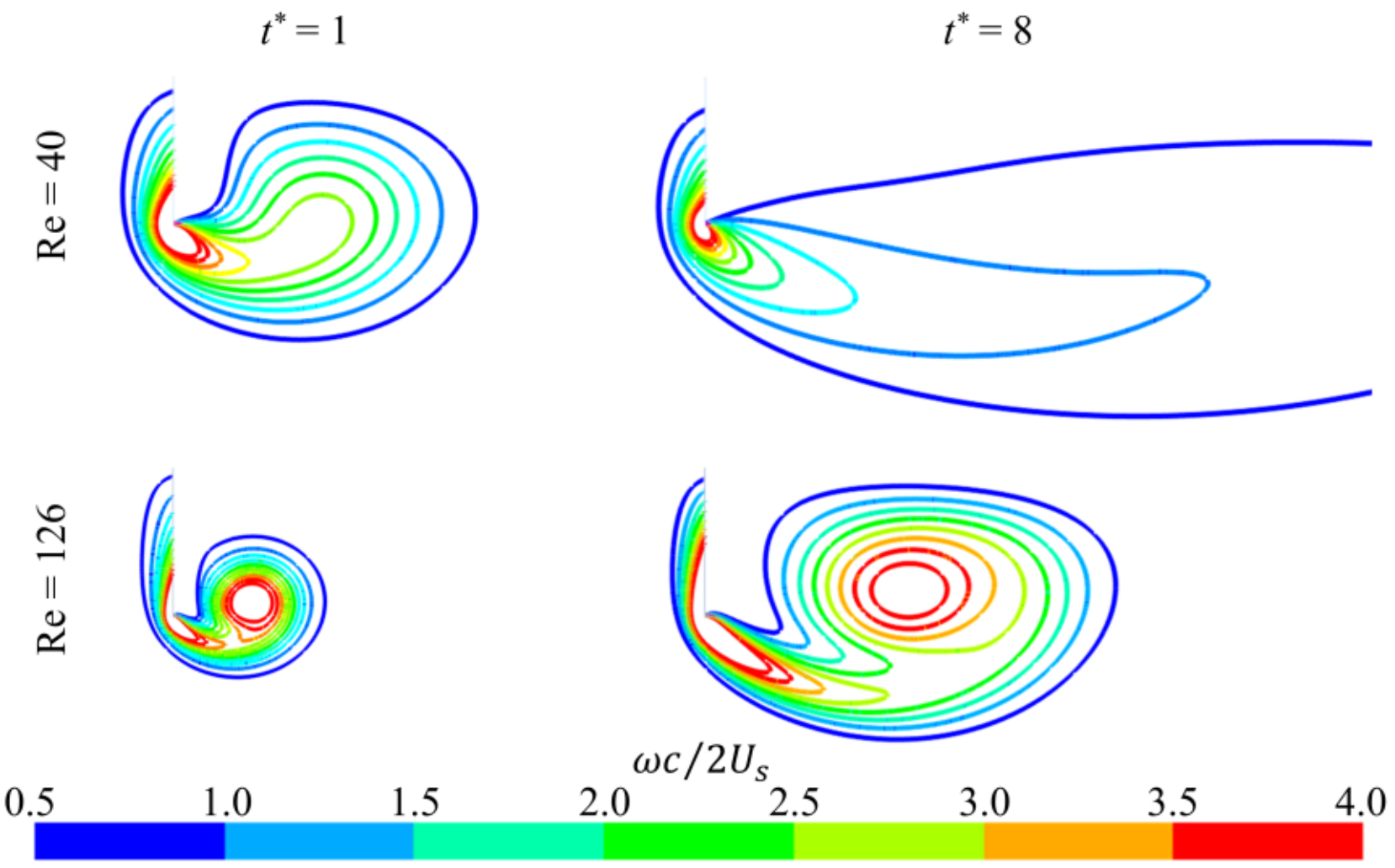

Figure B1. The normalized vorticity contour maps of an impulsive flat plate with $\alpha = 90^{\circ}$ for $Re = 40$ and $Re = 126$.

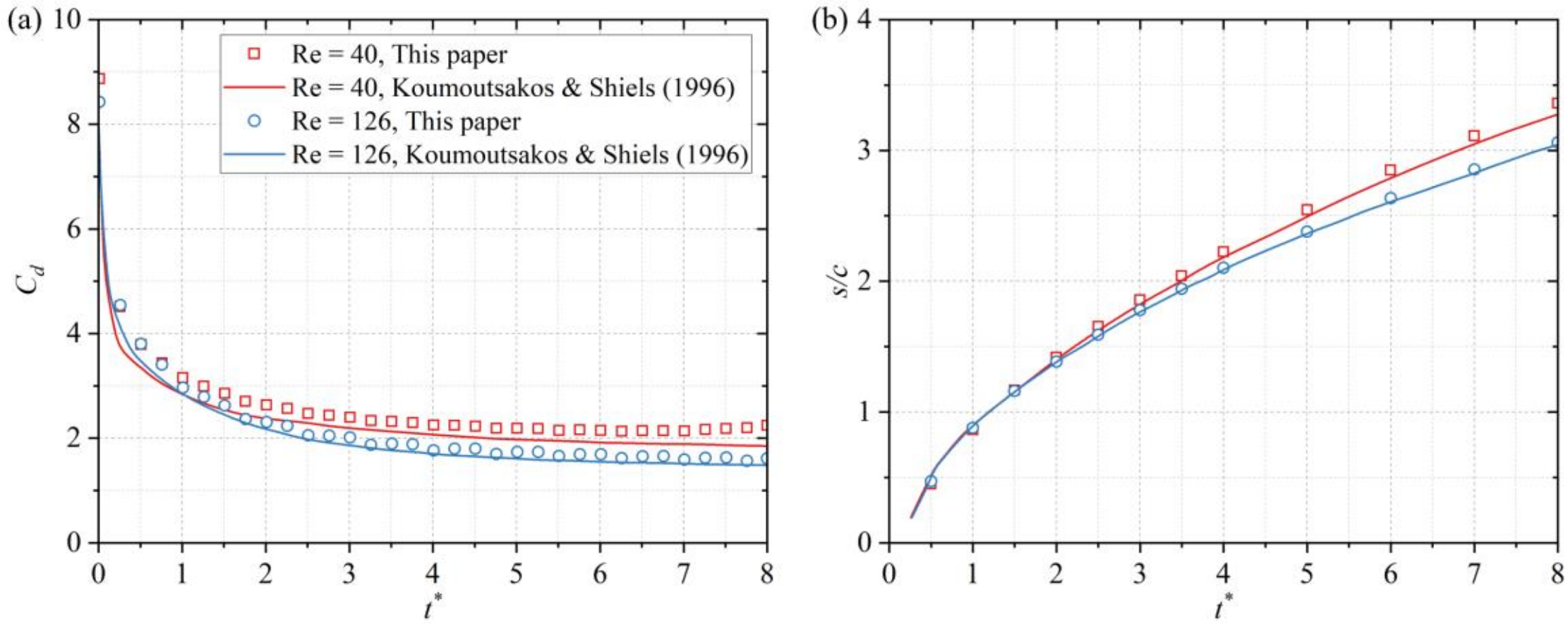

Figure B2. Time histories of (a) the drag coefficient and (b) the normalized wake vortex length, recorded during the translational motion of the flat plate with $\alpha = 90^{\circ}$ in the streamwise direction for $Re = 40$ and $Re = 126$.

## Disclosure statements

Acknowledgements

This work is supported by The John O. Hallquist Endowed Professorship and Presidential Innovation Professorship at Western Michigan University.

Funding statement

T.L. is supported by The John O. Hallquist Endowed Professorship and Presidential Innovation Professorship at Western Michigan University.

Competing Interests (compulsory)

There is no competing interest.

Data availability statement

Data will be available upon requesting.

Author ORCID

T.L.: 0000-0001-6297-1660

Author contributions

T.L. derived the theory. J.L. and S.W. performed numerical simulations.